\pdfoutput=1
\documentclass[a4paper,fleqn,usenatbib]{article}

\usepackage{amsmath}
\usepackage{abbreviations}
\usepackage{amssymb}
\usepackage{txfonts}
\usepackage{natbib}
\usepackage{hyperref}
\usepackage[mathcal]{euscript}
\usepackage[T1]{fontenc}
\usepackage{ae,aecompl}
\usepackage{graphicx}
\usepackage{subfig}
\usepackage{color}
\usepackage[utf8]{inputenc}
\usepackage{authblk}
\usepackage{multirow}
\usepackage{tensor}
\usepackage{mhchem}
\usepackage[a4paper,margin=2.5cm]{geometry}

\DeclareMathAlphabet{\mathpzc}{OT1}{pzc}{m}{it}
\newcommand{\pivec}{\ensuremath{\boldsymbol{\pi}}}
\newcommand{\xvec}{\ensuremath{\textbf{x}}}
\newcommand{\uvec}{\ensuremath{\textbf{u}}}
\newcommand{\Avec}{\ensuremath{\textbf{A}}}
\defcitealias{bacchini2018a}{Part I}

\begin{document}
\title{Generalized, energy-conserving numerical simulations of particles in general relativity. II. Test particles in electromagnetic fields and GRMHD}
\author[1]{F. Bacchini\thanks{E-mail: fabio.bacchini@kuleuven.be}}
\author[1,2]{B. Ripperda}
\author[2]{O. Porth}
\author[3]{L. Sironi}
\affil[1]{Centre for mathematical Plasma Astrophysics, Department of Mathematics, KU Leuven, Celestijnenlaan 200B, B-3001 Leuven, Belgium}
\affil[2]{Institut f\"{u}r Theoretische Physik, Max-von-Laue-Str. 1, D-60438 Frankfurt, Germany}
\affil[3]{Department of Astronomy, Columbia University, 550 W 120th St, New York, NY 10027, USA}
\renewcommand\Authands{ and }

\label{firstpage}
\maketitle

\begin{abstract}
Observations of compact objects, in the form of radiation spectra, gravitational waves from LIGO/Virgo, and direct imaging with the Event Horizon Telescope, are currently the main information sources on plasma physics in extreme gravity. Modeling such physical phenomena requires numerical methods that allow for the simulation of microscopic plasma dynamics in presence of both strong gravity and electromagnetic fields. In \cite{bacchini2018a} we presented a detailed study on numerical techniques for the integration of free geodesic motion. Here we extend the study by introducing electromagnetic forces in the simulation of charged particles in curved spacetimes. We extend the Hamiltonian energy-conserving method presented in \cite{bacchini2018a} to include the Lorentz force and we test its performance compared to that of standard explicit Runge-Kutta and implicit midpoint rule schemes against analytic solutions. Then, we show the application of the numerical schemes to the integration of test particle trajectories in general relativistic magnetohydrodynamic (GRMHD) simulations, by modifying the algorithms to handle grid-based electromagnetic fields. We test this approach by simulating ensembles of charged particles in a static GRMHD configuration obtained with the Black Hole Accretion Code (\texttt{BHAC}).
\end{abstract}

\section{Introduction}
\label{sec:intro}
Over the course of the last few decades, observational measurements of high-energy astrophysical environments have shed light on the complex dynamics of plasmas surrounding neutron stars and black holes. The information retrieved from observations has deeply contributed to the characterization of plasma phenomena in the magnetosphere and accretion disk of objects such as Sgr A*, the supermassive black hole at the Galactic Center (\citealt{doeleman2008}; \citealt{johnson2015}). Forthcoming observations will probe the event horizon scales, allowing for the direct imaging of the accretion flow around the black hole (\citealt{goddi2017}). The theoretical investigation of the mechanisms at the origin of the observed radiation spectra is typically carried out with computer codes that solve the equations of general relativistic ideal magnetohydrodynamics (GRMHD). Non-ideal effects can be introduced via extensions of the ideal GRMHD framework, e.g.\ by including resistivity (\citealt{dumbserzanotti2009}; \citealt{bucciantinidelzanna2012}; \citealt{palenzuela2013}; \citealt{dionysopoulou2015}; \citealt{qian2016}; Ripperda et al., in prep.), or radiation feedback (\citealt{mckinney2014}; \citealt{ryan2015}; \citealt{ryan2017}; \citealt{sadowski2017}). The GRMHD approach is adequate for the large-scale description of thermal plasmas, i.e.\ where the particle distribution is implicitly assumed to be Maxwellian (\citealt{moscibrodzkafalcke2013}).

GRMHD codes have proved extremely useful in the investigation of relativistic accretion flows and jets from compact objects from the macroscopic perspective. However, the dynamics of plasmas at the particle scale remains largely unexplored. Nonthermal processes associated to the dynamics of particles play an important role in determining specific observational features, e.g.\ strong X-ray flares with hard energy spectra, variability, and radio spectral slopes (\citealt{chael2018}). The GRMHD framework, by definition, cannot reproduce nonthermal phenomena associated with accelerated particles, since the particle distribution is implicitly assumed to be a Maxwellian. This implies that ad hoc prescriptions for the particle energy are needed, in order to avoid mismatches between observational measurements and the most advanced simulation results (see e.g.\ \citealt{sadowski2017}). Approaches based on the assumption of specific energy distributions have been successful in reproducing some of the nonthermal features (\citealt{porth2011}; \citealt{ressler2017}; \citealt{ball2018}; \citealt{chael2018}; \citealt{davelaar2018}), without however providing a first-principle description of the plasma dynamics at the microscopic level.

Investigating kinetic phenomena at the particle scales requires particle-based numerical methods. Numerical methods that evolve the particle distribution function directly (``Vlasov methods'') have been developed, although only for nonrelativistic calculations (see e.g.\ \citealt{palmroth2013}). However, these typically exhibit computational costs that far exceed those of algorithms that model the evolution of the distribution function by integrating the equations of motion of computational particles. General relativistic particle-based methods are gaining attention as the basis for the next generation of simulation codes for astrophysical plasma applications (\citealt{levinsoncerutti2018}). Simulating plasmas with kinetic approaches allows for the self-consistent description of nonthermal phenomena from first principles. The results of particles simulations can be used for studying phenomena such as magnetic reconnection in the collisionless regime (see e.g., \citealt{guo2014}; \citealt{sironispitkovsky2014}; \citealt{werner2017}; \citealt{werner2018}), or as a physically accurate input for GRMHD simulations (via heating prescriptions for electrons, or for radiative transport calculations), eliminating the need for assumptions on the particle energy spectra (\citealt{chael2018}). Advanced numerical methods are needed for particle simulations, ranging from test particle approaches, where the energy content of the particle population is considered to be negligible with respect to that of an underlying GRMHD background (\citealt{ripperda2017a}; \citealt{ripperda2017b}), to fully consistent Particle-in-Cell (PiC) simulations, where the feedback of the particles on the electromagnetic fields is taken into account. In all cases, it is necessary to find the most accurate algorithm to evolve populations of particles under the effect of both strong gravitational and electromagnetic fields. Applications to astrophysical environments range from pulsar magnetospheres and winds (\citealt{sironispitkovsky2011}; \citealt{philippov2015}) to accretion flows around supermassive black holes (\citealt{rowan2017}; \citealt{ball2018}; \citealt{werner2018}), with the aim of gaining insight into plasma phenomena at the microscopic scales.

In \cite{bacchini2018a} (Part I from here on) we presented a detailed study on numerical integrators for massive and massless particles under the effect of spacetime curvature. Here, we extend our study by including the effect of electromagnetic fields on the motion of charged particles. In our analysis, we consider standard explicit integration methods, implicit symplectic methods, and a new, implicit energy-conserving integrator that is a direct extension of that used in \citetalias{bacchini2018a}, modified to include the Lorentz force. Here, we also present a modification of the energy-conserving scheme that is suitable for implementation in GRMHD codes. We test each numerical integrator against analytic solutions in a variety of spacetimes and electromagnetic field configurations. Finally, we apply the numerical schemes to a representative example simulation of test particles in GRMHD, using a snapshot of a two-dimensional simulation of plasma accreting around a spinning black hole, produced with the Black Hole Accretion Code (\texttt{BHAC}, \citealt{porth2017}).

The paper is organized as follows: in Section \ref{sec:theory} we discuss the set of equations governing the motion of charged particles in curved spacetimes. In Section \ref{sec:schemes} we briefly review the characteristics of the aforementioned numerical schemes. In Section \ref{sec:tests} we test all integrators against analytic solutions in idealized setups. Finally, in Section \ref{sec:BHAC} we show an example simulation of test particles in GRMHD. Our main results and conclusions are discussed in Section\ref{sec:discussionsummary}.

\section{Charged particles in electromagnetic and gravitational fields}
\label{sec:theory}
The motion of charged, massive particles under the combined effect of strong spacetime curvature and external electromagnetic fields is governed by the equation of motion
\begin{equation}
 \frac{d^2 x^\mu}{d\tau^2} + \Gamma^\mu_{\lambda\sigma} \frac{dx^\lambda}{d\tau} \frac{dx^\sigma}{d\tau} = \frac{q}{m}g_{\nu\rho}F^{\mu\nu}\frac{dx^\rho}{d\tau},
 \label{eq:geodesic}
\end{equation}
where the four-position $x^\mu$ ($\mu=0,1,2,3$) evolves in proper time $\tau$ under the influence of two mechanisms, namely the geodesic motion (left-hand side) and the electromagnetic force (right-hand side). The Christoffel symbol $\Gamma^\mu_{\lambda\sigma}$ expresses the connection of a general Riemannian manifold associated to the metric tensor $g_{\mu\nu}$, and describes the motion of particles along geodesics. Deviations from geodesic motion are accounted for by the external force term $(q/m)F^{\mu\nu}dx^\nu/d\tau$, thus manifesting more strongly for particles with larger charge-to-mass ratio $q/m$. For static electromagnetic fields the covariant antisymmetric Maxwell tensor is
\begin{equation}
 F_{\mu\nu} = g_{\mu\alpha}g_{\nu\beta}F^{\alpha\beta} = \partial_\mu A_\nu - \partial_\nu A_\mu,
 \label{eq:maxwelltensor}
\end{equation}
obtained from the electromagnetic four-potential $A_\mu$. The derivative of the four-position with respect to $\tau$ is the four-velocity
\begin{equation}
 u^\mu:=\frac{dx^\mu}{d\tau},
\end{equation}
where $t:=x^0$ is the coordinate time, and therefore $u^0:=dt/d\tau$.

In this work, we adopt the 3+1 ADM formalism (e.g., \citealt{rezzollazanotti}) to define the lapse function $\alpha$, the shift three-vector $\beta^i$, and the spatial part of the metric $\gamma_{ij}$ (with $\gamma^{ij}$ its algebraic inverse). We choose a $(-,+,+,+)$ signature for the metric, and rewrite the second-order equation \eqref{eq:geodesic} into a set of first-order differential equations in the variables $x^i$ and $u_i=g_{i\mu}u^\mu$,
\begin{equation}
\frac{d x^i}{dt} = \gamma^{ij} \frac{u_j}{u^0} - \beta^i,
\label{eq:geodesic3p1x}
\end{equation}
\begin{equation}
\frac{du_i}{dt} = -\alpha u^0 \partial_i \alpha + u_k \partial_i \beta^k - \frac{u_j u_k}{2 u^0} \partial_i \gamma^{jk} + \frac{q}{m}F_{i\mu}\frac{u^\mu}{u^0},
\label{eq:geodesic3p1u}
\end{equation}
where 
\begin{equation}
u^0 = \left(\epsilon+\gamma^{jk} u_j u_k\right)^{1/2}/\alpha,
\label{eq:lfac}
\end{equation}
and $\epsilon=1$ for massive particles. Note that from equation \eqref{eq:geodesic3p1x}, the relation between $u_i$ and $u^i = dx^i/d\tau$ reads
\begin{equation}
 u_i = \gamma_{ij}u^j+u^0\beta_i.
 \label{eq:udownfromuup}
\end{equation}

The system of equations \eqref{eq:geodesic3p1x}-\eqref{eq:geodesic3p1u} can be more conveniently handled than the initial equation \eqref{eq:geodesic}. First, it reduces the problem to a system of six first-order equations, where there is no need to integrate temporal components. Second, it can be shown that the conservation of the norm of the four-velocity, $u_\mu u^\mu=-\epsilon$, is automatically satisfied. Third, expressing the evolution of the particle position and velocity with respect to coordinate time allows for matching the motion of particles with the time evolution of global quantities (e.g., electromagnetic fields).

Equation \eqref{eq:geodesic3p1u} differs from the case of pure geodesic motion by the force term $(q/m)F_{i\mu}u^\mu/u^0$. In the ADM framework, this term is often conveniently rearranged so that the electromagnetic contribution resembles the form of the Lorentz force in the special relativistic limit (equation \eqref{eq:lforceSR} below). For this purpose we define the quantities (\citealt{komissarov2011})
\begin{equation}
 D^i = \alpha F^{0i},
\end{equation}
\begin{equation}
 H_i = \frac{1}{2}\alpha e_{ijk} F^{jk},
\end{equation}
\begin{equation}
 E_i = F_{i0},
\end{equation}
\begin{equation}
 B^i = \frac{1}{2}e^{ijk} F_{jk},
\end{equation}
where the Levi-Civita (pseudo-)tensor $e_{ijk} = \varepsilon_{ijk} \sqrt{\gamma}$ is given by the determinant of the spatial three-metric, $\gamma = \textrm{det}(\gamma_{ij})$, and the antisymmetric tensor $\varepsilon_{ijk}$. Similarly, $e_{ijk} = \varepsilon^{ijk}/\sqrt{\gamma}$. The four field variables $D^i$, $H_i$, $E_i$, $B^i$ are not independent of each other, but are related by
\begin{equation}
 E_i = \alpha\gamma_{ij}D^j + e_{ijk}\beta^j B^k,
 \label{eq:EfromD}
\end{equation}
\begin{equation}
 H_i = \alpha\gamma_{ij} B^j - e_{ijk}\beta^j D^k,
 \label{eq:HfromB}
\end{equation}
or, alternatively, by the inverse relations
\begin{equation}
 \alpha D^i = \gamma^{ij}E_j - \gamma^{ij} e_{jkl}\beta^k B^l,
\end{equation}
\begin{equation}
 \alpha B^i = \gamma^{ij} H_j + \gamma^{ij} e_{jkl}\beta^k D^l.
\end{equation}
Through the relations \eqref{eq:EfromD}-\eqref{eq:HfromB}, Maxwell's equations can be effectively reduced to the time evolution of only two dynamic fields, $D^i$ and $B^i$ (\citealt{komissarov2011}), with the former being only a derived quantity (hence not needing an evolution equation) in the ideal MHD limit. This strategy is commonly adopted in GRMHD codes (e.g.\ \citealt{gammie2003}; \citealt{porth2017}), and is suitable for the numerical integration of time-varying fields.

The definitions above can be employed to express the Lorentz force term in equation \eqref{eq:geodesic3p1u} in terms of the dynamic fields $D^i$ and $B^i$. By expanding the tensor product we have
\begin{equation}
\begin{aligned}
  F_{i\mu} \frac{u^\mu}{u^0} & = F_{i0} + F_{ij}\frac{u^j}{u^0} \\
  & = E_i + e_{ijk} \frac{u^j}{u^0}B^k \\
  & = \alpha\gamma_{ij} D^j + e_{jik} \frac{\gamma^{jl}u_l}{u^0} B^k,
\end{aligned}
\end{equation}
where we have used equations \eqref{eq:udownfromuup} and \eqref{eq:EfromD}. The equation of motion \eqref{eq:geodesic3p1u} for a charged particle now reads
\begin{equation}
 \frac{du_i}{dt} = -\alpha u^0 \partial_i \alpha + u_k \partial_i \beta^k - \frac{u_j u_k}{2 u^0} \partial_i \gamma^{jk} + \frac{q}{m}\left(\alpha\gamma_{ij}D^j+e_{ijk}\frac{\gamma^{jl}u_l}{u^0}B^k\right),
\label{eq:geodesic3p1uEB}
\end{equation}
and combined with equation \eqref{eq:geodesic3p1x} for the position, it forms a set of six coupled, nonlinear equations explicitly involving only $x^i$, $u_i$, and the dynamic fields $D^i$, $B^i$. The special relativistic limit is retrieved by setting $\alpha=1$, $\beta^i=0$, $\gamma^{ij}=\eta^{ij}$ (where $\eta^{ij}$ is the Minkowski three-metric for flat spacetime), as
\begin{equation}
 \frac{d\textbf{u}}{dt} = \frac{q}{m}\left(\textbf{D}+\frac{\textbf{u}}{\Gamma}\times\textbf{B}\right).
 \label{eq:lforceSR}
\end{equation}
where $\Gamma=u^0$ is the special relativistic Lorentz factor.

For numerical calculations, one can freely choose the formulation of the equations to be integrated depending on the preferred computational scheme. In this work, we will mainly consider two expressions of the equations of motion. When simulating charged particles in analytically defined electromagnetic fields (i.e., known at every position in space), for accuracy it is desirable to employ a formulation involving the four-potential. In this case one can employ equation \eqref{eq:geodesic3p1u}, with the Maxwell tensor calculated from \eqref{eq:maxwelltensor}. If the four-potential is not available, e.g.\ when the electromagnetic field data is defined on a numerical grid (as is the case for GRMHD simulations) one can make use of equation \eqref{eq:geodesic3p1uEB}.

In this work, we consider only stationary metrics and static fields with no dependence on the coordinate time $t$, i.e.\ the metric functions $\alpha$, $\beta^i$, $\gamma^{ij}$, and the four-potential $A_\mu$ are functions of the position $x^i$ only. This choice allows us to test the robustness of numerical integrators with respect to energy conservation, but it does not represent a limitation of their applicability, which extends to time-varying electromagnetic and gravitational fields. Time-invariance of the metric and four-potential implies, in all cases, the existence of at least one Killing vector field, $K^\mu=(1,0,0,0)$, representing symmetry with respect to time-translations. As a consequence, for any particle there exists a conserved quantity $-K^\mu \pi_\mu = -\pi_0 = E$ which we label as the total energy. Here, $\pi_0$ is the 0-th component of the (normalized) conjugate momentum, $\pi_0 = u_0+qA_0/m$, with $u_0=-\alpha^2 u^0+\beta^i u_i$ (see Section \ref{sec:ham}). A shown in \citetalias{bacchini2018a}, conservation of energy plays an important role in numerical simulations of geodesic motion. Here we consider the additional effect of electromagnetic fields, and we therefore expect that the choice of computational method and the associated numerical errors similarly impact the simulation results, as will be demonstrated in the next Sections. This is also motivated by the  well-known properties of numerical integrators for special relativistic charged particles, that can be heavily affected by spurious non-conservation of energy due to numerical errors (\citealt{ripperda2018}).

\section{Numerical methods}
\label{sec:schemes}
The motion of charged particles in electromagnetic fields is a particularly difficult challenge from the numerical point of view. The main difficulty is generally represented by the separation of time scales between the motion of the gyro-center along magnetic field lines and the gyration around them. This requires particularly robust numerical methods that can describe both the gyro-motion and the acceleration of the gyro-center accurately.

At the same time, in curved spacetimes, the motion of massive particles is influenced by the gravitational field. The description of the resulting motion is in general complicated, as are the equations describing it. Typical motion around compact objects, for instance, is represented by several types of bound (possibly unstable) orbits. For such a periodic motion, numerical methods that are capable of keeping a particle on the correct orbit for long times are ideal, as they produce more physically accurate results (see \citetalias{bacchini2018a}).

When the effects of gravitational and electromagnetic fields are combined, it is necessary to employ a method that is both robust in handling different time scales, and that exhibits long-term accuracy and reliability. In the next Sections, we briefly review standard available numerical methods, and we introduce a new, exactly energy-conserving method derived from the Hamiltonian formalism.

\subsection{Explicit non-symplectic methods}
\label{sec:RK4}
Explicit methods for ordinary differential equations (ODEs) advance the numerical solution in a finite number of non-iterative steps. One of the most successful explicit methods in scientific computing is the fourth-order Runge-Kutta method (RK4 from now on, see e.g \citealt{press}). At each time step, the scheme requires four evaluations of the right-hand side of the discretized ODE, resulting in an error of order $O(\Delta t^5)$. Explicit methods such as RK4 are generally incapable of  preserving first integrals of the system, such as the associated Hamiltonian, if this exists. Another well-known issue of explicit, non-symplectic schemes is the non-conservation of phase space volume (see e.g.\ \citealt{hairer}; \citealt{fengqin}). The error in these quantities accumulates unboundedly, and for integration over long times the resulting computed solution becomes unacceptably inaccurate. However, for a widely used method such as the RK4 scheme, the scaling of errors associated with a reduction in $\Delta t$ is satisfactory enough to be generally acceptable. In \citetalias{bacchini2018a}, the RK4 scheme was tested against simulations of pure geodesic motion. In that case, long-term simulations required extreme reductions of the time-step in order to preserve accuracy, especially for unstable orbital motion.

For applications to the motion of special relativistic particles in electromagnetic fields, explicit non-symplectic schemes are typically discarded due to the fast degradation of the description of the gyro-motion (\citealt{qin2013}; \citealt{ripperda2018}). In Section \ref{sec:wald}, we will show that this problem manifests in general relativistic simulations as well, hence we expect that in practical applications, the RK4 scheme cannot be applied without unacceptable loss of accuracy in the results.

\subsection{Implicit symplectic methods}
\label{sec:IMR}
Simple explicit schemes such as RK4 lack symplecticity, i.e.\ the capability of an integrator to preserve trajectories in the phase space. A symplectic scheme presents the additional advantage of preserving first integrals of motion (such as the energy) with an error that is bounded in time, i.e.\ no secular growth of energy errors is observed (\citealt{hairer}). The remaining error depends on the order of the method, and it decreases exponentially with the integration step.

While explicit symplectic schemes can be constructed, they are generally not applicable to systems characterized by inseparable Hamiltonians (see Section \ref{sec:ham}), resulting  in energy errors that are not bounded in time (although the increase in error is generally very slow, see \citealt{tao2016}). Instead, one can rely on implicit symplectic schemes, such as the implicit midpoint rule (IMR from now on), which is the simplest second-order, symplectic, implicit integration scheme (\citealt{hairer}).

Implicit schemes such as (or slight variations of) the IMR  are typically applied to the motion of charged particles in special relativistic simulations. This is motivated by the fact that the form of the Lorentz force in the special relativistic limit allows, for specific discretization choices, for the formal inversion of the equation of motion (\citealt{boris1970}; \citealt{vay2008}; \citealt{higueracary2017}). As a consequence, even though the discretization scheme is implicit, the solution procedure is actually explicit (non-iterative). There also exist discretizations that are non-invertible, and that require an iterative solution procedure (hence a higher computational cost); these can present desirable features such as energy conservation in global simulations, e.g.\ for Particle-in-Cell (PiC) codes (\citealt{lapentamarkidis2011}; see \citealt{ripperda2018} for a review of available special relativistic particle integrators). Generally speaking, iterative or non-iterative implicit schemes exhibit stability and bounded energy errors, hence they are usually preferred to explicit schemes such as the RK4.

For charged particles in general relativistic contexts, the nonlinear nature of equation \eqref{eq:geodesic3p1u} prevents the formal inversion when the IMR scheme is applied. Therefore, iterative algorithms are the only possible choice for advancing the solution to the next time step. In \citetalias{bacchini2018a}, the IMR scheme was applied to pure geodesic motion. In such a context, the results are generally accurate, but exhibit errors typically one order of magnitude larger than those affecting results obtained with the RK4 and the Hamiltonian schemes, at least for integration over short times or very unstable orbital paths. For integration over very long times, instead, both the IMR and Hamiltonian schemes can keep energy errors bounded, proving superior to non-symplectic explicit schemes.

\subsection{Implicit energy-conserving methods}
\label{sec:ham}
The symplectic nature of implicit schemes such as the IMR implies conservation of first integrals of the motion (e.g., the total energy) to a degree that scales with the integration step. Energy errors, although bounded in time, are nonzero, and can prove detrimental in some situations. For pure geodesic motion (see \citetalias{bacchini2018a}), we demonstrated how unstable photon orbits are strongly affected by such errors, which prevent schemes like the IMR from producing accurate results unless decreasing the time step by orders of magnitude. In such cases, a second-order numerical scheme that can conserve energy outperforms the second-order IMR, and even the higher-order RK4 scheme. In \citetalias{bacchini2018a}, an energy-conserving scheme has been derived, for pure geodesic motion, based on a Hamiltonian formalism.

Here, we extend the same argument to the motion of charged particles in general relativistic gravitomagnetic fields. The 3+1 Hamiltonian for a particle of mass $m$ and charge $q$, subjected to electromagnetic and gravitational forces, is written
\begin{equation}
 H(\textbf{x},\pivec) = \alpha\sqrt{1 + \gamma^{ij} \left(\pi_i-\frac{q}{m}A_i\right) \left(\pi_j-\frac{q}{m}A_j\right)} - \beta^k\left(\pi_k-\frac{q}{m}A_k\right) - \frac{q}{m}A_0,
 \label{eq:ham}
\end{equation}
where $\pi_i=u_i+qA_i/m$ is the conjugate momentum counterpart of $u_i$. As mentioned above, we consider static fields such that $A_\mu$ is a function of $x^i$ only. The Hamiltonian $H$ represents the total energy of the particle, and is therefore conserved in time. This can be shown by using the definition of the conjugate momentum,
\begin{equation}
\begin{aligned}
 \pi_0 &= u_0+\frac{q}{m}A_0 \\
 & = -\alpha^2 u^0 + \beta^i u_i + \frac{q}{m}A_0,
\end{aligned}
\end{equation}
and by substituting with the expression of $u^0$ we retrieve $|H|=|\pi_0|=E$ as implied by Killing's equation. In absence of electromagnetic fields, $A_\mu=0$, we retrieve the Hamiltonian for pure geodesic motion,
\begin{equation}
 \tilde{H}(\textbf{x},\textbf{u}) = \alpha\sqrt{1 + \gamma^{ij} u_i u_j} - \beta^k u_k.
 \label{eq:hamgeo}
\end{equation}
The special relativistic limit of the Hamiltonian \eqref{eq:ham} is also easily retrieved, for flat spacetime, as
\begin{equation}
 H(\textbf{x},\pivec) = \sqrt{1 + \left(\pivec-\frac{q}{m}\Avec\right)^2}+\frac{q}{m}\phi,
 \label{eq:hamSR}
\end{equation}
for an electrostatic scalar potential $\phi$ and a vector potential $\Avec$.

The equations of motion in terms of $x^i$ and $\pi_i$ are derived in the usual way,
\begin{equation}
 \frac{dx^i}{dt} = \frac{\partial H(\textbf{x},\pivec)}{\partial\pi_i} = \frac{\gamma^{ij} (\pi_j-qA_j/m)}{u^0} - \beta^i,
\end{equation}
\label{eq:geodesic3p1xpi}
\begin{equation}
\begin{aligned}
 \frac{d\pi_i}{dt} = -\frac{\partial H(\textbf{x},\pivec)}{\partial x^i} & = -\alpha u^0 \partial_i \alpha + (\pi_k-qA_k/m) \partial_i \beta^k - \frac{(\pi_j-qA_j/m) (\pi_k-qA_k/m)}{2 u^0} \partial_i \gamma^{jk}  \\
 & + \frac{q}{m}\partial_i A_0 + \frac{q}{m}\left[\frac{\gamma^{jk}(\pi_j-qA_j/m)}{u^0} - \beta^k\right]\partial_i A_k,
\end{aligned}
\label{eq:geodesic3p1pi}
\end{equation}
where the definition of $u^0$ is now
\begin{equation}
u^0 =\frac{1}{\alpha} \sqrt{1 + \gamma^{jk} \left(\pi_j-\frac{q}{m}A_j\right) \left(\pi_k-\frac{q}{m}A_k\right)}. 
\end{equation}
In equation \eqref{eq:geodesic3p1pi} the Lorentz force appears via the terms that include spatial derivatives of the components of $A_\mu$. The conservation of the Hamiltonian \eqref{eq:ham} in time follows immediately by application of the chain rule,
\begin{equation}
 \begin{aligned}
 \frac{d H(\xvec,\pivec)}{dt} & =\frac{\partial H(\textbf{x},\pivec)}{\partial x^i}\frac{d x^i}{dt}+\frac{\partial H(\textbf{x},\pivec)}{\partial \pi_i}\frac{d \pi_i}{dt} \\
 & = -\frac{d\pi_i}{dt}\frac{dx^i}{dt}+\frac{dx^i}{dt}\frac{d\pi_i}{dt} \\
 & =0.
 \end{aligned}
 \ \label{eq:hamenergycondcont}
\end{equation}

A numerical scheme constructed to conserve the Hamiltonian (hence the energy) exactly, should fulfill a discrete equivalent of the condition above. For a discrete time increment  $\Delta t$ between time levels $n$ and $n+1$, a numerical scheme advances the position and momentum by the increments $x^{i,n+1}-x^{i,n}=\Delta x^i$, $\pi_{i,n+1}-\pi_{i,n}=\Delta \pi_i$. If energy is preserved during the update, then $H(\xvec^{n+1},\pivec^{n+1})=H(\xvec^n,\pivec^n)$ to machine precision. Therefore, a discrete equivalent of the condition for energy conservation above reads
\begin{equation}
 \frac{\Delta H(\xvec,\pivec)}{\Delta t} = \frac{\Delta^x_i H(\textbf{x},\pivec)}{\Delta  x^i}\frac{\Delta x^i}{\Delta t}+\frac{\Delta_\pi^i H(\textbf{x},\pivec)}{\Delta \pi_i}\frac{\Delta \pi_i}{\Delta t} = 0.
 \label{eq:hamenergyconddisc}
\end{equation}
The first equality holds assuming that it is possible to define discrete operators $\Delta_i^x,\Delta^i_\pi$ such that the discrete time increment of $H$ on the left-hand side can be expanded in terms of increments of $H$ with respect to the single variables $x^i,\pi_i$. This requirement essentially corresponds to finding discrete operators $\Delta_i^x,\Delta^i_\pi$ that act, in the numerically-defined condition \eqref{eq:hamenergyconddisc}, as partial derivatives act in the continuous case, equation \eqref{eq:hamenergycondcont}. In other words, such operators must respect a discrete equivalent of the chain rule when applied to a generic function $f(x^i,\pi_i)$,
\begin{equation}
 \Delta f(\xvec,\pivec) = f(\xvec^{n+1},\pivec^{n+1})-f(\xvec^n,\pivec^n) = \Delta^x_i f(\xvec,\pivec) \Delta x^i + \Delta_\pi^i f(\xvec,\pivec) \Delta \pi_i,
\end{equation}
which mimics the continuous analogue
\begin{equation}
 df(\xvec,\pivec) = \partial_{x^i} f(\xvec,\pivec) dx^i + \partial_{\pi_i} f(\xvec,\pivec) d\pi_i.
\end{equation}

The definition of the discrete operators can be found in full in Appendix \ref{app:ham}. With these definitions, it is sufficient to employ the discrete equations of motion
\begin{equation}
 \frac{\Delta x^{i}}{\Delta t} = \frac{\Delta_\pi^i H(\xvec,\pivec)}{\Delta \pi_i},
 \label{eq:hampos}
\end{equation}
\begin{equation}
 \frac{\Delta \pi_{i}}{\Delta t} = -\frac{\Delta^x_i H(\xvec,\pivec)}{\Delta x^i},
 \label{eq:hammom}
\end{equation}
to ensure that the condition \eqref{eq:hamenergyconddisc} is automatically satisfied. The resulting numerical scheme is second-order like the IMR, and involves in general the solution of a system of nonlinear, implicit equations, that has to be carried out iteratively. The additional computational cost however is counterbalanced by exact conservation of energy in time, regardless of the simulation parameters such as $\Delta t$.

Energy-conserving schemes of this type have been applied for simpler Hamiltonian systems in many contexts (\citealt{fengqin}; \citealt{chatziioannouvanwalstijn2015}). These schemes generally do not preserve phase-space trajectories, as is the case instead for symplectic schemes such as the IMR. Identifying the most suitable integrator is therefore case-dependent, but conservation of energy certainly plays an important role in the accuracy of the results, as shown in \citetalias{bacchini2018a}. There, an energy-conserving scheme was successfully applied to pure geodesic motion, exhibiting higher robustness than the RK4 and the IMR schemes, especially for motion along unstable orbits.

Note that for charged particle motion, the scheme above requires that the four-potential $A_\mu$ be available analytically at all time steps. In the next Section, we discuss the application of such a scheme when only the dynamic fields $D^i, B^i$ are known at discrete grid points, which is generally the case in grid-based global simulation codes.

\subsubsection{Modified Hamiltonian scheme for grid-defined electromagnetic fields}
\label{sec:schemesBHAC}

\texttt{BHAC} (\citealt{porth2017}) is a GRMHD code that solves Maxwell's equations coupled to the plasma fluid equations in a form that does not involve the four-potential (as is the case for other codes that instead evolve $A_\mu$, see e.g.\ \citealt{etienne2015}). In fact, only two of the dynamic fields from the formulation presented in Section \ref{sec:theory} are used, namely the three-vectors $D^i$ and $B^i$. In a way, these can be identified with the electric and magnetic field, although the definition of such fields in general relativity takes a more intricate meaning (see Section \ref{sec:theory}). If only $D^i$ and $B^i$ are available in place of $A_\mu$, we can solve the momentum equation for charged particles in the form \eqref{eq:geodesic3p1uEB}. The RK4 and IMR schemes can be directly applied to this equation. However, the Hamiltonian scheme in the form presented in the previous Section intrinsically requires the four-potential $A_\mu$.

In order to render the Hamiltonian scheme applicable to particle simulations that do not make use of $A_\mu$, we present here a slightly modified algorithm which relies on $D^i$ and $B^i$ instead. Because the conservation properties of the original Hamiltonian scheme are based on the availability of $A_\mu$, the most immediate consequence of such a modification is that exact preservation of the invariants in generally lost. However, we will show that this ``modified Hamiltonian'' scheme retains energy conservation in some form. The new numerical scheme is constructed by imposing two main properties. First, for physical consistency we can demand that the numerically-computed magnetic force acting on the particles exerts no work. Second, in the limit of vanishing electromagnetic fields, if the metric is available analytically, the scheme must converge to that presented in \citetalias{bacchini2018a} for pure geodesic motion, hence retaining exact energy conservation. Additionally, we can demand that the method remains second-order accurate like the original Hamiltonian scheme.

The first condition follows from the time component of the equation of motion \eqref{eq:geodesic3p1uEB}. This reads
\begin{equation}
 \frac{du_0}{dt} = \frac{q}{m} \frac{u^i}{u^0} E_i,
\end{equation}
where the change in energy of the particle is associated only to the electric field term $E_i$. The expression above holds by using the definitions of $E_i=F_{i0}$ and $B^i=(1/2)e^{ijk}F_{jk}$, and due to the antisymmetry of the Levi-Civita tensor, $u^i e_{ijk}u^jB^k=0$ (or in standard vector notation, $\textbf{u}\cdot(\textbf{u}\times\textbf{B}) = 0$). In other words, the three-momentum $u^i$ should always be perpendicular to the magnetic force $e_{ijk}u^j B^k$. Since $u^i/u^0=dx^i/dt$, this implies that also the displacement in the position is perpendicular to the magnetic force.

The energy-conserving properties above (no spurious work from magnetic fields and from the curvature terms) can be imposed on the numerical scheme by an appropriate choice of discretization. First, we take the discrete position equation to be
\begin{equation}
 \frac{\Delta x^{i}}{\Delta t} = \frac{\Delta_u^i \tilde{H}(\xvec,\uvec)}{\Delta u_i},
 \label{eq:hamposmod}
\end{equation}
which is similar to equation \eqref{eq:hampos}, but with the energy-conserving discretization discussed in Appendix \ref{app:ham} applied to the Hamiltonian for pure geodesic motion \eqref{eq:hamgeo}, $\tilde{H}(\xvec,\uvec) = \alpha\sqrt{1+\gamma^{ij}u_i u_j} - \beta^k u_k$. Then, the discrete momentum equation is taken as
\begin{equation}
 \frac{\Delta u_{i}}{\Delta t} = -\frac{\Delta^x_i \tilde{H}(\xvec,\uvec)}{\Delta x^i} + \frac{q}{m}\left(\alpha\gamma_{ij}D^j + e_{ijk}\beta^j B^k + e_{ijk} \frac{\Delta x^j}{\Delta t} B^k\right),
 \label{eq:hammommod}
\end{equation}
where the first term corresponds again to the energy-conserving discretization of the Hamiltonian for geodesic motion. The second term, expressing the Lorentz force, is such that the magnetic force $e_{ijk}\Delta x^j B^k/\Delta t$ is now always perpendicular to the three-momentum $u^i/u^0$, i.e.\ perpendicular to the displacement $\Delta x^i/\Delta t$, as required.

The scheme above respects both requirements and, in fact, retains energy conservation in the limit of zero electric fields and for pure geodesic motion. If the electric field term $E_i = \alpha \gamma_{ij}D^j + e_{ijk} \beta^j B^k$ vanishes in equation \eqref{eq:hammommod}, contracting the two discrete equations of motion gives
\begin{equation}
 \frac{\Delta x^i}{\Delta t} \frac{\Delta u_i}{\Delta t} = - \frac{\Delta x^i}{\Delta t}\frac{\Delta^x_i \tilde{H}(\xvec,\uvec)}{\Delta x^i} 
  = \frac{\Delta_u^i \tilde{H}(\xvec,\uvec)}{\Delta u_i} \frac{\Delta u_i}{\Delta t},
\end{equation}
or alternatively,
\begin{equation}
 \frac{\Delta^x_i \tilde{H}(\xvec,\uvec)}{\Delta x^i}\frac{\Delta x^i}{\Delta t} + \frac{\Delta_u^i \tilde{H}(\xvec,\uvec)}{\Delta u_i} \frac{\Delta u_i}{\Delta t} = \frac{\Delta \tilde{H}(\xvec,\uvec)}{\Delta t} = 0,
\end{equation}
following the same argument used to write the condition \eqref{eq:hamenergyconddisc}. In practice, in case of vanishing electric fields, with this choice of discretization the magnetic field exerts no work and energy (i.e.\, the Hamiltonian $\tilde{H}$) is still conserved exactly.

Aside from non-conservation of energy in the most general case, the angular momentum $L$, which is a constant of the motion in specific setups (e.g.\ along the $\theta$-direction in the Wald solution with aligned magnetic field, see Section \ref{sec:wald}) is not conserved exactly by the modified scheme. In those setups, the original Hamiltonian integrator, which is formulated in terms of $x^i$ and $\pi_i$, automatically respects the condition $d\pi_\varphi/dt = dL/dt = 0$. However, since the modified scheme is formulated in terms of $x^i$ and $u_i$, this property does not hold anymore.

Finally, note that no condition is specified for the position at which the Lorentz force term in equation \eqref{eq:hammommod} must be evaluated. In order to retain second-order accuracy, we choose this position to be the time-average between two consecutive time levels, $(x^{i,n+1}+x^{i,n})/2$. We also emphasize that the scheme above is valid even when $E^i$ and $B^i$ are obtained via interpolation at the particle position (e.g.\ from a computational grid), and energy conservation is still ensured in the limit of vanishing electric fields. In Section \ref{sec:interpBHAC} we analyze the effect of interpolation (which acts as an additional source of numerical error) on the accuracy of the integrators.

\section{Tests in analytic electromagnetic fields}
\label{sec:tests}
In this Section, we test all integrators described above by simulating bound orbits of charged particles around compact objects. As physically meaningful examples, we consider the Schwarzschild and Kerr spacetimes, coupled to several configurations of test electromagnetic fields. Because of the non-integrability of the resulting system, analytic solutions do not exist in general, hence we rely on other types of diagnostics to assess the performance of the integrators (conservation of first integrals, preservation of gyration, and other case-dependent requirements). Then, for quantitative comparison of the numerical results with theoretical predictions, we consider analytically-derived orbits in the electromagnetic field of a Kerr-Newman black hole, where the equations of motion are integrable. In this context, we apply all integrators to the simulation of several unstable spherical orbits. In all cases we use geometrized units $c=G=1$ (hence time, mass, and distances are measured with the same units). Additionally, everywhere we consider a black hole mass $M=1$.

\subsection{Black holes in external electromagnetic fields}
\label{sec:extfields}
As a first test, we consider the motion of charged particles around Schwarzschild and Kerr black holes. The spacetime metric reads, in Boyer-Lindquist coordinates,
\begin{equation}
ds^2 = -\left(1-\frac{2Mr}{\rho^2}\right)dt^2 - \frac{4Mra\sin^2\theta}{\rho^2}d\varphi dt + \frac{\rho^2}{\Delta}dr^2 + \rho^2 d \theta^2 + \left(r^2 + a^2 + \frac{2Mra^2 \sin^2 \theta}{\rho^2}\right)\sin^2 \theta d \varphi^2,
\label{eq:kerr}
\end{equation}
with $\rho^2 \equiv r^2 + a^2 \cos^2 \theta$ and $\Delta \equiv r^2 - 2Mr + a^2$. This solution presents metric singularities at the inner and outer event horizons, corresponding to the condition $\Delta=0$, and located at
\begin{equation}
 r_\pm = M \pm \sqrt{M^2-a^2}.
\end{equation}
In the non-rotating limit $a=0$, the Schwarzschild solution is retrieved, and the only event horizon is the spherical surface at the Schwarzschild radius $r_S=2M$.

In addition to the geodesic motion governed by the spacetime curvature, we consider the effect of external electromagnetic fields on the motion of charged particles. In this case, the source of the electromagnetic fields is not the black hole itself, but rather a distant magnetized object, or external current loops from accreting plasma. This configuration does not exhibit a sufficient number of invariants characterizing the motion of test particles. As a consequence, the equations of motion are non-integrable, and chaotic behavior may be observed in the particle trajectories. Chaoticity characterizes trajectories which do not remain within limited equi-potential region, but rather fill the phase space uniformly without evident recursion (see e.g.\ \citealt{kopacek2010b}).

Non-integrability also implies that these systems lack analytic solutions that can be directly compared to numerical results. Hence, in order to analyze the performance of the numerical integrators here presented, we rely on alternative measures of the computational error. For the two configurations presented below, we can monitor the error in the conserved energy $E$ and angular momentum $L$ (the latter not in all cases); additionally, we can verify the capability of each scheme to keep the particles on prescribed bound orbits (without spurious escape) without unphysically suppressing the gyration around magnetic fields lines.

\subsubsection{The Wald solution}
\label{sec:wald}
The Wald solution of Maxwell's equations in curved spacetime is an electromagnetic configuration consisting of a Schwarzschild or Kerr black hole immersed in an external magnetic field (\citealt{wald1974}; \citealt{alievozdemir2002}). Such a field is supposedly originated by a distant object (e.g.\ a magnetar). This configuration allows for the appearance of chaotic scattering orbits, hence it represents a strong test for the robustness of the integration schemes.

The four-potential corresponding to a uniform magnetic field with arbitrary inclination with respect to the black hole spin axis reads (\citealt{kopacekkaras2014})
\begin{equation}
\begin{aligned}
 A_0 & = \frac{arM B_z}{\Sigma}(1+\cos^2\theta)-aB_z \\
 & + \frac{aMB_x\sin\theta\cos\theta}{\Sigma}(r\cos\psi-a\sin\psi) - \frac{rQ}{\Sigma},
\end{aligned}
\end{equation}
\begin{equation}
 A_r = -B_x(r-M)\cos\theta\sin\theta\sin\psi,
\end{equation}
\begin{equation}
\begin{aligned}
 A_\theta & = -aB_x(r\sin^2\theta+M\cos^2\theta)\cos\psi \\
 & -B_x(r^2\cos^2\theta-rM\cos2\theta+a^2\cos2\theta)\sin\psi,
\end{aligned}
\end{equation}
\begin{equation}
\begin{aligned}
 A_\varphi & = B_z\sin^2\theta\left[ \frac{r^2+a^2}{2}-\frac{a^2rM}{\Sigma}(1+\cos^2\theta)\right] \\
 & -B_x\sin\theta\cos\theta\left[\Delta\cos\psi+\frac{(r^2+a^2)M}{\Sigma}(r\cos\psi-a\sin\psi)\right] + \frac{arQ\sin^2\theta}{\Sigma},
\end{aligned}
\end{equation}
where $B_z$ and $B_x$ represent the asymptotic strength of the magnetic field along the directions parallel and perpendicular to the magnetic axis, and
\begin{equation}
 \psi = \varphi+\frac{a}{r_+ - r_-}\log\frac{r-r_+}{r-r_-}.
\end{equation}
Here, $Q$ represents the asymptotic charge of the black hole that is accumulated via the infall of charged particles through the even horizon (\citealt{wald1974}). The inclination of the magnetic field lines with respect to the spin axis is quantified by the ratio $B_x/B_z$.

As mentioned above, the equations of motion in this configuration are non-integrable. In fact, the only quantity that is always conserved (aside from $g^{\mu\nu}u_\mu u_\nu=-1$) is the energy, $E=-u_0-qA_0/m$, since both $A_\mu$ and $g_{\mu\nu}$ are independent of $t$. The four-potential, $A_\mu$, is not independent of the coordinate $\varphi$ in general, but becomes so in the aligned configuration $B_x/B_z=0$. Therefore, the angular momentum, $L=u_\varphi+qA_\varphi/m$, is not a constant of the motion for non-aligned cases. 

Here we consider several bound orbits taken from various reference works (\citealt{takahashikoyama2009}; \citealt{kopacek2010a}; \citealt{kopacek2010b}; \citealt{kopacekkaras2014}; \citealt{kolos2015}; \citealt{stuchlikkolos2016}; \citealt{tursunov2016}; \citealt{kopacekkaras2018}). The parameters and the numerical values of the initial conditions characterizing each orbit are listed in Table \ref{tab:wald}. In all cases, we set $q/m=1$, an initial angle $\varphi=0$, and an initial component $u_r=0$.

\begin{table}[!h]
\centering
\begin{tabular}{|c|c|c|c|c|c|c|c|c|}
\hline
Orbit name & $B_z$ & $B_x$ & $a$ & $Q$ & $L$ & $E$ & $(r,\theta)$ & $(u_\theta,u_\varphi)$ \\ 
\hline 
RSA1 & 0.2 & 0 & 0 & 0 & 4.5 & 0.873 & $(4,\pi/2)$ & $(0,2.9)$ \\ 
\hline
RSA2 & 0.2 & 0 & 0 & 0 & 4.5 & 0.834 & $(5,\pi/2)$ & $(0,2.0)$ \\ 
\hline
RSA3 & 0.2 & 0 & 0 & 0 & 8 & 0.9 & $(9.5,1.6)$ & $(0,-1.024)$ \\ 
\hline
RSA4 & 2 & 0 & 0 & 0 & 7.44 & 0.5 & $(2.6,\pi/2+0.1)$ & $(0,0.799)$ \\ 
\hline
RSA5 & -2 & 0 & 0 & 0 & 68 & 14.8 & $(8.5,1.06)$ & $(0,122.983)$ \\ 
\hline
RKA1 & -2 & 0 & 0.998 & -3.992 & -14.9 & 0.07 & $(3.37,\pi/2+0.07)$ & $(0,-2.020)$ \\ 
\hline
RKA2 & 2 & 0 & 0.7 & 0 & 16.25 & 1.81 & $(4.2,\pi/2-0.1)$ & $(0,-1.930)$ \\ 
\hline
RKA3 & 2 & 0 & 0.9 & 0 & 18.1 & 2.2 & $(4,\pi/2-0.2)$ & $(0,1.565)$ \\ 
\hline
RKA4 & 1 & 0 & 0.9 & 1 & 6 & 1.58 & $(3.68,1.18)$ & $(0.341,-0.322)$ \\ 
\hline
RKA5 & 1 & 0 & 0.9 & 1 & 6 & 1.65 & $(3.68,1.18)$ & $(1.745,-0.322)$ \\ 
\hline
RKA6 & 2 & 0 & 0.5 & 2 & 5 & 1.78 & $(3.11,\pi/4)$ & $(0.648,-0.040)$ \\ 
\hline
RKA7 & 1 & 0 & 0.9 & 1 & 6 & 1.6 & $(3.68,1.18)$ & $(0.962,-0.322)$ \\ 
\hline
RKA8 & 1 & 0 & 0.9 & 0 & 5 & 1.24 & $(3,\pi/2)$ & $(0.497,0.365)$ \\ 
\hline
CKA1 & 1 & 0 & 0.9 & 1 & 6 & 1.75 & $(3.68,1.18)$ & $(2.779,-0.322)$ \\ 
\hline
CKA2 & 1 & 0 & 0.9 & 1 & 6 & 1.8 & $(3.68,1.18)$ & $(3.205,-0.322)$ \\ 
\hline
CKA3 & 1 & 0 & 0.9 & 1 & 6 & 1.75 & $(3.68,1.18)$ & $(2.779,-0.129)$ \\ 
\hline
CKI1 & 1 & 0.07 & 0.9 & 1 & 6 & 1.58 & $(3.68,1.18)$ & $(0.135,0.132)$ \\ 
\hline
CKI2 & 1 & 0.15 & 0.9 & 1 & 6 & 1.75 & $(3.68,1.18)$ & $(2.698,0.429)$ \\ 
\hline
CKI3 & 1 & 0.05 & 0.9 & 0 & 5 & 1.24 & $(3,\pi/2)$ & $(0.497,0.365)$ \\ 
\hline
CKI4 & 1 & 0.1 & 0.9 & 0 & 5 & 1.24 & $(3,\pi/2)$ & $(0.497,0.365)$ \\ 
\hline
\end{tabular} 
\caption{Parameters and initial conditions for the orbits in the Wald configuration. From left to right we list the inclination of the asymptotic magnetic field $B_z,B_x$, the black hole spin $a$, the inductive charge $Q$, and the particle angular momentum and energy $L$, $E$; we also report the numerical values of the initial position $(r,\theta)$ and momentum $(u_\theta,u_\varphi)$ (we set an initial $\varphi=0$ and $u_r=0$ everywhere). The orbit names denote their character (R = regular, C = chaotic), the metric (S = Schwarzschild, K = Kerr), and the magnetic field inclination (A = aligned, I = inclined).}
\label{tab:wald}
\end{table}

\begin{figure}[!h]                                                                                                                                                                                                                                                                                                                                                                                                                                                                                                                                                                                                                                                                                                                                                                                                                                                                                                                                                                                                                                                                                                  
\centering
\includegraphics[width=1\columnwidth]{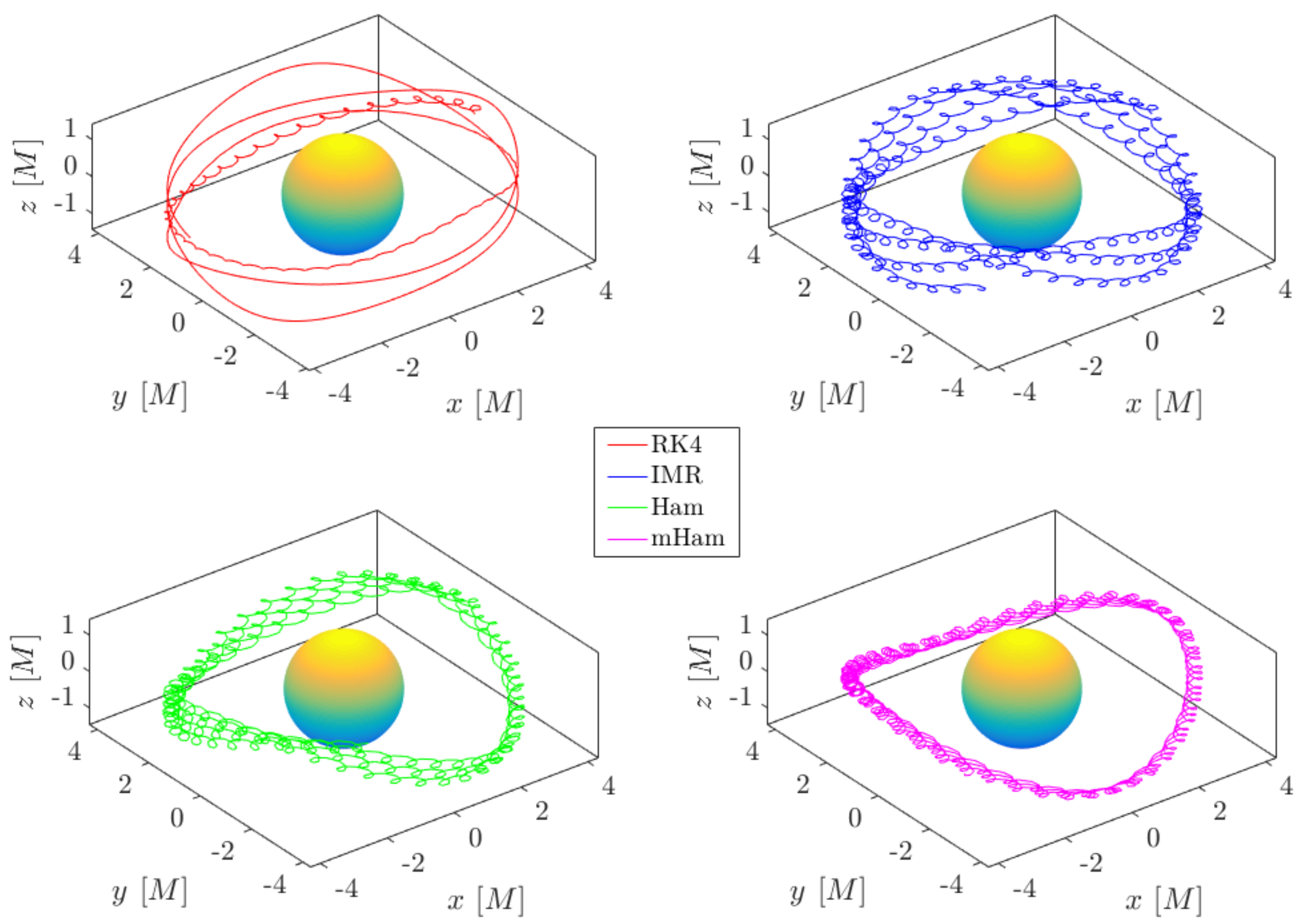}
\caption{Simulation of orbit RKA3 from Table \ref{tab:wald} with the RK4 method (red line, top-left), IMR scheme (blue line, top-right), Hamiltonian (green line, bottom-left), and modified Hamiltonian (magenta line, bottom-right) schemes. The central Kerr black hole, rotating from west to east, is shown as a colored sphere of radius $r_+$. The plots clearly show that the RK4 method introduces numerical errors resulting in a damping of the particle gyration. The other schemes, instead, preserve this feature until the end of the run.}
\label{fig:waldgyr}
\end{figure}

\begin{figure}[!h]                                                                                                                                                                                                                                                                                                                                                                                                                                                                                                                                                                                                                                                                                                                                                                                                                                                                                                                                                                                                                                                                                                  
\centering
\includegraphics[width=1\columnwidth]{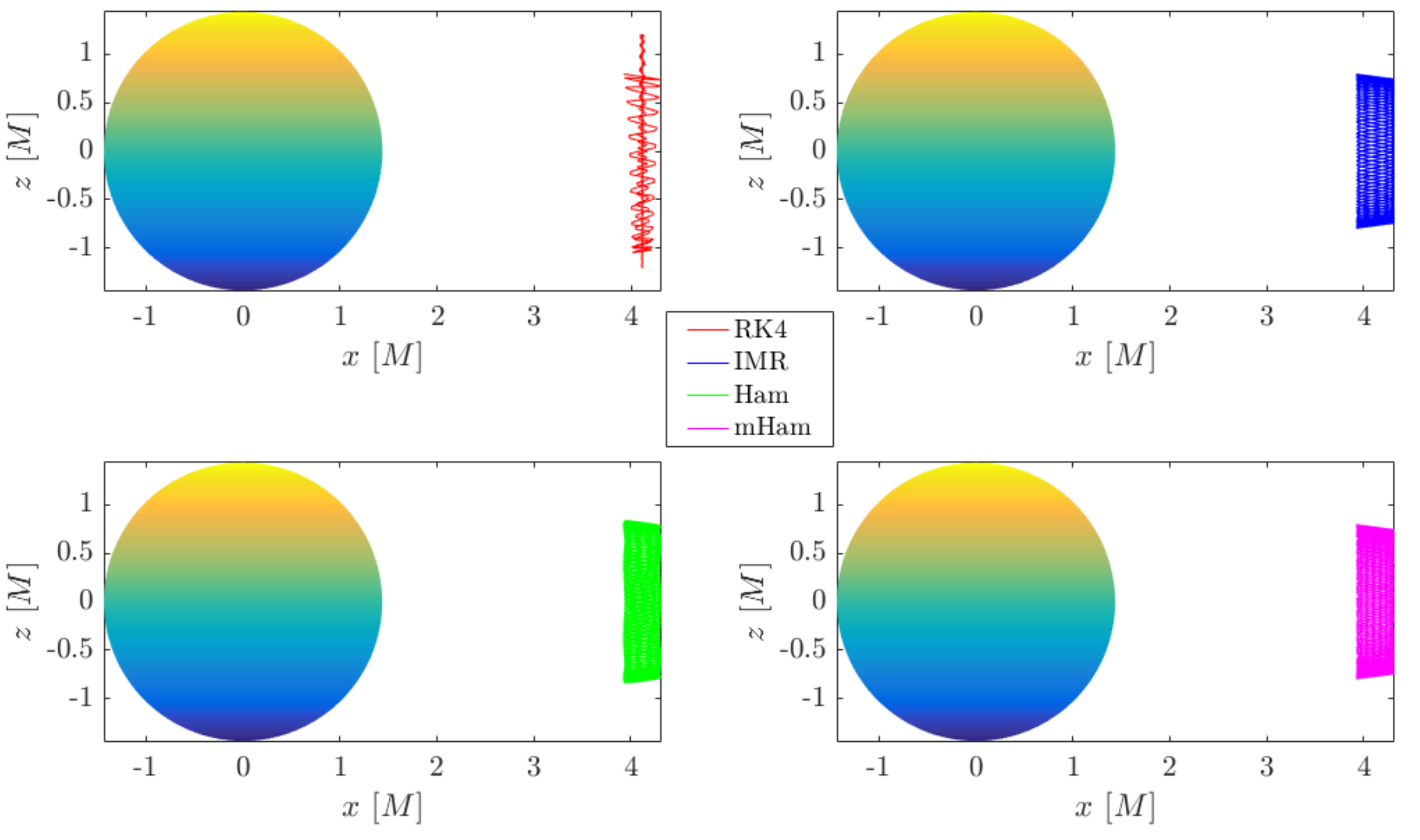}
\caption{Simulation of orbit RKA3 as in Figure \ref{fig:waldgyr}, projected onto the $\varphi=0$ plane, obtained with the four integration methods. A spurious damping of the particle gyration is observed in the RK4 run, due to unbounded energy errors and lack of symplecticity.}
\label{fig:waldgyrproj}
\end{figure}

As a first test, we evaluate the ability of each scheme to preserve the particle gyration around magnetic field lines. Figure \ref{fig:waldgyr} shows the regular orbit RKA3 (see Table \ref{tab:wald} for an explanation of the naming convention used here), where the gyration motion is clearly visible. The orbit was simulated up to $t=1000$ with $\Delta t=1$. The four panels show the trajectory obtained with the RK4 scheme (top-left), IMR scheme (top right), Hamiltonian scheme (bottom-left) and modified Hamiltonian scheme (bottom-right). The same orbits are shown projected on the poloidal plane in Figure \ref{fig:waldgyrproj}. Both figures show clearly that, in the results of the RK4 integrator, the particle gyration is progressively damped until it disappears completely. The IMR scheme, instead, preserves the gyration indefinitely, due to its symplectic character. The Hamiltonian and modified Hamiltonian schemes do not possess symplecticity, but nevertheless perform as well as the IMR, and preserve the gyration correctly. This is a feature observed also for certain non-symplectic special relativistic integrators (\citealt{ripperda2018}), which can be attributed to the highly geometric character of the schemes, intrinsic to their energy-conservation properties. 
Analogous results are obtained for each orbit in Table \ref{tab:wald}. For our choice of test cases, we find that in order to avoid the rapid loss of gyro-motion when using the RK4 scheme, the value of $\Delta t$ must be reduced by one or two orders of magnitude with respect to that allowed by the IMR and Hamiltonian schemes.

\begin{figure}[!h]
\centering
\subfloat{\includegraphics[width=.5\columnwidth]{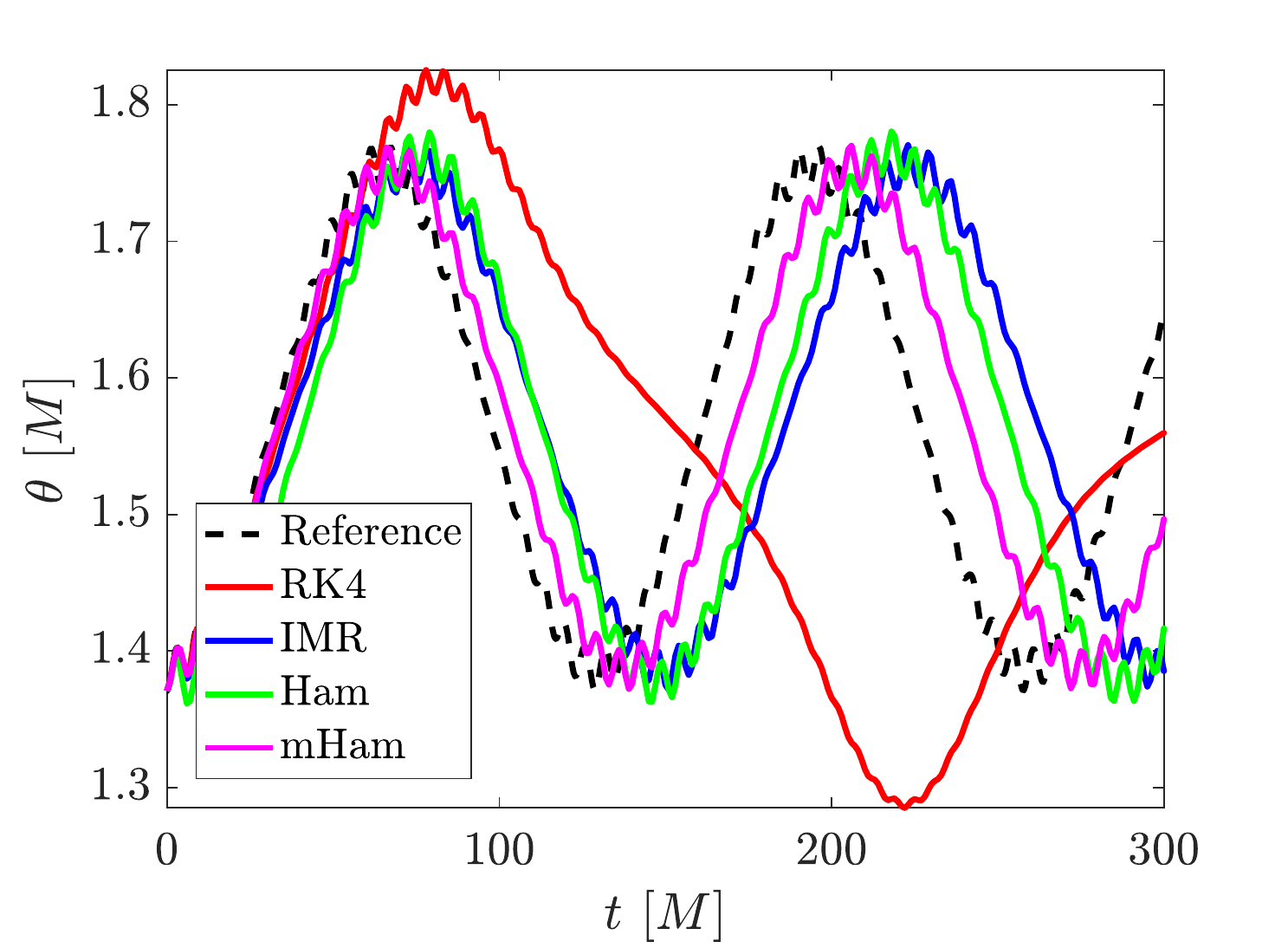}}
\subfloat{\includegraphics[width=.5\columnwidth]{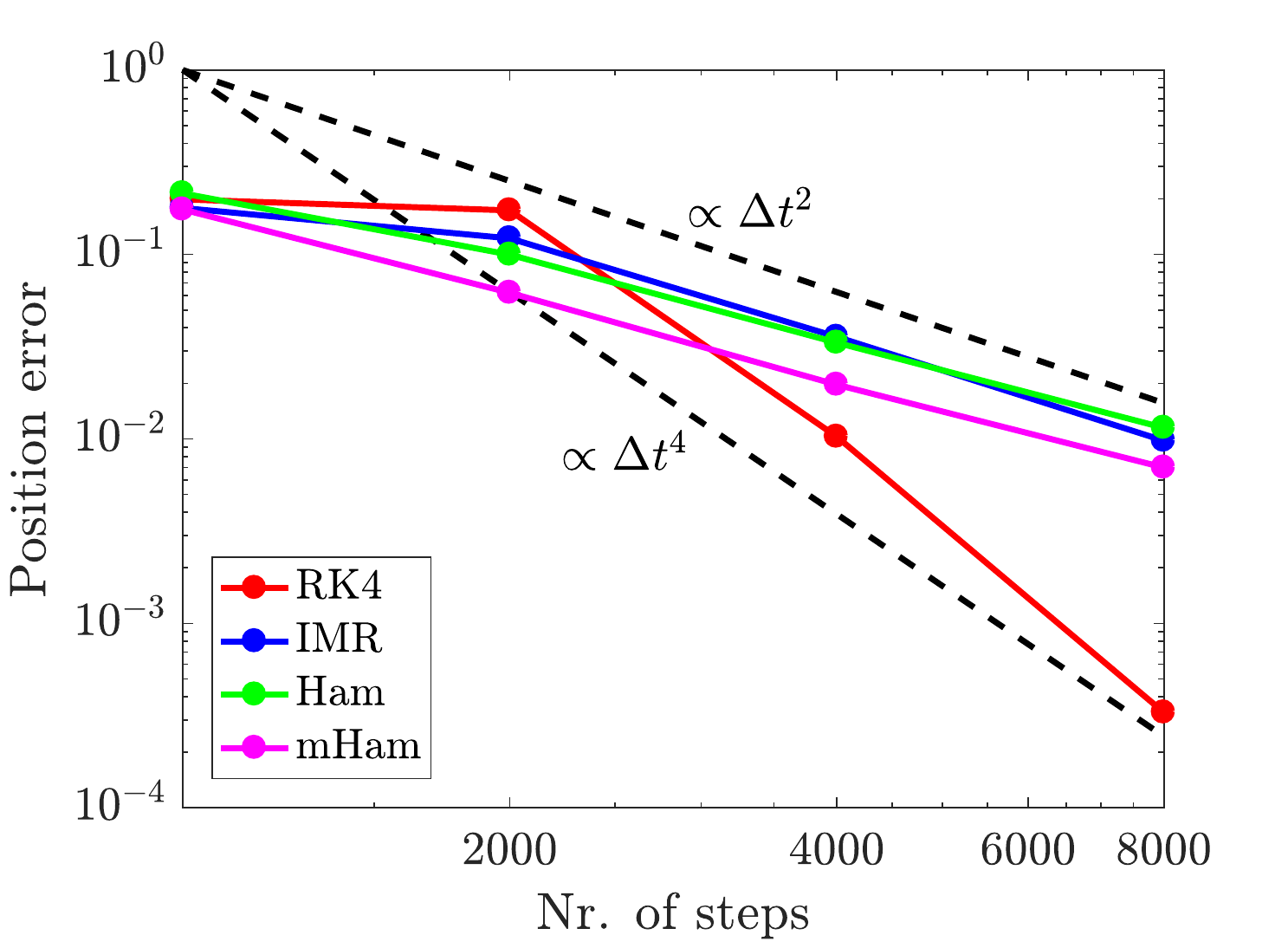}}
\caption{Trajectory in the $\theta$-coordinate for orbit RKA3 simulated with the four methods and $\Delta t=1$ (left panel), compared to a high-resolution reference run (dashed line). The error trends in terms of the root-mean-square difference with respect to the reference simulation are reported for increasing numbers of integration steps (right panel), confirming the second-order character of the IMR, Hamiltonian, and modified Hamiltonian schemes.}
\label{fig:waldgyrerr}
\end{figure}

The phase difference between the oscillating trajectories in Figure \ref{fig:waldgyr} suggests that the various methods introduce different errors in the particle position. Since no analytic solutions are available to compare the calculations with exact results, we measure the convergence properties of each scheme by evaluating the relative difference between the outcome of each method and a reference high-resolution run (obtained with a sufficiently small $\Delta t$ such that all integrators converge to the same solution). By varying the number of integration steps, we can then monitor the error reduction trend that is characteristic of each integrator. The results are shown in Figure \ref{fig:waldgyrerr} (left panel), where we compare part of the trajectory in the $\theta$-coordinate obtained with the four methods by selecting $\Delta t=1$. A different phase lag with respect to the reference run is clearly observable for each method. The error trend of each scheme (in terms of the root-mean-square difference in the $\theta$-coordinate) is shown in the right panel for different numbers of total integration steps. In all cases, the modified Hamiltonian method exhibits a slightly smaller error than the IMR and original Hamiltonian methods. The RK4 method performs far worse than all other schemes at larger time steps, exhibiting higher accuracy only when increasing the number of integration steps four-fold. The error trends also confirm the second-order character of the IMR, Hamiltonian, and modified Hamiltonian schemes.

The results support the conclusions drawn in \citetalias{bacchini2018a}, confirming that explicit integrators such as RK4 are not suitable for simulating the motion of charged particles, and prove inferior (at least for large time steps) to lower-order implicit methods. While this issue is well-known for special relativistic particle simulations (\citealt{qin2013}), here we prove that the same problem arises in general relativistic calculations, in which explicit integrators are often used for geodesic motion (see e.g.\ \citealt{chan2017}). When the effect of spacetime curvature is combined with the action of the Lorentz force, features such as the particle gyration around magnetic field lines can be artificially eliminated by explicit integrators, unless the simulation is run with small $\Delta t$, becoming slow and expensive.

In order to more quantitatively evaluate the numerical error affecting each integrator, we analyze the accuracy of preservation of invariants of the motion. We consider orbits RSA5 and CKI2, a regular and a chaotic orbit respectively. For orbit RSA5, the asymptotic magnetic field is aligned with the black hole spin axis, hence $E$ and $L$ are both conserved quantities. For orbit CKI2, instead, $B_x/B_z\neq0$ and only $E$ is conserved.

\begin{figure}[!h]
\centering
\subfloat[Trajectory in three-dimensional space.]{\includegraphics[width=0.5\columnwidth]{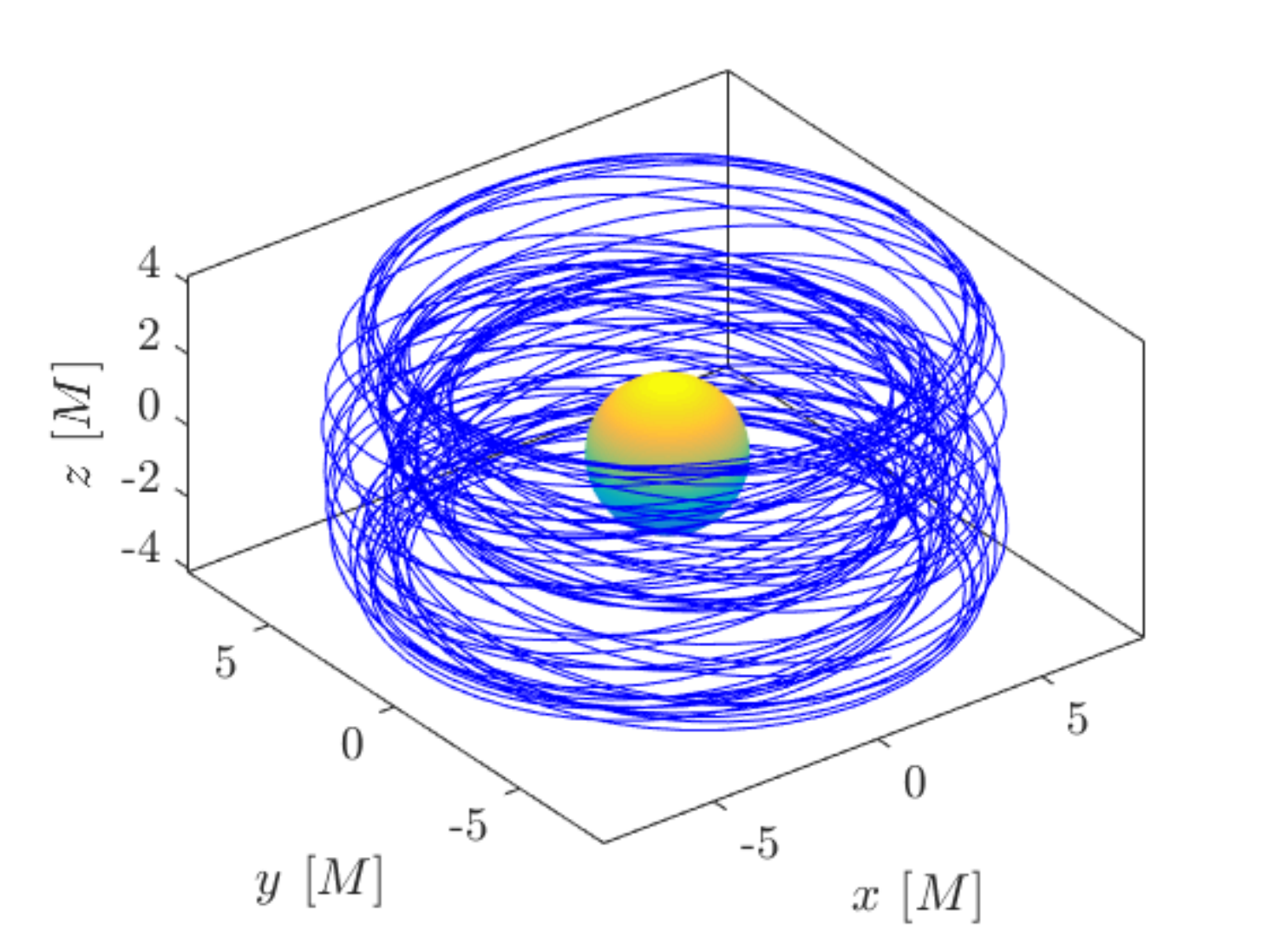}}
\subfloat[Evolution of the relative error in $E$ and $L$.]{\includegraphics[width=0.5\columnwidth]{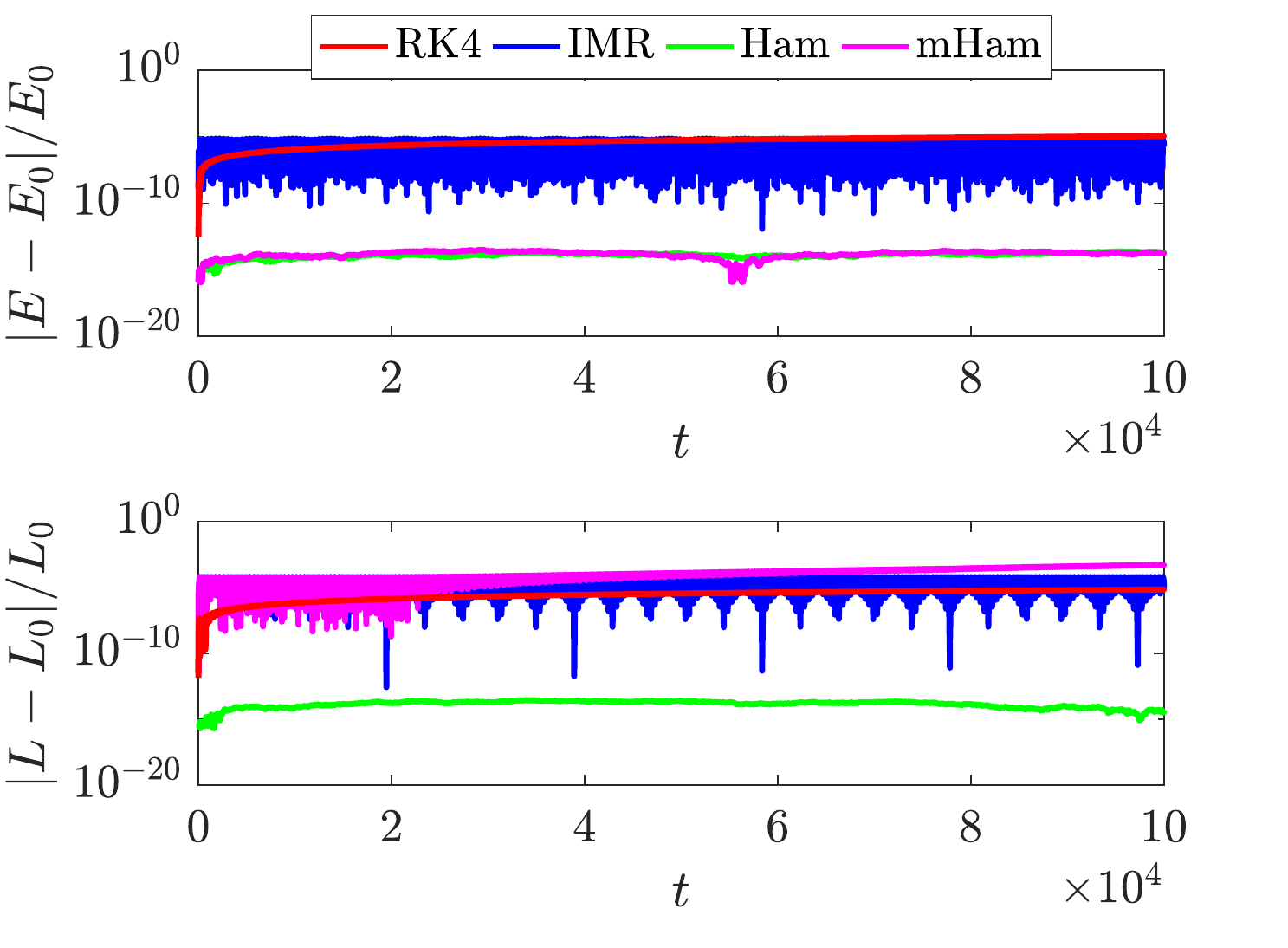}}
\caption{Simulation of the regular orbit RSA5 with $\Delta t=1$ until $t=100000$. The left panel shows part of the particle trajectory. The right panel shows the evolution of the relative error in the conserved quantities for the RK4 (red line), IMR (blue line), Hamiltonian (green line), and modified Hamiltonian (magenta line) integrators. Both Hamiltonian schemes preserve the energy to machine precision at all times. The original Hamiltonian scheme also conserves momentum to machine precision, with the other three integrators showing worse accuracy (similar, among them).}
\label{fig:waldenRSA5}
\end{figure}

Figure \ref{fig:waldenRSA5} shows the numerical results for orbit RSA5 simulated with $\Delta t=1$ until $t=10^5$. The left panel shows part of the trajectory. In the right panel, the evolution of the relative error on $E$ and $L$ is reported for all four integrators. The RK4 results (red lines) are clearly characterized by a secular growth of the error in both conserved quantities. Eventually, such drift leads to the particle escaping the bound trajectory, either ending up captured by the black hole or travelling to infinity. The error in the IMR results (blue lines), instead, is bounded for both $E$ and $L$, and allows for keeping the particle on the right orbit indefinitely. The results obtained with the Hamiltonian integrator (green lines) are characterized by exact conservation (to machine precision) of the energy, as expected; more interestingly, also the angular momentum $L$ is conserved up to a precision far exceeding that of the IMR results. This is due to the formulation of the equation of motion \eqref{eq:geodesic3p1pi} in terms of the conjugate momentum $\pi_i$: for $i=3$, this naturally reduces to $d\pi_3/dt=dL/dt=0$, which reflects in the discrete equations. This feature clearly makes the Hamiltonian integrator more reliable in terms of physical accuracy. Finally, the modified Hamiltonian integrator (magenta lines) shows the capability of conserving energy exactly at all times for this case, as expected, since no electric fields are present. However, due to the different formulation with respect to the original Hamiltonian scheme, exact conservation of the angular momentum is lost. The measured error in $L$ is similar to or slightly above that of the IMR scheme. Such an error appears to be unbounded for the modified Hamiltonian scheme, although in our tests we observed a mildly growing trend that causes the error to double every ~100000 steps. For lower values of $\Delta t$, the error in $E$ of the RK4 and IMR schemes improves, but it is necessary to reduce the time step by several orders of magnitude for these schemes to reach the level of accuracy of the Hamiltonian integrators.

\begin{figure}[!h]
\centering
\subfloat[Trajectory in three-dimensional space.]{\includegraphics[width=.5\columnwidth]{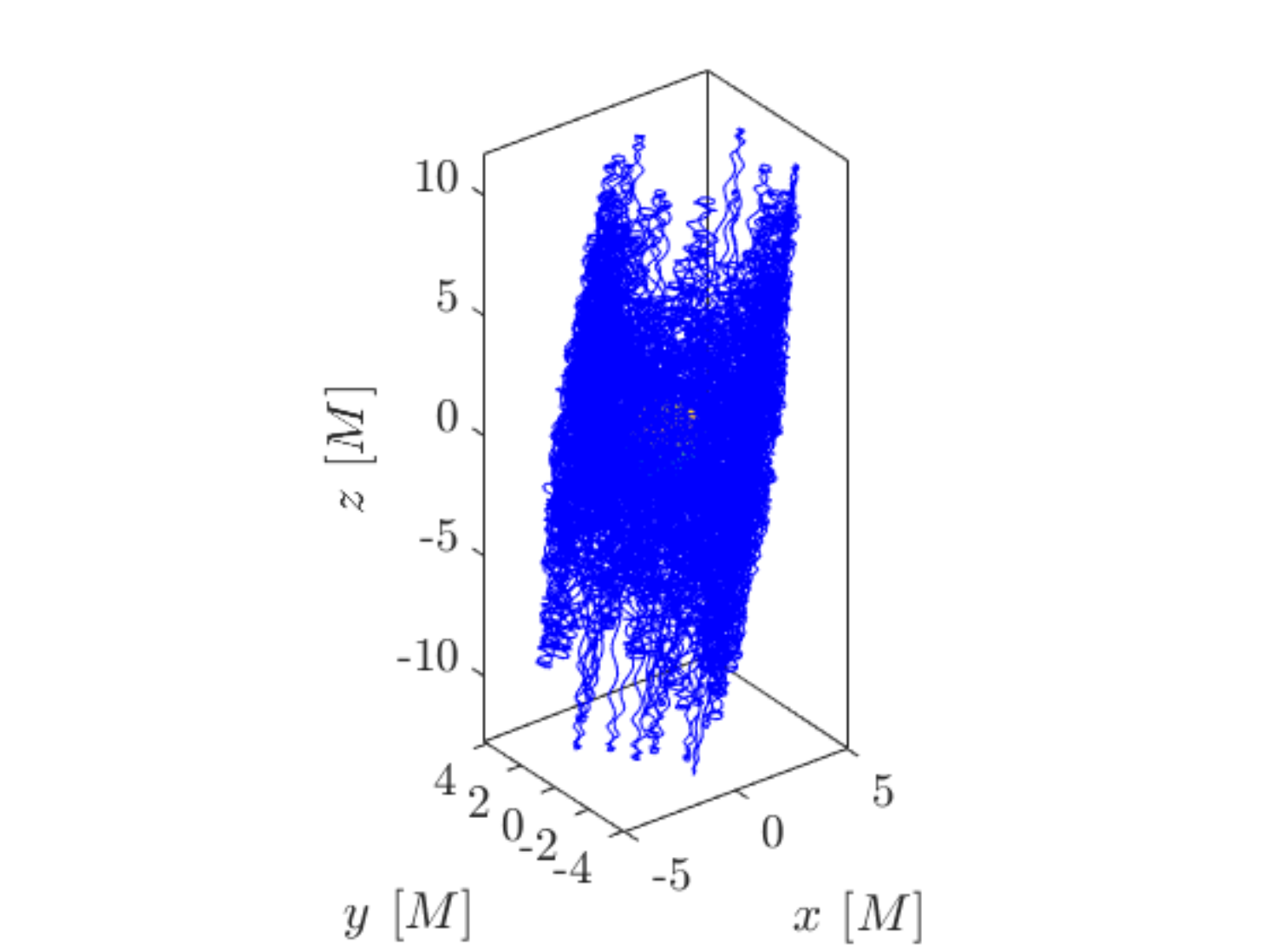}}
\subfloat[Evolution of the relative error in $E$.]{\includegraphics[width=.5\columnwidth]{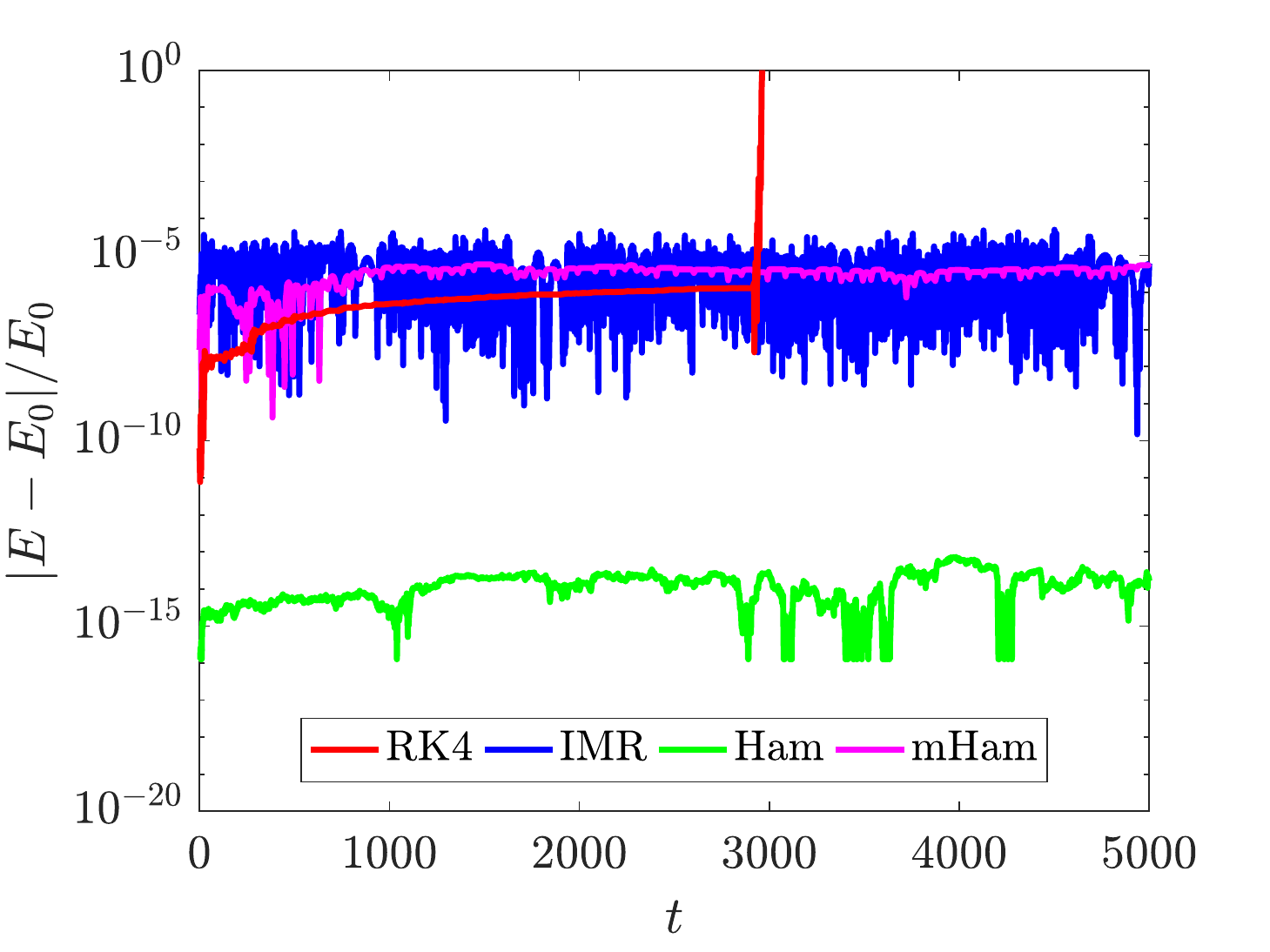}}
\caption{Simulation of the chaotic orbit CKI2 with $\Delta t=0.1$ until $t=5000$. The left panel shows part of the particle trajectory. The right panel shows the evolution of the relative error in the conserved energy for the RK4 (red line), IMR (blue line), Hamiltonian (green line), and modified Hamiltonian (magenta line) integrators. The RK4 integrator causes the spurious escape of the particle from the orbit. The Hamiltonian scheme, instead, keeps the particle bounded and preserves the energy to machine precision at all times. The IMR and modified Hamiltonian schemes perform equally well in keeping energy errors bounded.}
\label{fig:waldenCKI2}
\end{figure}

Finally, Figure \ref{fig:waldenCKI2} shows the simulation results for orbit CKI2 calculated with $\Delta t=0.1$ until $t=5000$. Part of the orbit is shown in the left panel. The evolution of the relative error in the only conserved quantity $E$ is shown in the right panel for all four integrators. Here, we observe that the chaoticity of the orbit further worsens the performance of the RK4 integrator, with its associated accumulation of error rapidly causing the spurious escape of the particle from the bound trajectory. The IMR and Hamiltonian integrators, instead, keep the particle bounded until the end of the run, with the energy error of the Hamiltonian remaining of order machine precision at all times. The modified Hamiltonian integrator does not preserve energy to machine accuracy, as expected, due to the non-vanishing electric field. However, energy errors remain bounded and of the same order of those observed in the IMR results.

\subsubsection{The dipole solution}
\label{sec:DP}
A more realistic electromagnetic configuration is represented by a black hole surrounded by a dipolar field, e.g.\ created by toroidal plasma currents flowing outside of the event horizon. Such a situation could be common in astrophysical black holes in presence of accretion disks, which provide the necessary current for the dipolar field (no-ingrown hair theorem). In this case, the metric \eqref{eq:kerr} describing a Schwarzschild or Kerr black hole is coupled to the four-potential $A_\mu=(A_0,0,0,A_\varphi)$ (\citealt{takahashikoyama2009}), where
\begin{equation}
 A_0 = \frac{3a\mathcal{M}}{2\zeta^2\Sigma} \left\{ \left[r(r-M)+(a^2-Mr)\cos^2\theta\right] \frac{1}{2\zeta}\log\left(\frac{r-r_-}{r-r_+}\right) - (r-M\cos^2\theta) \right\},
\end{equation}
\begin{equation}
 A_\varphi = \frac{3\mathcal{M}\sin^2\theta}{4\zeta^2\Sigma} \left\{ (r-M)a^2\cos^2\theta + r(r^2+Mr+2a^2) - \left[r(r^3-2Ma^2+a^2r) + \Delta a^2\cos^2\theta\right] \frac{1}{2\zeta}\log\left(\frac{r-r_-}{r-r_+}\right) \right\},
\end{equation}
where $\zeta=(M^2-a^2)^{1/2}$, and $\mathcal{M}$ is the dipole moment.

Since the source of the electromagnetic field is not the black hole itself, the field is considered ``external'' to the black hole, similarly to the Wald configuration above. In this case, the dipole field and the black hole spin axis are aligned, and the energy $E=-u_0-qA_0/m$ and the angular momentum $L=u_\varphi+qA_\varphi/m$ are conserved quantities of the particle motion. Overall, the system is still non-integrable, hence chaotic motion can be observed in the particle trajectories.

Charged particles traveling in such a configuration are subjected to drift in the azimuthal direction, gyration around the magnetic field lines, and cross-equatorial oscillation, being trapped in the magnetic bottle appearing near the poles. As meaningful examples of such trajectories, here we simulate two bound orbits, one regular and one chaotic, taken from \cite{takahashikoyama2009}. In both cases we assume $q/m=1$, $\mathcal{M}=70$, and we initialize a particle at $(r,\theta,\varphi)=(10.5,\pi/2,0)$. After specifying the values of the energy $E$ and angular momentum $L$, we prescribe an initial four-velocity characterized by the ratio $u^\theta/u^r=\tan\chi$. Then, the initial components of $u_i$ are completely determined by the normalization condition, $g^{\mu\nu}u_\mu u_\nu=-1$. We run both tests with $\Delta t=1$ until $t=100000$.

\begin{figure}[!h]
\centering
\subfloat[Trajectory in three-dimensional space.]{\includegraphics[width=.5\columnwidth]{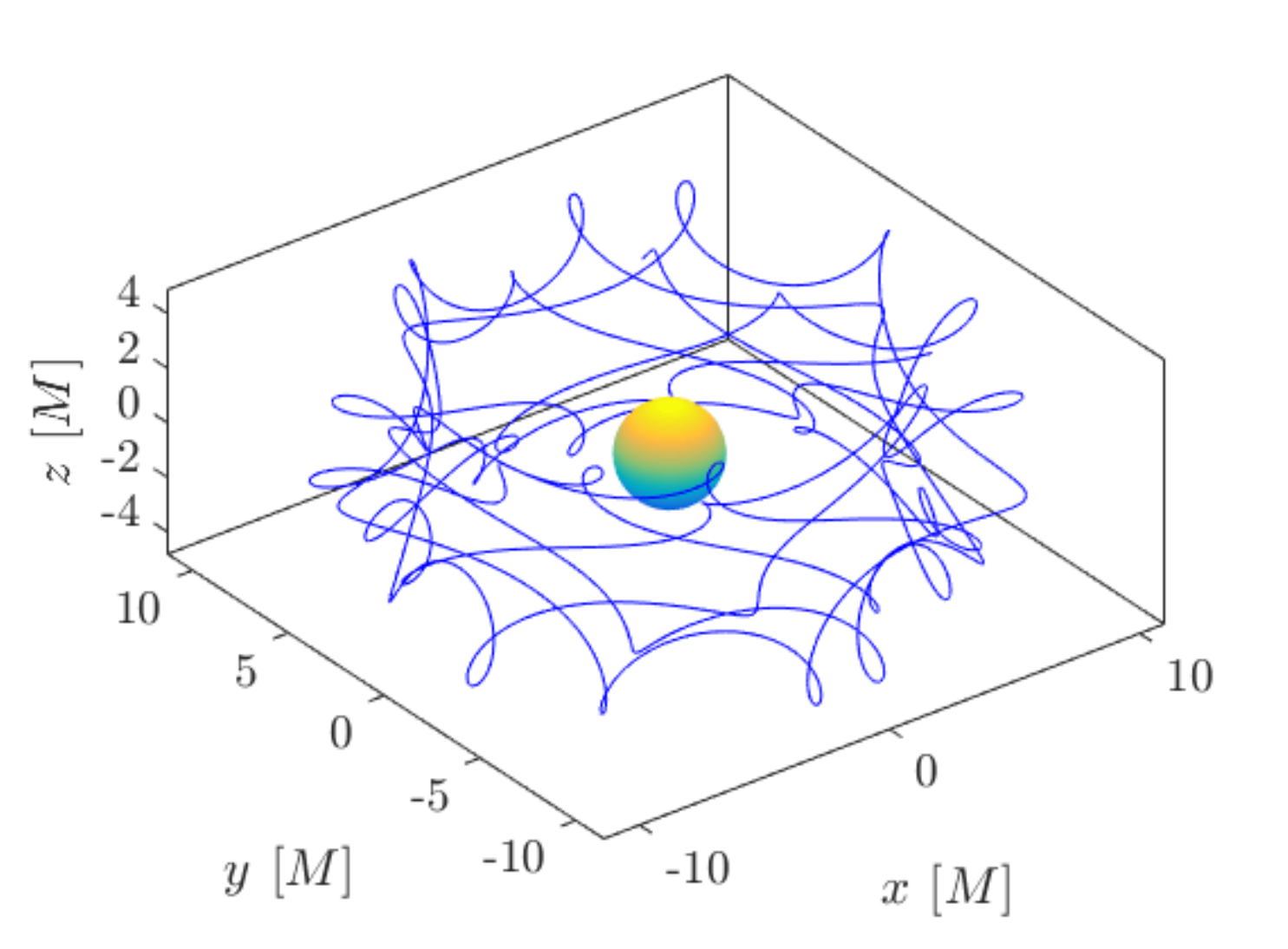}}
\subfloat[Projection in the poloidal plane.]{\includegraphics[width=.5\columnwidth]{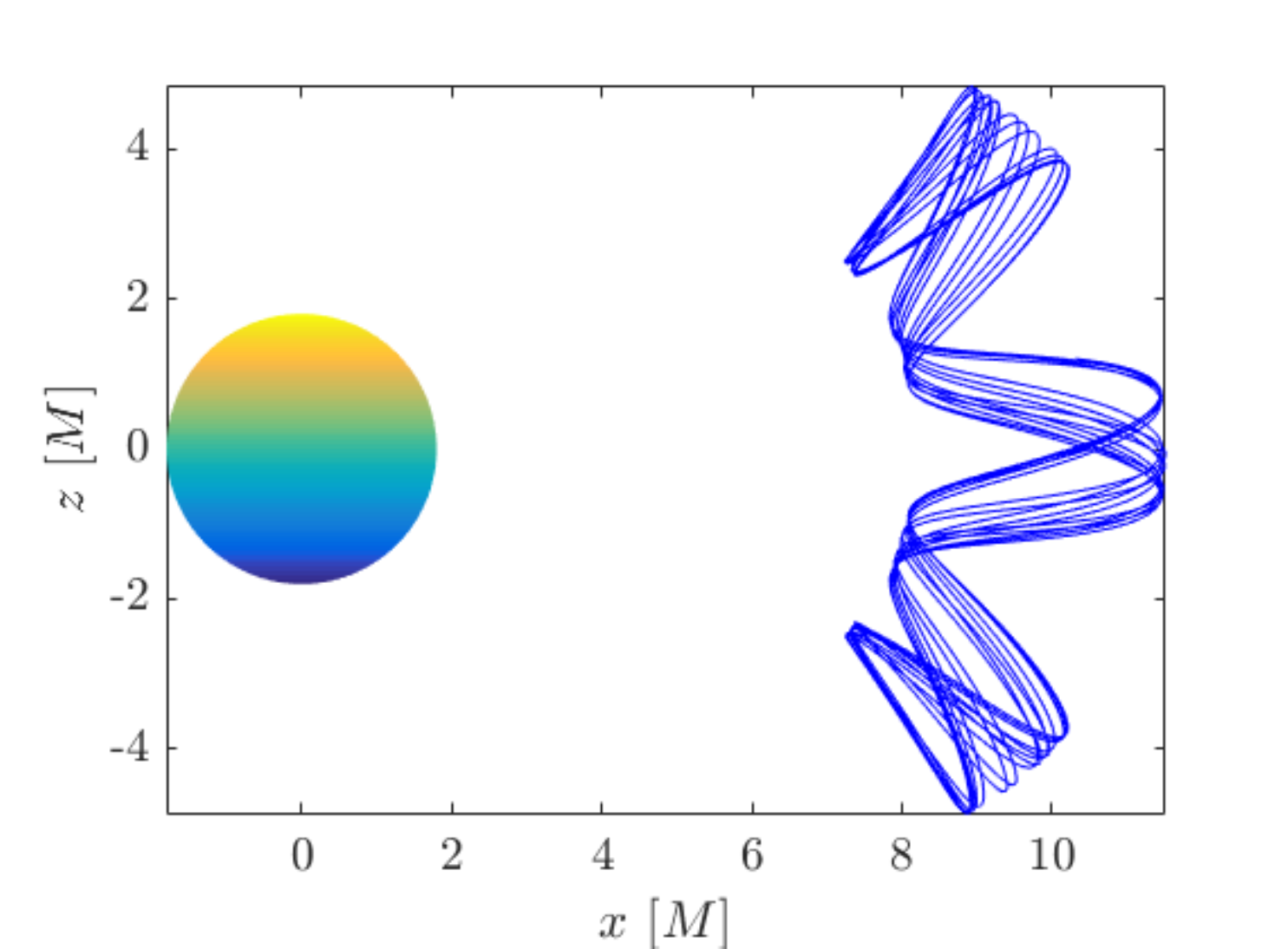}} \\
\caption{The regular orbit of a charged particle around a Kerr black hole in a dipolar field. The orbit is characterized by $E=0.885$ and $L=-7$. The initial four-velocity is such that $\chi=-0.01\pi$. The left panel shows the trajectory in three dimensions. In the right panel, a projection of the orbit in the poloidal plane shows the non-chaotic nature of the trajectory.}
\label{fig:DPorbit1}
\end{figure}

\begin{figure}[!h]
\centering
\subfloat[Trajectory in three-dimensional space.]{\includegraphics[width=.5\columnwidth]{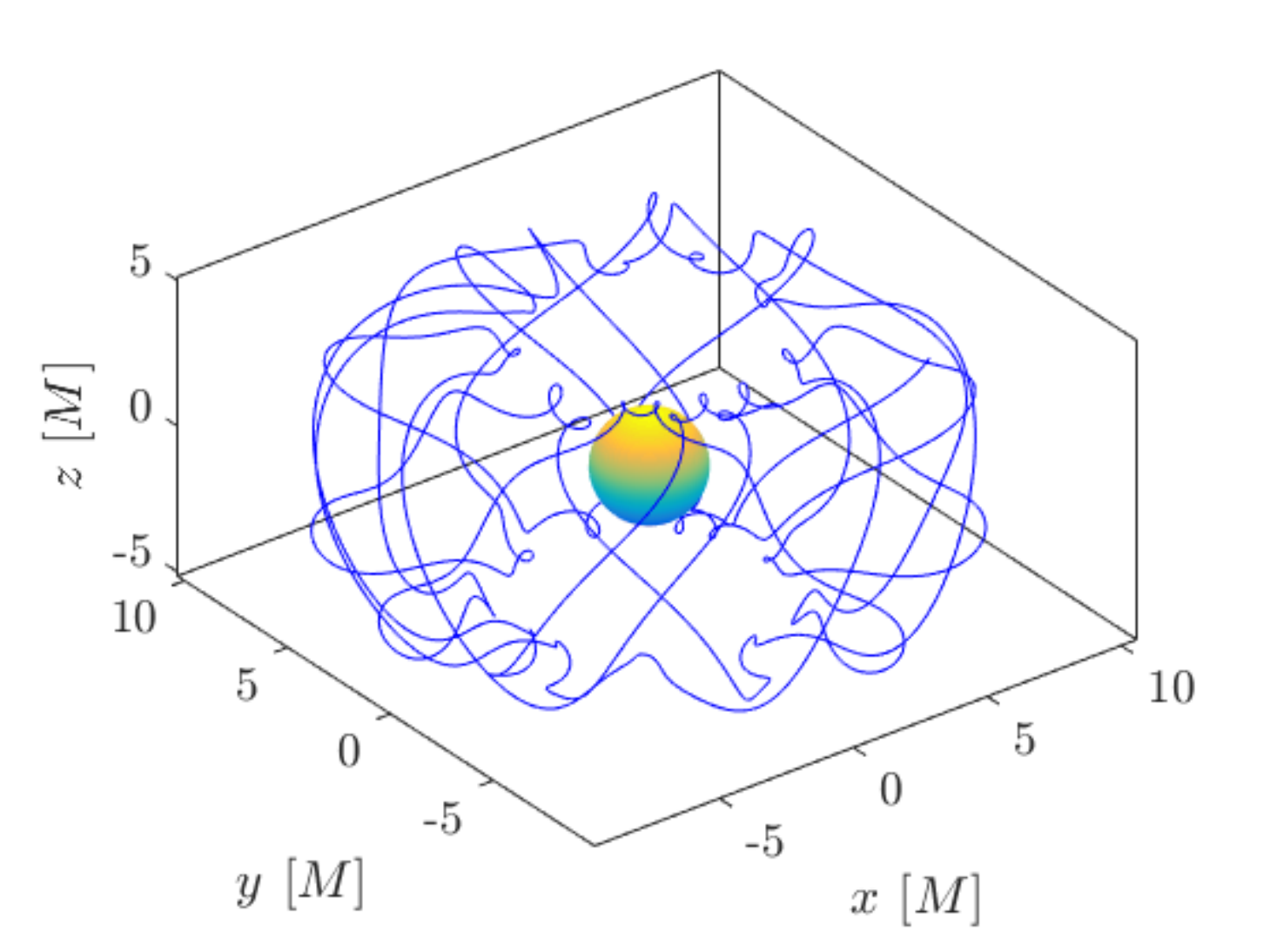}}
\subfloat[Projection in the poloidal plane.]{\includegraphics[width=.5\columnwidth]{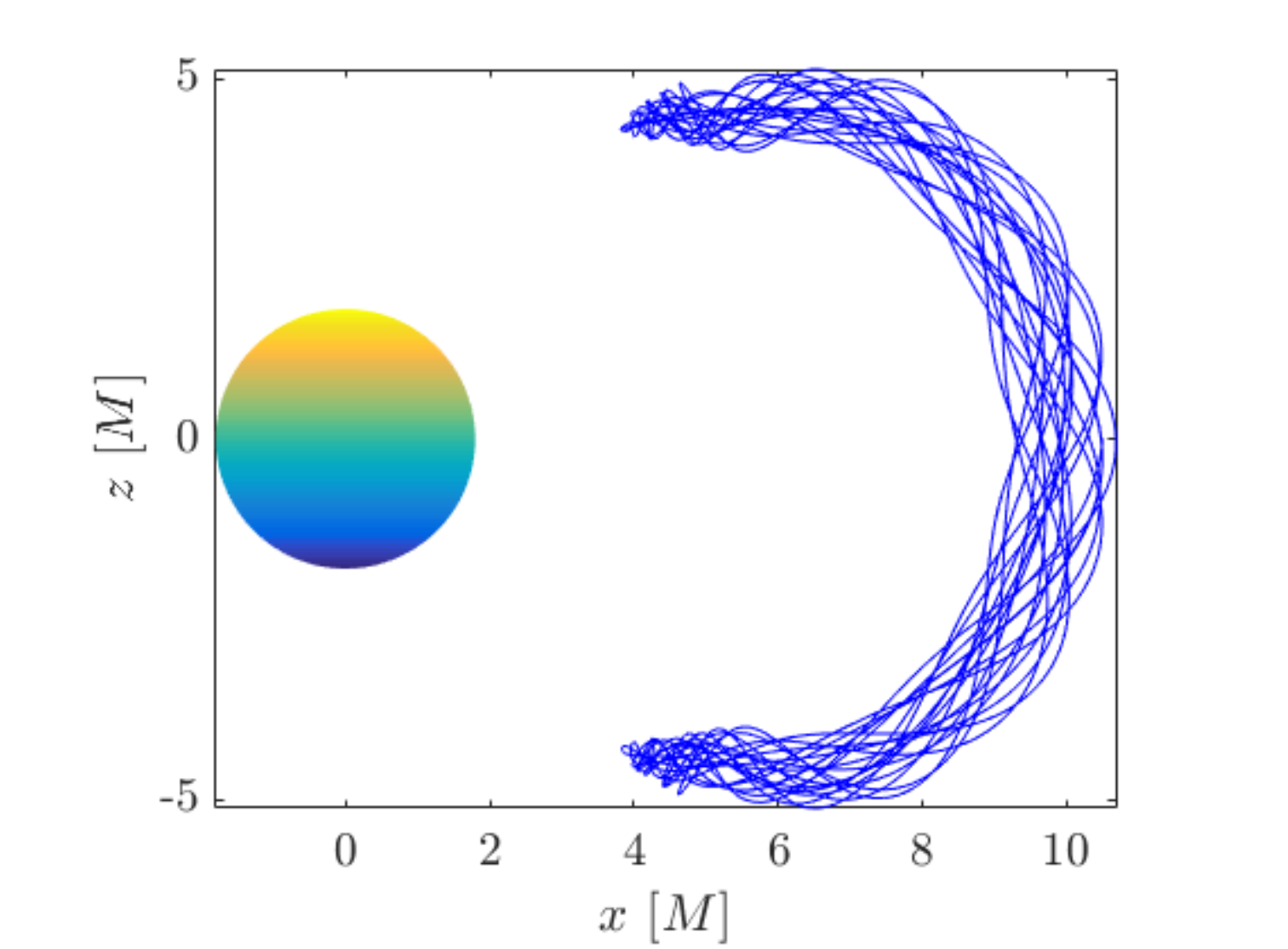}} \\
\caption{The chaotic orbit of a charged particle around a Kerr black hole in a dipolar field. The orbit is characterized by $E=0.89$ and $L=-7$. The initial four-velocity is such that $\chi=-0.3\pi$. The left panel shows the trajectory in three dimensions. In the right panel, a projection of the orbit in the poloidal plane shows the chaotic nature of the trajectory.}
\label{fig:DPorbit2}
\end{figure}

Figure \ref{fig:DPorbit1} shows the regular orbit characterized by $E=0.885$ and $L=-7$, around a Kerr black hole of spin $a=0.9$. The initial four-velocity corresponds to a parameter $\chi=-0.01\pi$. In the left panel, part of the trajectory is shown in three-dimensional space. In the right panel, a projection of the orbit on the poloidal plane evidences the regularity of the orbit. Figure \ref{fig:DPorbit2} analogously shows the chaotic orbit characterized by $E=0.89$, $L=-7$. Here, $a=0.6$ and $\chi=-0.3\pi$. The poloidal projection shown in the right panel clearly shows the chaotic nature of the motion, which uniformly fills the torus-like region associated to the effective potential determined by $E$ and $L$.


\begin{figure}[!h]
\centering
\subfloat[$E=0.885,L=-7$ regular orbit.]{\includegraphics[width=.5\columnwidth]{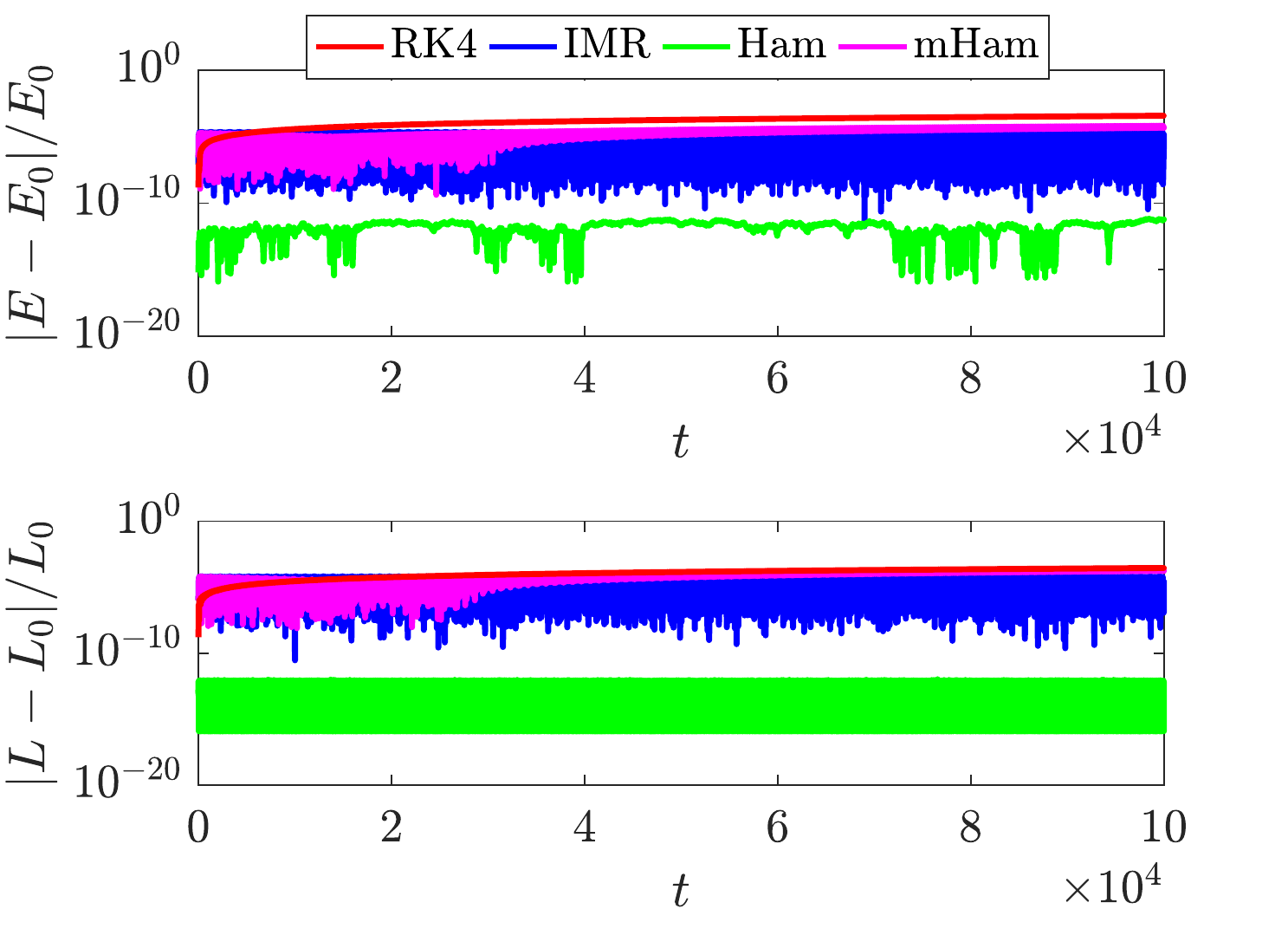}}
\subfloat[$E=0.89,L=-7$ chaotic orbit.]{\includegraphics[width=.5\columnwidth]{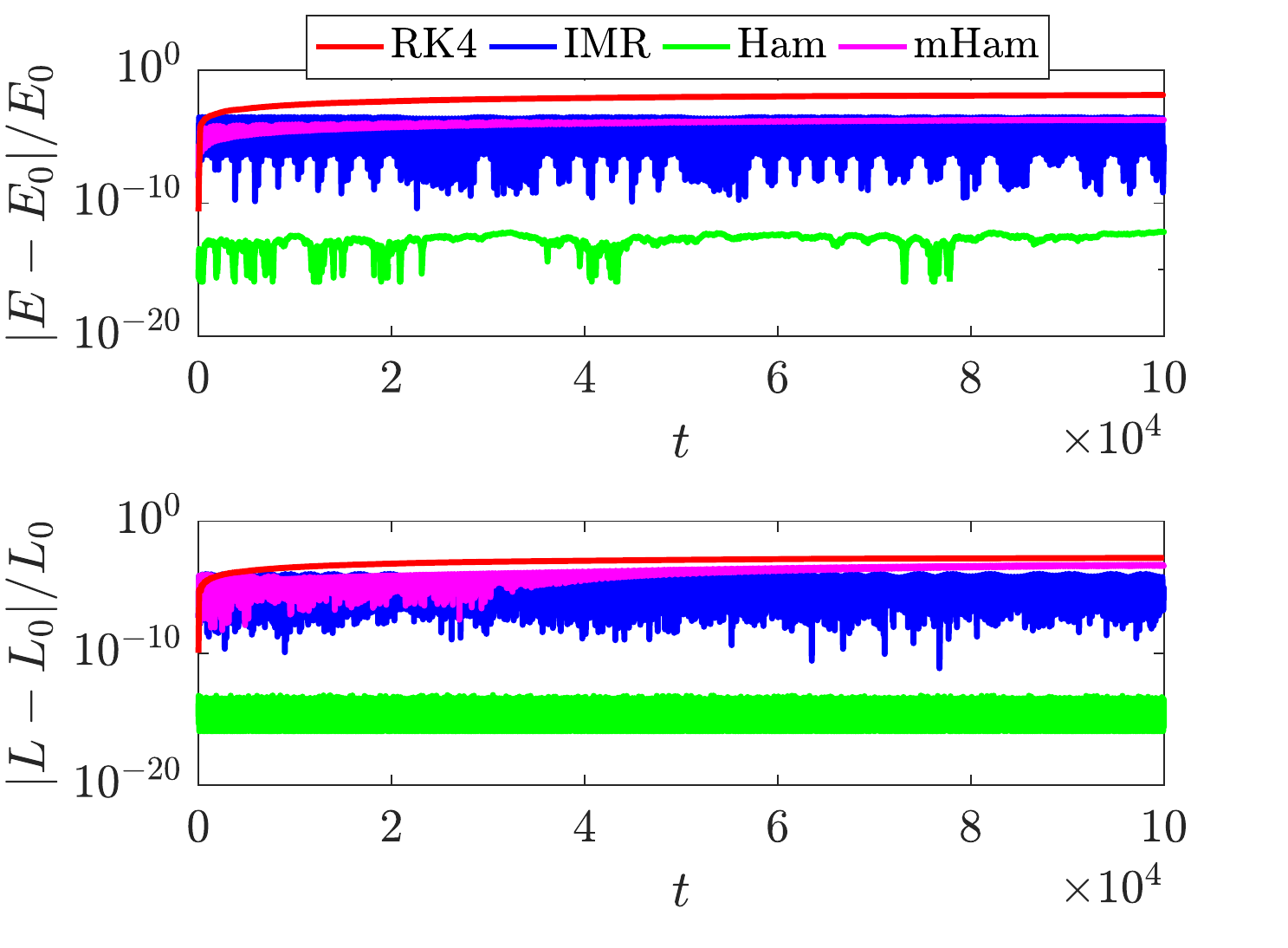}}
\caption{Evolution of the relative error in the conserved energy (top panels) and angular momentum (bottom panels) for the $E=0.885$, $L=-7$ regular orbit (left panels) and the $E=0.89$, $L=-7$ chaotic orbit (right panels). The IMR (blue lines) and modified Hamiltonian (magenta lines) schemes keep errors in energy and momentum bounded at all times. The original Hamiltonian scheme (green lines) preserves both quantities to machine precision, while the RK4 scheme (red lines) introduces a secular unbounded growth in the error leading to the spurious release of the particle from the orbit.}
\label{fig:DPorbiterr}
\end{figure}

As for the Wald configuration, we monitor the error in the conserved quantities of the motion $E$ and $L$. The results are shown in Figure \ref{fig:DPorbiterr}, where we report the evolution of the error in both energy and angular momentum for the $E=0.885,L=-7$ orbit (left panels) and the $E=0.89,L=-7$ orbit (right panels), for all four integrators. The behavior of the numerical error is very similar to the previous case, with the RK4 scheme showing an unbounded secular growth of the error for both quantities, that eventually leads to the spurious release of the particle from the orbit. The IMR and modified Hamiltonian schemes, instead, preserve the bounded trajectory and keep the errors limited. A mildly growing trend is observed again in the angular momentum error for the modified Hamiltonian method. The original Hamiltonian integrator further improves the conservation of both energy and angular momentum, with errors of the order of machine accuracy at all times.

\subsection{Unstable orbits in the Kerr-Newman spacetime}
\label{sec:KN}
While the combination of Schwarzschild or Kerr black holes in an external electromagnetic field represents a nonintegrable system, this is not the case when the source of electromagnetic fields is the black hole itself, via a nonzero electric (or magnetic) charge. In this case, the charge parameter intrinsically links the electromagnetic potential with the metric functions. As a result, the system presents a number of conserved quantities for the motion of test particles equal to the number of degrees of freedom, and the equations of motion can be solved analytically.

Charged black holes are considered purely theoretical objects which can hardly be observed in reality. In general, it is thought that the charge of any such object would be quickly counterbalanced by infalling particles of opposite charge, attracted from the surrounding environment. Hence, the vast majority of black holes in the universe are expected to be essentially neutral. Nevertheless, an idealized charged black hole in vacuum can be used as a suitable test-ground for the numerical integration of charged particles. Since the solution of the equations of motion can be found analytically, the numerical results can be compared directly to theoretical predictions, in order to obtain quantitative measures of the numerical error.

Here, we refer to the work by \cite{hackmannxu2013} to test the four numerical schemes against analytically known orbits in the Kerr-Newman spacetime of a charged, spinning black hole. The associated line element reads, in Boyer-Lindquist coordinates,
\begin{equation}
 ds^2 = \frac{\rho^2}{\Delta}dr^2 + \rho^2 d\theta^2 + \frac{\sin^2\theta}{\rho^2}\left[ (r^2+a^2)d\varphi - a dt\right]^2 - \frac{\Delta}{\rho^2} \left[ a \sin^2\theta d\varphi -dt\right]^2,
\end{equation}
where $\rho^2 = r^2+a^2\cos^2\theta$, $\Delta = r^2-2Mr+a^2+Q^2+P^2$, and $a$ is the black hole spin. The most general formulation for the Kerr-Newman spacetime includes $Q$ and $P$, the electric and magnetic charges of the black hole. In general, it is assumed that a magnetic charge could never manifest in classical physics. The black hole spin, mass, and charge are related by $a^2+Q^2+P^2\le M^2$, due to cosmic censorship.

The Kerr-Newman charge is the source for the intrinsic four-potential $A_\mu=(A_0,0,0,A_\varphi)$, where
\begin{equation}
 A_0 = \frac{Qr+aP}{\rho^2}\cos\theta,
\end{equation}
\begin{equation}
 A_\varphi = \frac{1}{\rho^2} \left[-aQr\sin^2\theta - (r^2+a^2)P\cos\theta \right],
\end{equation}
which define the conserved energy and angular momentum, respectively $-E = \pi_0 = u_0 + qA_0/m$ and $L = \pi_\varphi = u_\varphi + qA_\varphi/m$. An additional constant of the motion arises from the separability of the Hamilton-Jacobi equations in the form of the Carter constant $C$, given by (\citealt{hiscock1981})
\begin{equation}
 C = u_\theta^2 + a^2\cos^2\theta + \frac{1}{\sin^2\theta}\left(aE\sin^2\theta-L+\frac{qP}{m}\cos\theta\right)^2 - \left(L-aE\right)^2.
\end{equation}

The motion of particles in the Kerr-Newman spacetime is integrable, and the equation of motion \eqref{eq:geodesic} can be solved analytically. Here, we quantify the numerical error introduced by each integration scheme by simulating analytically-derived unstable orbits. The derivation of the orbit parameters is explained in detail in Appendix \ref{app:KN}, where we show the step-by-step construction of arbitrary spherical orbits in this spacetime. With this procedure, we identify the orbits summarized in Table \ref{tab:KNorbits}. Even though we give here precise numerical values for the initial simulation parameters, because of their unstable nature the orbits are extremely sensitive to the values of position and momentum. Hence, to reproduce the orbits shown here, one should rather follow the procedure in Appendix \ref{app:KN} to retrieve the values of the initial parameters with sufficient precision.

\begin{table}[!h]
\centering
\begin{tabular}{|c|c|c|c|c|}
\hline
Orbit name & $qQ/m$ & $L$ & $E$ & $r_0$ \\ 
\hline 
A & 0.9 & 1 & 1.00885 & 2.61044 \\ 
\hline
B & 1.1 & 1 & 1.09032 & 2.11159 \\ 
\hline
C & 1.1 & 1.5 & 1.16215 & 1.84050 \\ 
\hline
D & 2 & 1 & 1.53422 & 1.69458 \\ 
\hline
E & 10 & 10 & 8.12266 & 1.49386 \\
\hline
F & 10 & -10 & 3.69550 & 1.77958 \\
\hline
\end{tabular} 
\caption{Parameters for the unstable spherical orbits in the Kerr-Newman spacetime. In all cases we fix $a=0.6$, $K = 1$, $Q^2+P^2=0.4$ and we impose $P=Q$. Then, we choose values for $qQ/m$ and $L$ and we solve $u^r=du^r/d\tau=0$ from \eqref{eq:urKN} for $E$ and $r_0$.}
\label{tab:KNorbits}
\end{table}

\begin{figure}[!h]
\centering
\subfloat[Orbit B.]{\includegraphics[width=0.5\columnwidth]{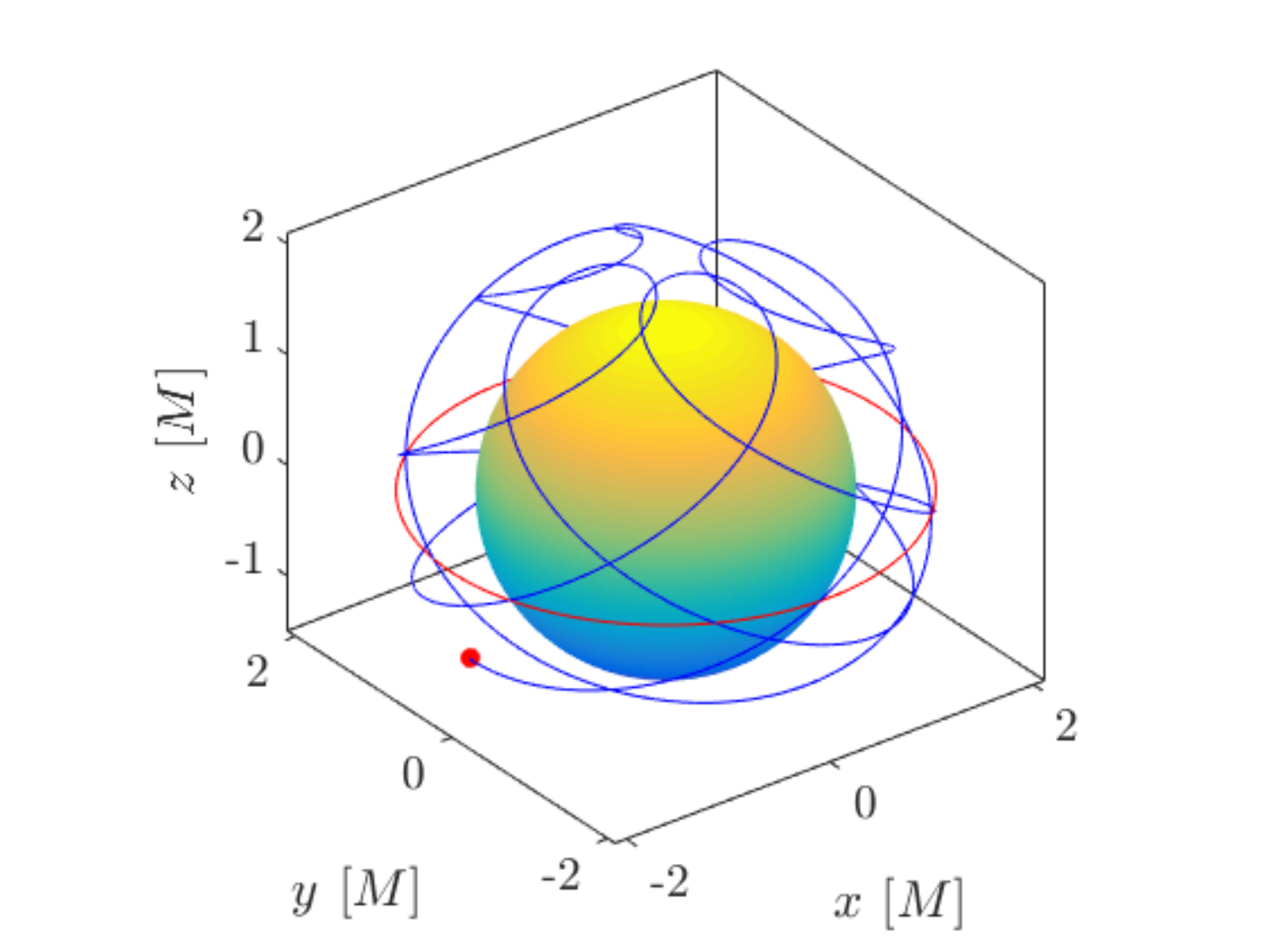}}
\subfloat[Orbit D.]{\includegraphics[width=0.5\columnwidth]{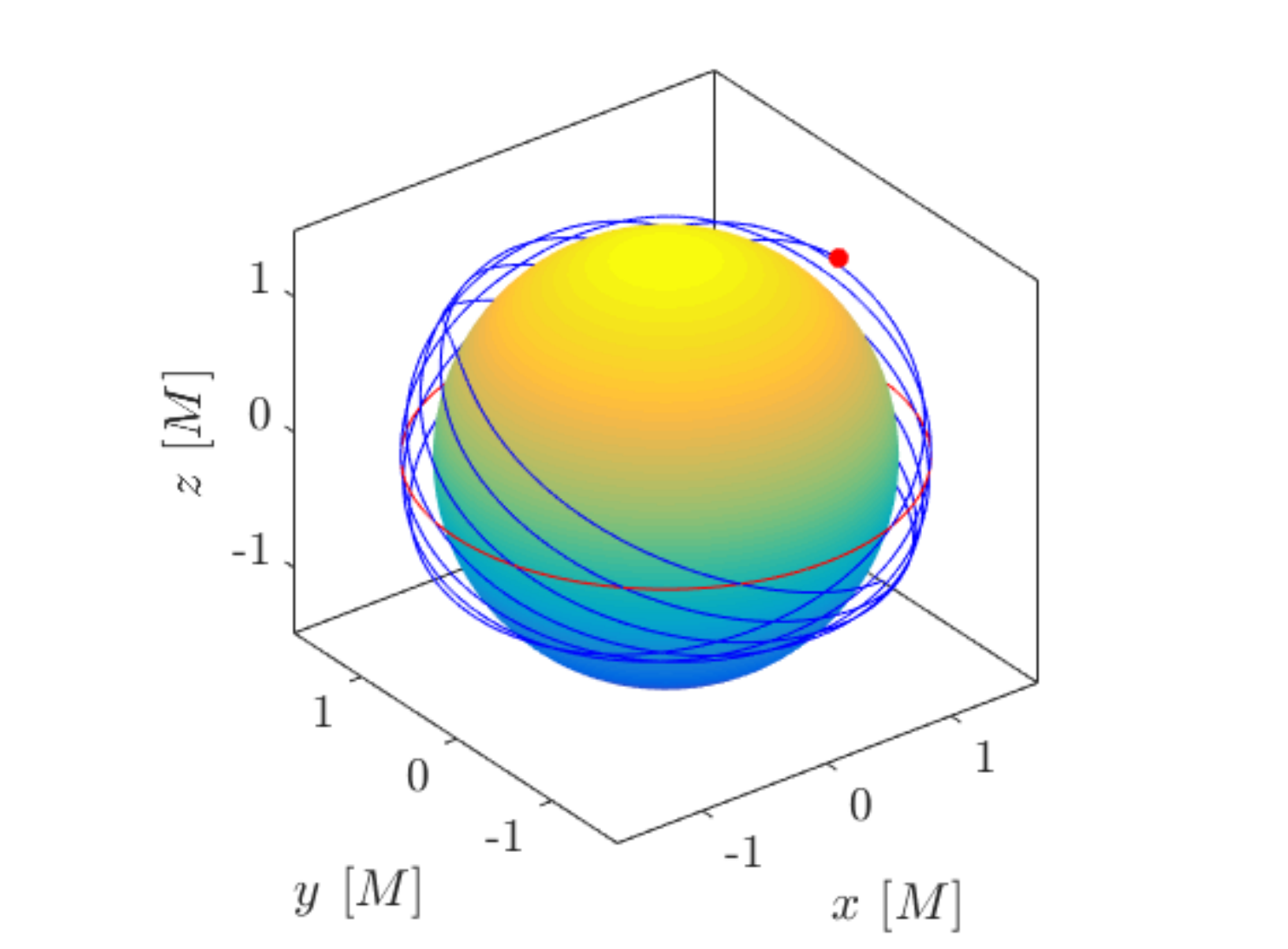}}
\caption{Representation in three-dimensional space of the unstable spherical orbits B (left panel) and D (right panel) from Table \ref{tab:KNorbits}. The outer event horizon is shown as a colored sphere of radius $r_+ = M \pm \sqrt{M^2-a^2-Q^2-P^2}$. The equatorial red circle indicates the constant radius $r_0$ characterizing each orbit. The starting point of the orbits is marked by a red dot.}
\label{fig:KNorbitBD}
\end{figure}

The simulation of each orbit is carried out until $t=1000$ (for orbits E and F, which are bound to smaller regions of space and thus require smaller time steps, until $t=500$ and $t=75$ respectively). Each spherical path is calculated with all four integrators for a range of $\Delta t$. Throughout the computation, we monitor the deviation of the radial position from the initial value $r_0$, as well as the error in the conserved quantities $E$, $L$, and $C$. Figure \ref{fig:KNorbitBD} shows a part of orbit B (left panel) and D (right panel). The outer event horizon is represented as a colored sphere of radius $r_+$. The sense of rotation of the black hole is from left to right. The red circle of radius $r_0$ indicates the theoretical value of $r$ at which the orbit should remain in the absence of perturbations. The initial starting point of each orbit is marked with a red dot.

\begin{figure}[!h]
\centering
\subfloat[Orbit B.]{\includegraphics[width=0.5\columnwidth]{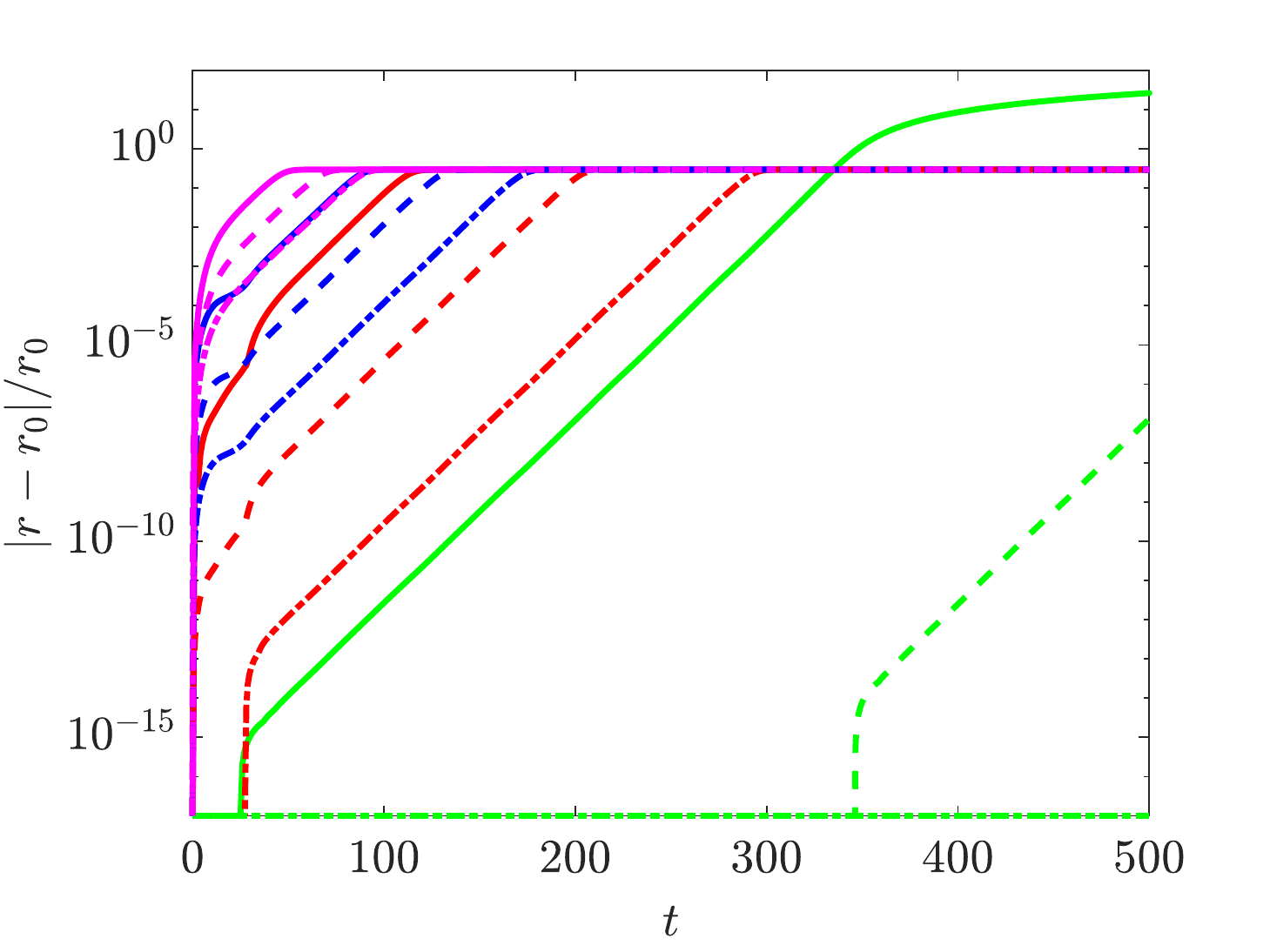}}
\subfloat[Orbit D.]{\includegraphics[width=0.5\columnwidth]{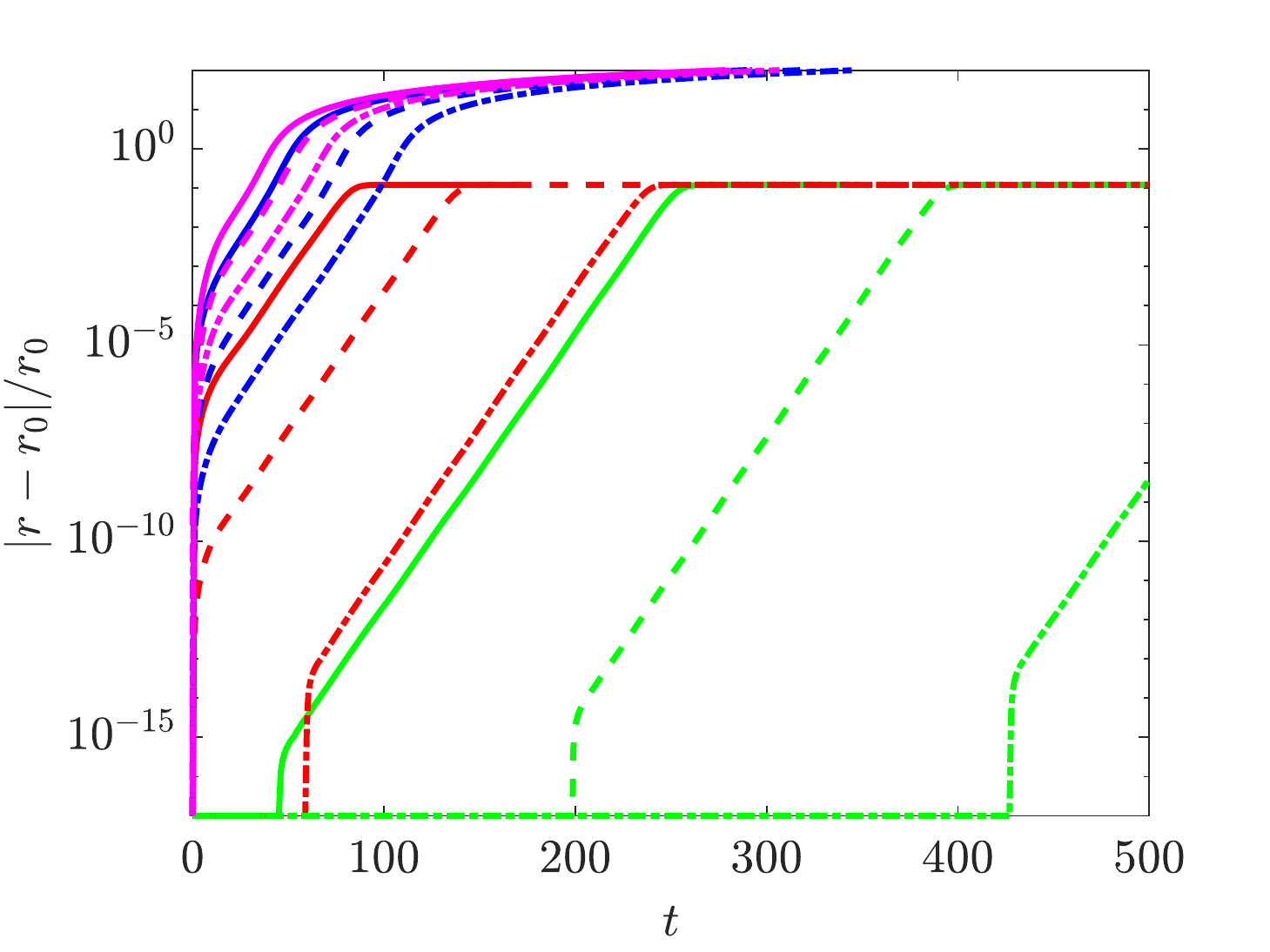}}
\caption{Evolution of the relative error in the initial radius for orbit B (left panel) and D (right panel) simulated with $\Delta t=1$ (solid lines), $\Delta t=0.1$ (dashed lines), and $\Delta t=0.01$ (dash-dotted lines) until $t=500$. The Hamiltonian scheme (green lines) performs better than the RK4 (red lines), IMR (blue lines), and modified Hamiltonian (magenta lines) schemes, triggering the linear growth of error and the release of the particle from the (physically unstable) bound orbits at much later times.}
\label{fig:KNorbitBDr}
\end{figure}

Figure \ref{fig:KNorbitBDr} shows the time evolution, until $t=500$, of the relative error in the radius, measuring deviations from the theoretical value $r_0$, for orbits B (left panel) and D (right panel), for $\Delta t=1$ (solid lines), $\Delta t=0.1$ (dashed lines), and $\Delta t=0.01$ (dash-dotted lines). The results are analogous to those from the unstable spherical photon orbits analyzed in \citetalias{bacchini2018a}: an initial exponential growth of the error in $r_0$ is observed, until the particle is released from the bound orbit. The deviation from the physically unstable path is triggered by numerical errors, with less accurate schemes causing larger perturbations and therefore an earlier release from the orbit. In all cases, the Hamiltonian scheme preserves the motion on the initial radius $r_0$ far longer than all other schemes, which instead introduce much larger perturbations at earlier times. Even with the smallest time step $\Delta t=0.01$, the RK4 scheme performs worse than the Hamiltonian scheme at the largest time step $\Delta t=1$, despite the former being of higher order than the latter.

\begin{figure}[!h]
\centering
\subfloat[Orbit B.]{\includegraphics[width=0.5\columnwidth]{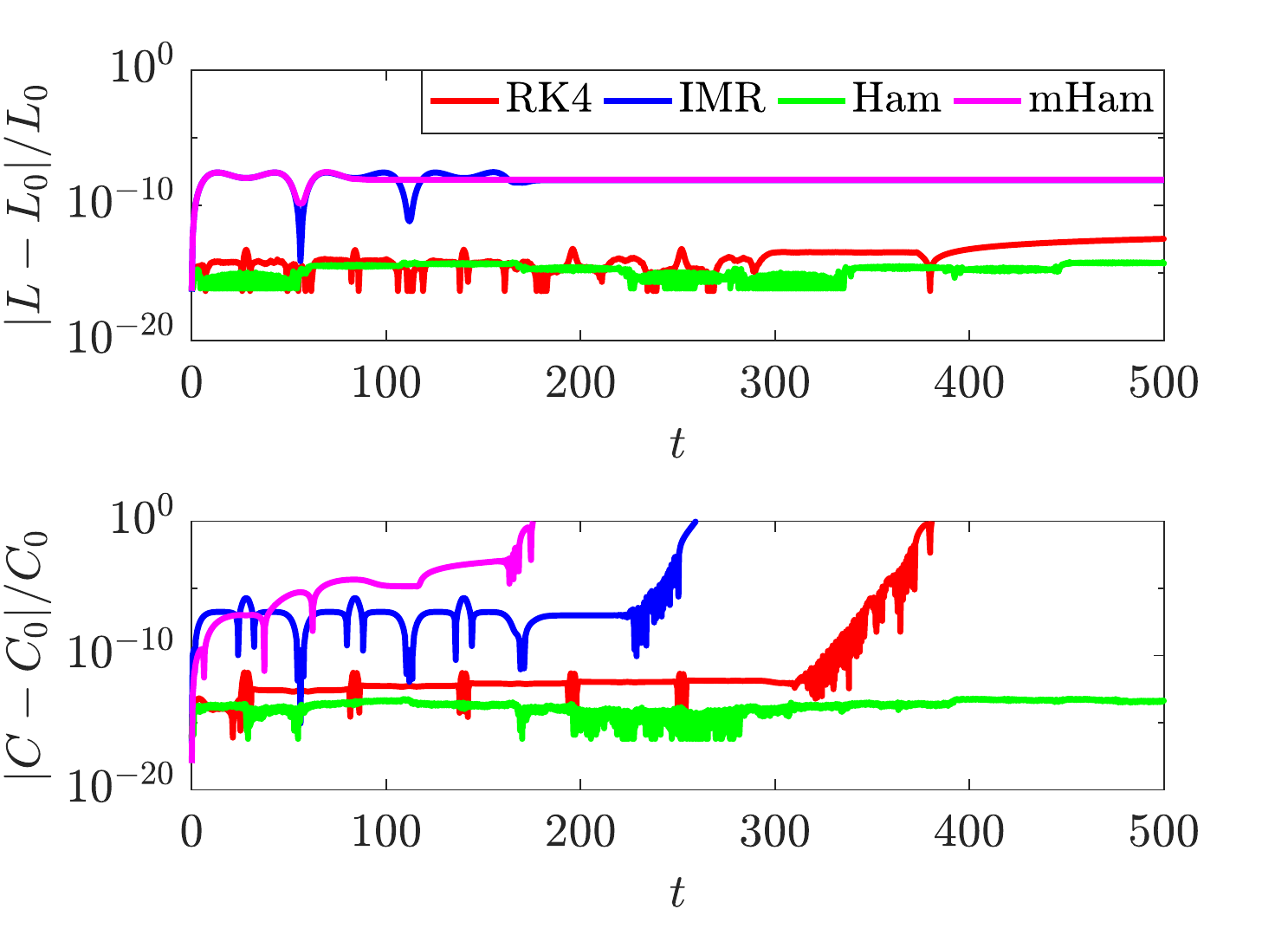}}
\subfloat[Orbit D.]{\includegraphics[width=0.5\columnwidth]{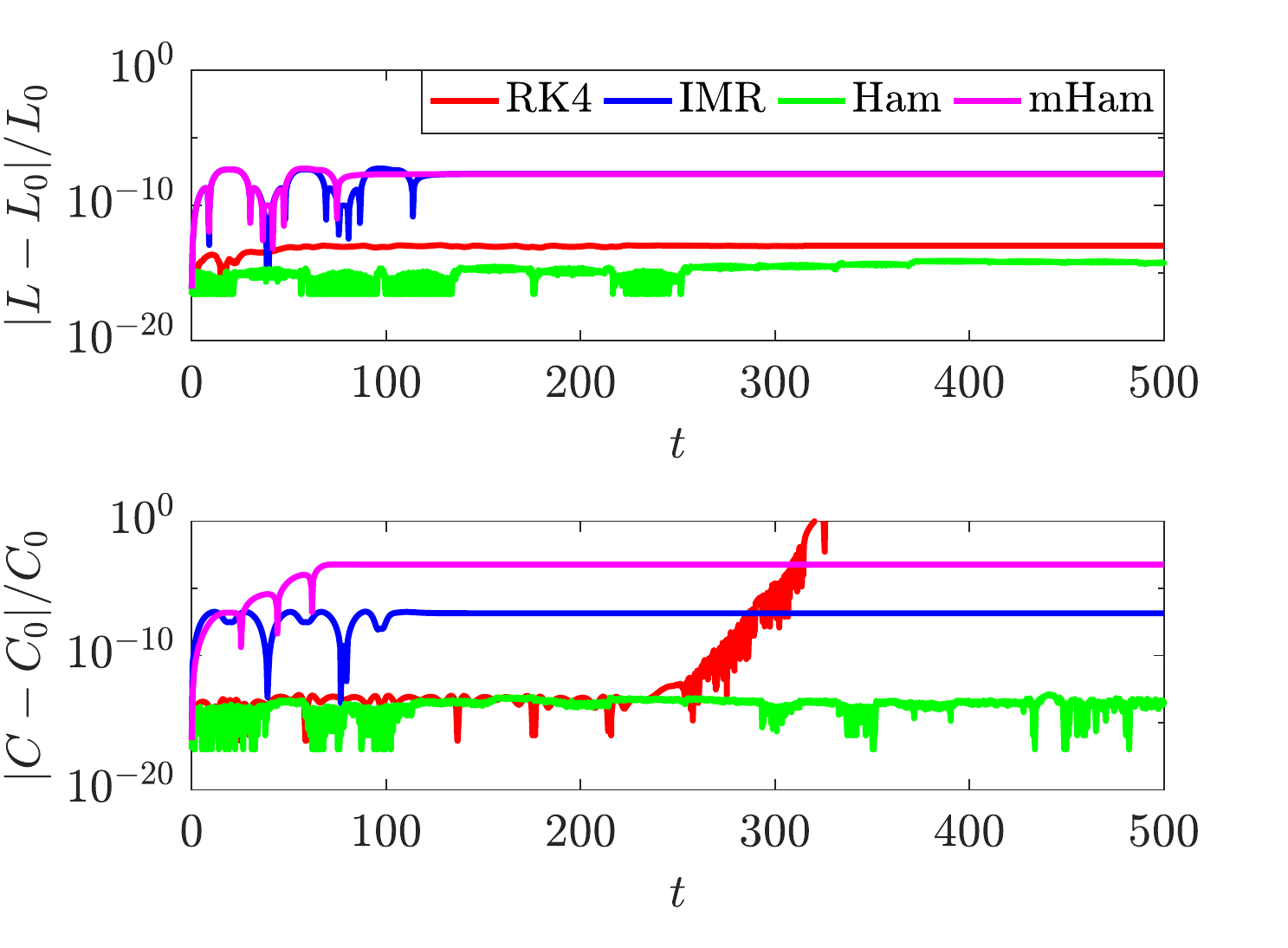}}
\caption{Evolution of the relative error in the conserved angular momentum (top panels) and the Carter constant (bottom panels) for orbit B (left panels) and D (right panels) simulated with $\Delta t=0.01$ until $t=500$. The Hamiltonian scheme (green line) conserves both quantities, together with the energy, to machine precision. The RK4 (red line), IMR (blue line), and modified Hamiltonian (magenta lines) schemes, instead, introduce errors which are larger by several orders of magnitude.}
\label{fig:KNorbitBDerr}
\end{figure}

Finally, in Figure \ref{fig:KNorbitBDerr} we show the time evolution of the relative error in $L$ and $C$ for the smallest $\Delta t=0.01$, for the same orbits B and D. The results clearly show that the energy-preserving character of the Hamiltonian integrator is also associated, in this case, to the exact conservation of all other invariants of the motion, even after the release of the particle from the $r=r_0$ unstable orbit. The performance of the Hamiltonian scheme is therefore superior to that of all other integrators, which exhibit larger (by several orders of magnitude) errors in all conserved quantities and the release of the particle from the bound orbit at much earlier times.

For all orbits in Table \ref{tab:KNorbits}, we observe that the Hamiltonian scheme produces better-quality results in terms of conservation of all invariants ($E$, $L$, and $C$), while showing much less sensitivity to radial perturbations that drive the particle trajectory away from the initial radius $r_0$. The RK4 scheme performs in general better than the IMR scheme, but both require extreme reductions of the time step in order to achieve the same performance of the Hamiltonian scheme, greatly increasing the computational costs. The modified Hamiltonian scheme exhibits errors in $L$ comparable to those of the IMR scheme. However, numerically-induced perturbations to the spherical motion seem to arise earlier. This feature can be attributed to the chosen formulation of the method, which is obtained as a combination of the original Hamiltonian method and the IMR scheme. The resulting ``mixed'' character of the integrator allows for the properties discussed in Section \ref{sec:schemesBHAC} (exact energy conservation for vanishing electric fields), but also introduces stronger deviations in the particle trajectories. This result confirms the findings presented in Part I, where unstable geodesic paths were studied. While the Kerr-Newman solution does not reflect realistic astrophysical situations, the results presented here and in the previous Sections clearly confirm the higher physical reliability of energy-preserving schemes such as the Hamiltonian integrator. Compared to that of standard, same-order symplectic schemes such as the IMR, or even higher-order explicit schemes such as the popular RK4, the new Hamiltonian scheme exhibits higher accuracy at large time steps. This allows for inexpensive but reliable simulations of charged particles under the combined action of static electromagnetic fields and stationary curved spacetimes.

\section{Application to test particle simulations in GRMHD}
\label{sec:BHAC}
The main target application for general relativistic charged particle integrators is the modeling of ensembles of particles in physically realistic electromagnetic field configurations. State-of-the-art simulations of plasmas commonly model the dynamics of accretions disks and jet formation processes within the general relativistic magnetohydrodynamic (GRMHD) framework (\citealt{rezzollazanotti}). Such a model implicitly considers quasi-neutral, collision-dominated plasmas and evolves the fluid equations to simulate the global dynamics of the fluid and electromagnetic quantities. Data from such simulations are then used to reconstruct emission spectra (with e.g.\ post-processing tools, see \citealt{dexter2016}; \citealt{chan2017}; \citealt{porth2017}; \citealt{bronzwaer2018}) by solving the radiation transfer equation and calculating very large numbers of geodesic paths for the emitted photons. The final outcome consists of synthetic radiation maps that can be compared to forthcoming observations of accretion flows (e.g.\ in the context of the Event Horizon Telescope). The main limitations of this model include the absence of temperature decoupling between the various particle species which in reality constitute the plasma (\citealt{ressler2017}; \citealt{ryan2017}; \citealt{chael2018}), as well as the need for the assumption of a non-thermal particle distribution necessary to match the observed radiation spectra (\citealt{porth2011}).

A more realistic approach to the problem is to inject charged particles in the electromagnetic field configuration produced by GRMHD simulations and let the particles evolve under the combined influence of the gravitational and electromagnetic fields. This strategy is based on the assumption that the energy content of the plasma is mainly associated to thermal particles. The injected particles then represent a small small, non-thermal population with negligible effect on the electromagnetic fields. Such a ``test particle'' approach can be used to produce more realistic particle distributions where no assumption is made on the nonthermal energy spectrum associated to particle acceleration. With this data, synthetic radiation maps could be constructed more accurately for better matching observational data.

A further step on this path consists of calculating the particle feedback on the electromagnetic fields, as is the case for e.g.\ special relativistic simulations with Particle-in-Cell (PiC) codes (e.g.\ \citealt{spitkovsky2005}; \citealt{cerutti2013}). Currently, general relativistic PiC (GRPiC) algorithms (that take into account the spacetime curvature) are being developed (\citealt{levinsoncerutti2018}; \citealt{parfrey2018}) and will hopefully self-consistently produce interesting insight into the microscopic dynamics of plasmas around compact objects. The charged particles integrators presented in this work, as well as in \citetalias{bacchini2018a}, are a necessary component of the GR-PiC approach.

In the next Sections, we consider test particle simulations where information on the fields is taken from GRMHD calculations. These are obtained with the particle integrators presented in this work. Here, we show the results of a test run obtained using data from the GRMHD code \texttt{BHAC} (\citealt{porth2017}).

\subsection{Effect of interpolation}
\label{sec:interpBHAC}
GRMHD codes such as \texttt{BHAC} model the fluid dynamics of plasmas and the evolution of the electromagnetic fields on a computational grid. In applying the particle integrators presented in this work, interpolation of the field quantities from the grid points onto the particle position becomes necessary. Here, we evaluate the effect of interpolation on the accuracy of the four integrators. In typical test particle simulations, we apply trilinear interpolation of $D^i$, $B^i$, and $A_\mu$ in the three spatial directions, such that e.g.\ the electric field at the particle location is given by 
\begin{equation}
 D^i(\xvec_p) = \sum_g D^i_g W_1(x_p^1-x_g^1)W_2(x_p^2-x_g^2)W_3(x_p^3-x_g^3),
\end{equation}
where the directional interpolation functions $W_i$ relate the values of $D^i$ at the grid points (subscript $g$) with the interpolated value at the particle position (subscript $p$).

Aside from an additional interpolation step, the integration algorithm remains unchanged. The interpolated fields are used in the equations of motion according to the chosen numerical scheme, in place of analytically-derived values. The interpolation procedure introduces an additional source of error, which is reduced as the spatial resolution of the grid increases. Note that the energy preservation properties of the original Hamiltonian scheme, based on the assumption that the four-potential $A_\mu$ is available analytically, in this case are inevitably lost. The modified Hamiltonian scheme, instead, retains its properties (exact energy conservation in the limit of vanishing electric fields) even when grid-defined quantities are employed.

In order to assess the effect of interpolation on the performance of the four integrators applied to the equations of motion, we consider again the particle orbits discussed in Section \ref{sec:tests}. As representative examples, we choose orbit RSA5 in the Wald solution from Table \ref{tab:wald} in Section \ref{sec:wald} and the chaotic orbit with $E=0.89, L=-7$ in the dipole solution from Section \ref{sec:DP}. We simulate both orbits with $\Delta t=1$ until $t=100000$, monitoring the conservation of both energy $E$ and angular momentum $L$. The physical domain extends over $r \in [r_+, 15]$, $\theta\in[0,\pi]$, $\varphi\in[0,2\pi]$ and it is discretized with an increasing number of grid points, from $16\times16\times32$ up to $64\times64\times128$.

\begin{figure}[!h]
\centering
\subfloat[Regular orbit RSA5.]{\includegraphics[width=.5\columnwidth]{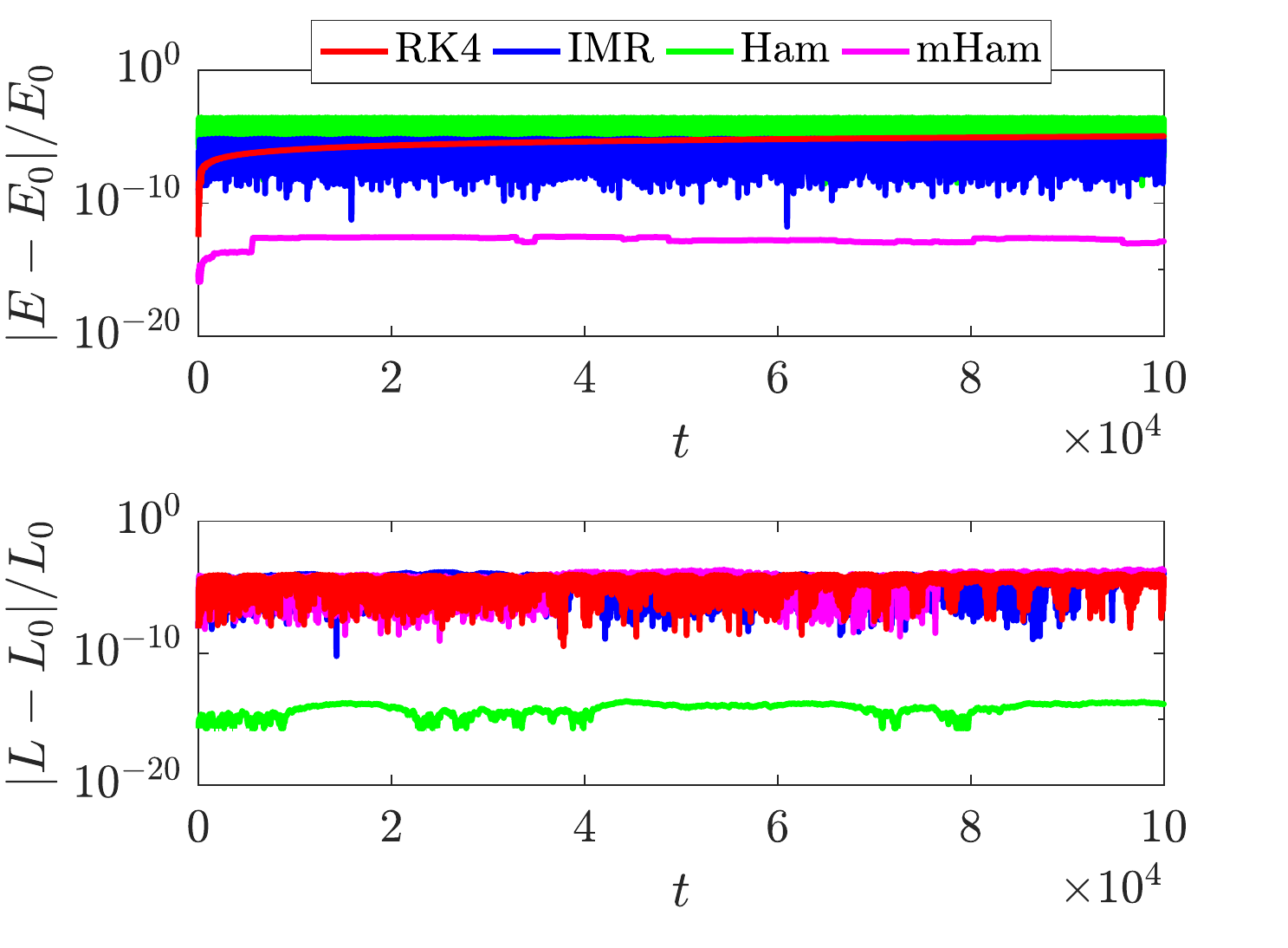}}
\subfloat[$E=0.89,L=-7$ chaotic orbit.]{\includegraphics[width=.5\columnwidth]{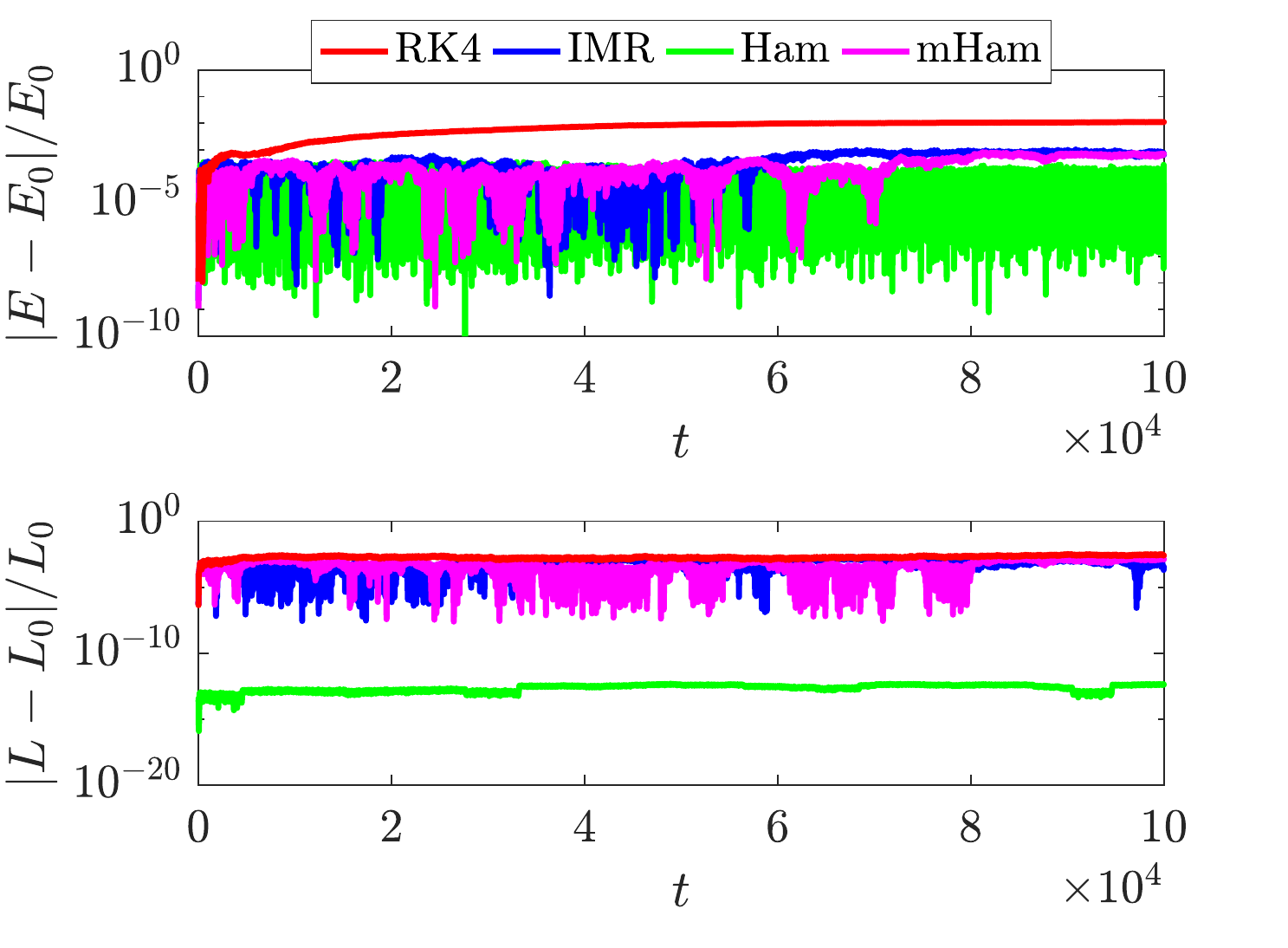}}
\caption{Evolution of the relative error in the conserved energy (top panels) and angular momentum (bottom panels) for the regular orbit RSA5 from section \ref{sec:wald} (left panels) and the $E=0.89,L=-7$ chaotic orbit from section \ref{sec:DP} (right panels), using interpolated values of the electromagnetic fields. The simulation was run on a $64\times64\times128$ grid with $\Delta t=1$ until $t=100000$. The modified Hamiltonian scheme (magenta lines) preserves the energy to machine precision ($\sim$ $10^{-15}$ in our simulation) in case of vanishing electric fields (top left panel), while the original Hamiltonian scheme (green lines) loses its energy-conserving character due to the interpolation step. In all cases, the RK4 scheme introduces a secular unbounded growth in the energy error, while the IMR scheme keeps energy errors bounded and with the same accuracy of the Hamiltonian schemes. The original Hamiltonian method retains exact momentum conservation in all cases, while the IMR and modified Hamiltonian schemes perform equally well in keeping errors in $L$ bounded in time.}
\label{fig:orbitsinterr}
\end{figure}

The results are shown in Figure \ref{fig:orbitsinterr} for the highest spatial resolution. The left panels shows the evolution of the relative error in $E$ and $L$ for the regular Wald orbit RSA5. This configuration is characterized by zero electric fields in the whole domain. Hence, as expected, the modified Hamiltonian method retains energy conservation to machine precision, thus much better than the RK4 and IMR schemes by orders of magnitude. The original Hamiltonian scheme, now employing grid-defined values of $A_\mu$, inevitably loses the energy-conserving character, and exhibits bounded energy errors slightly above those affecting the IMR scheme. The right panels show the results for the chaotic dipole orbit with $E=0.89$, $L=-7$, where an electric field is present. As predicted, energy conservation is lost in the results of the Hamiltonian integrator, as well as in those of the modified Hamiltonian scheme. Both now presents an error in $E$ of the same order of that shown by the IMR scheme, without observed secular trends. For both simulations, the RK4 scheme retains an unbounded growth in the energy error which eventually leads to the spurious escape of the particle from the bound orbit. In the simulation of both orbits, exact conservation of momentum is retained by the original Hamiltonian scheme, due to the formulation of the equations of motion in terms of the conjugate momentum. This property is not retained in the the modified Hamiltonian scheme, which exhibits a relative error in $L$ of the same order as the other methods, and bounded in time. We repeat the simulations varying the spatial resolution as explained above, and for all schemes we detect an improvement in both energy and angular momentum conservation by approximately one order of magnitude each time the grid resolution is doubled.

These results, based on the use of grid-defined electromagnetic fields, confirm that explicit schemes such as RK4 are prone to unbounded energy errors which both suppress physical features such as the particle gyration around magnetic field lines, and ultimately cause an unphysical drift in the particle trajectory. The symplectic nature of the IMR scheme, instead, preserves gyration and keeps errors in the conserved quantities bounded over very long times. The modified Hamiltonian scheme performs as well as the IMR scheme, with the additional advantage of conserving energy exactly when the electric field vanishes. The original Hamiltonian scheme inevitably loses exact energy conservation, due to the interpolation step. Finally, both the IMR and the modified Hamiltonian schemes, which are second-order accurate, can preserve additional invariants of the motion such as the angular momentum as accurately as the fourth-order accurate RK4 method. The original Hamiltonian method retains exact conservation of $L$, due to the formulation of the equations of motion in terms of the conjugate momentum.

The fact that no spurious work due to magnetic fields and curvature terms is introduced by the modified Hamiltonian scheme also implies that all work done on the particles in the simulation is only attributed to the electric field. This is a desirable property for numerical integrators of charged particles, which is nontrivial to achieve even in special relativistic calculations (see \citealt{ripperda2018}). In Particle-in-Cell (PiC) codes, this is an essential requirement for attaining exact conservation of the total energy, in terms of the sum of kinetic and electromagnetic contributions: the total variation in the kinetic energy of the particles corresponds exactly (in absolute value) to the total variation in electromagnetic energy. Globally energy-conserving PiC codes exhibit high stability and the elimination of numerical instabilities that affect state-of-the-art traditional PiC codes (\citealt{lapentamarkidis2011}; \citealt{lapenta2017a}; \citealt{lapenta2017b}). The possibility of achieving such a result in general relativistic calculations is an attractive one, which deserves more in-depth investigations in the future. Note that conservation of the global energy does not require conservation of the Hamiltonian at the single-particle level, pursued in this work through the original Hamiltonian integrator. Rather, the conservation of the Hamiltonian in a particle simulation, if achievable, imposes that each single particle exchanges the physically correct amount of energy with the electromagnetic field. Compared to the conservation of the global energy, this involves a higher degree of accuracy: not only is global energy conserved, but also the variation of energy of each particle becomes more accurate.

In the next Section, we apply the integrators to an example simulation of test particles in \texttt{BHAC}, which employs a formulation of the electromagnetic fields in terms of the quantities $D^i$, $B^i$. Because the four-potential $A_\mu$ is not available in this code, numerical schemes relying on a vector potential formulation, such as the Hamiltonian integrator presented here, cannot be employed. Hence, for the test-case considered, we only employ the RK4, IMR, and modified Hamiltonian integrators (with the latter being specifically constructed for the implementation in GRMHD codes).

\subsection{Test particles in a fixed GRMHD background}
\label{sec:testBHAC}
\begin{figure}[!h]
\centering
\subfloat{\includegraphics[width=0.3944\columnwidth, trim={15mm 0mm 15mm 0mm}, clip]{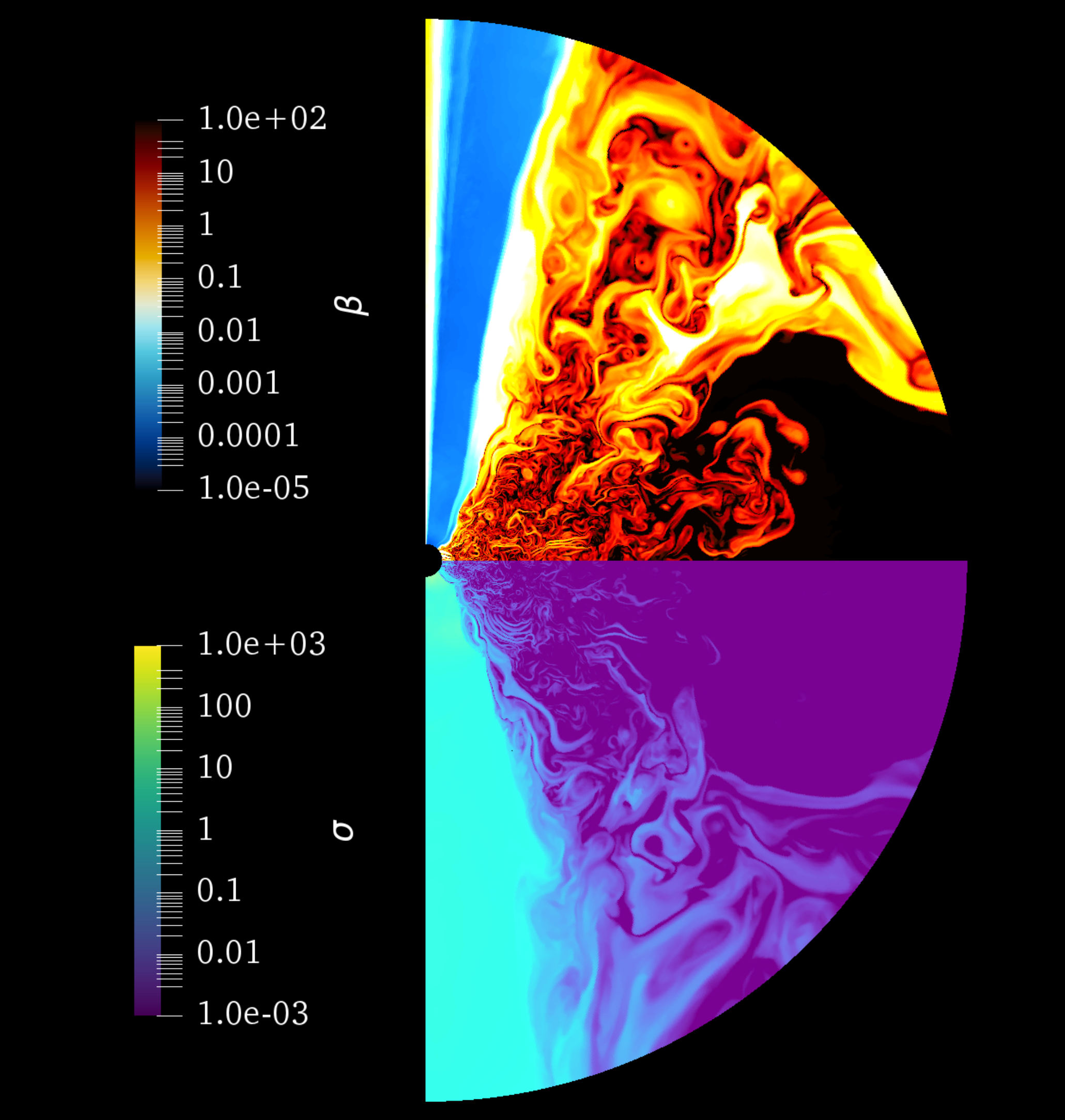}}
\subfloat{\includegraphics[width=0.6056\columnwidth, trim={0mm 0mm 0mm 0mm}, clip]{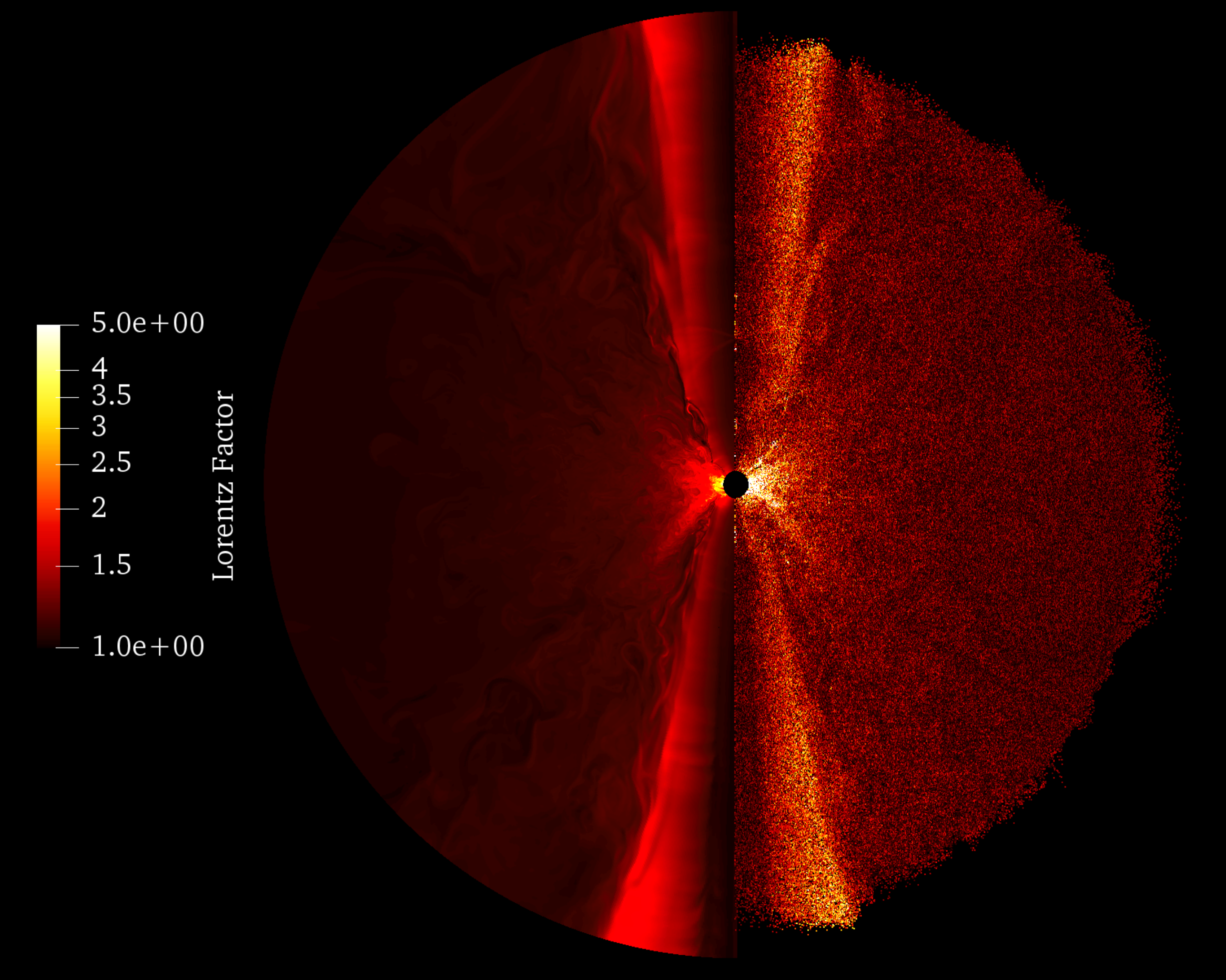}}
\caption{Spatial distribution of the dimensionless $\beta=2p/B^2$ and $\sigma=B^2/\rho$ (left panel) from the GRMHD simulation with \texttt{BHAC}, at the late stage of the development of the MRI. The Lorentz factor of the fluid from the GRMHD simulation is also shown (right panel, left half) and compared to that of $10^6$ particles (right half) with $q/m=10^4$, integrated with the modified Hamiltonian method until $t=5$.}
\label{fig:FMfar}
\end{figure}

As an example of application of the particle integrators tested in the previous sections, we consider the motion of charged particles in a two-dimensional ideal GRMHD snapshot obtained with \texttt{BHAC} (\citealt{porth2017}). The GRMHD simulation was initialized with a stable Fishbone-Moncrief (FM) plasma torus in equilibrium around a Kerr black hole of spin $a=0.9375M$ (\citealt{fishbonemoncrief1976}). The domain extends over the $r-\theta$ plane, with axisymmetry along the $\varphi$-direction. The equilibrium is perturbed such that the magneto-rotational instability (MRI) develops in the accretion disk. The instability causes the progressive accumulation of magnetic field flux in the vicinity of the event horizon, and a jet of plasma is launched from the polar region (Figure 13 in \citealt{porth2017}). In this regime, several interesting features manifest in the plasma flow surrounding the black hole. Regions of high turbulence, shock fronts, and magnetic flux tubes (``plasmoids'') are clearly visible in Figure \ref{fig:FMfar} (left panel), where we show the late-stage spatial distribution of typical GRMHD dimensionless quantities: the plasma $\beta=2p/B^2$ (with $p$ the thermal pressure and $B^2$ the magnitude of the magnetic field; top half) and the magnetization $\sigma=B^2/\rho$ (with $\rho$ the rest-mass density; bottom half).

In this setup, typically describing an accretion disk surrounding a supermassive black hole as in the Galactic Center, we consider the motion of ensembles of test particles. For simplicity, we assume that the  the evolution of the fields is much slower compared to the particle dynamics. Hence, we keep the electromagnetic fields fixed in time and we evolve ensembles of $10^4$, $10^5$, and $10^6$ particles until $t=10$ (in units of $GM/c^3$). The particles are initially distributed uniformly in the domain, with a random velocity drawn from a uniform distribution. The normalization of all quantities is such that the units of length, time, and mass reflect the typical parameters of Sgr A*, the supermassive black hole at the Galactic Center. This directly impacts the value of the particle charge-to-mass ratio $q/m$ used in our simulation: for our choice of parameters, protons are characterized by $q/m\sim10^8$, while for electrons $q/m\sim10^{11}$, in code units.

Such large values for the $q/m$ ratio pose a challenge for the numerical integrators, since the typical time scales of the gyro-motion around field lines scale as the gyro-frequency $\Omega_c\sim q/m$ (assuming a typical Lorentz factor $\Gamma\sim$1). Hence, resolving the gyro-motion typically requires a time step $\Delta t \sim m/q$. Employing such small time steps can become prohibitive for realistic charge-to-mass ratios; however, underresolving the gyration can introduce large errors in the calculations. One should note that these values of $q/m$ only apply to our specific choice of parametrization for the time, length, and mass units; different parametrizations could lead to less prohibitive values. Additionally, the scaling $\Delta t \sim m/q$ only holds as long as the particle Lorentz factor remains of order $\sim$1; when $\Gamma$ increases due to particle acceleration, the gyro-motion takes place over slower time scales, and the gyro-radius becomes larger (i.e.\ up to the scale of an MHD cell size), allowing for less restrictive values of $\Delta t$. For simplicity, in our example we assume a range of values $q/m=10^2,10^4,10^6$ and employ decreasing time steps $\Delta t=10^{-3},10^{-4},10^{-5}$ which ensure convergence of the iterative solution procedure for the IMR and modified Hamiltonian integrators. Note that, with this choice of parameters, the largest value of $q/m$ is two orders of magnitude smaller than the realistic charge-to-mass ratio for protons.

Figure \ref{fig:FMfar} shows the spatial distribution of the Lorentz factor computed from the GRMHD quantities (right panel, left half). This is compared to the Lorentz factor of $10^6$ particles with $q/m=10^4$ at time $t=5$ during the integration (right panel, right half), obtained with the modified Hamiltonian method. It can be seen that the distribution of the particle Lorentz factor matches that of the fluid, with the fastest particles found close to the central black hole. This can be attributed both to the presence of stronger electromagnetic fields closer to the black hole, and to the gravitational pull that attracts material more strongly in this region, ultimately causing the infall through the event horizon. Fast particles ($\gamma\sim 3$) are also found in the jets propagating along the central axis, where the Poynting-flux extracted from the black hole magnetosphere is converted into fluid kinetic energy by the action of the MHD Lorentz force. The overall match between particle and fluid Lorentz factor can be attributed to the  electric field $\textbf{E}=-\textbf{v}\times\textbf{B}$ calculated from the fluid velocity. Because the particle velocity does not initially equal the thermal velocity associated to the fluid bulk energy, this electric field acts on the particle motion until the particles adjust to the fluid velocity. The results obtained with the three integrators do not show significant differences, at least for this proof-of-principle application, proving that convergence is reached correctly in the description of the thermal motion of the particles. More evident differences are expected in long production runs, due to the secular growth of errors characterizing explicit methods.

\begin{figure}[!h]
\centering
\includegraphics[width=1\columnwidth, trim={0mm 0mm 0mm 0mm}, clip]{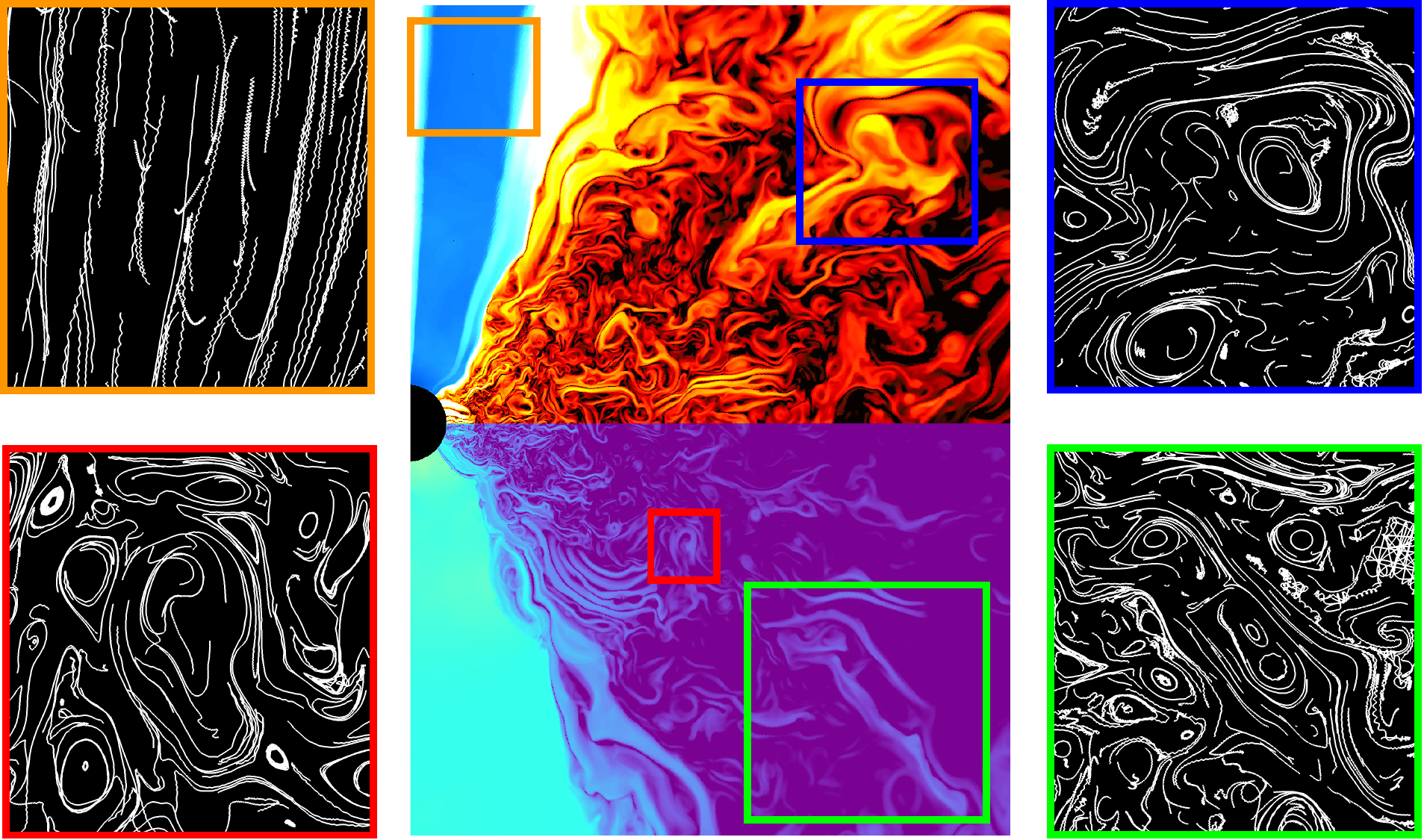}
\caption{Close-up view of the central region near the event horizon of the compact object. The blow-up panels show individual particle trajectories (calculated with the modified Hamiltonian scheme) in the highly turbulent region of the accretion flow and in the jet. Gyration around magnetic field lines can be seen more clearly for the faster particles traveling in the jet emerging from the black hole.}
\label{fig:FMclose}
\end{figure}

Finally, we monitor the motion of individual particles throughout the simulation. Figure \ref{fig:FMclose} presents a close-up view of the region closer to the central object, with several blow-up panels showing particle trajectories inside the turbulent accretion disk and in the jet emerging from the black hole. It can be seen how particles traveling through the highly turbulent region gyrate around the closed field lines of magnetic flux tubes. For these particles the gyro-motion is almost imperceptible, due to the large difference in scales between the size of the gyro-radius and that of the spatial region considered. For particles travelling in the jet, however, the gyration can be seen more clearly. This is confirmed by quantitative analysis of the simulation data. From the GRMHD data we measure typical values of the magnetic field strength in the disk and in the jet of $B_0^\mathrm{disk}\sim 10^{-1}$ and $B_0^\mathrm{jet}\sim 10^{-3}$, respectively. The test particle data shows typical Lorentz factors in the disk and in the jet of $\Gamma^\mathrm{disk}\sim 5$ and $\Gamma^\mathrm{jet}\sim 1.5$. As a result, the ratio of the typical gyrofrequency values $\Omega_C=qB_0/(m\Gamma)$ between particles in the disk and in the jet is $\Omega_C^\mathrm{disk}/\Omega_C^\mathrm{jet}\sim 30$, or equivalently, the characteristic gyroradius of particles in the disk region is $\sim 30$ times smaller than that of particles in the jet region.

The runs for $q/m=10^4$ are repeated with each integrator for $10^4$, $10^5$ and $10^6$ particles, in order to monitor the computational cost of the calculation. All runs are performed on 360 CPUs, measuring the time taken by the simulation to advance all particles for 1000 time steps. The results are presented in Table \ref{tab:FMtime}, where we list the run-time for 1000 integration steps, corresponding to 0.1 MHD times. For this choice of parameters, the measured run-time of the IMR scheme is very close to that of the RK4 scheme, being $\sim$1.5 times larger only in the $10^5$ particles case, and equal for the other runs. The modified Hamiltonian scheme, instead, exhibits a cost approximately 10 times large than the IMR method. When increasing the number of particles, we observe slightly super-linear scaling in the run-time. The run-time measured for the RK4 and IMR schemes in the $10^5$ particle run is comparable to the run-time of the MHD evolution, which we measure corresponding to approximately 1 minute per $0.1$ MHD times (in units of light-crossing time).

\begin{table}[!h]
\centering
\begin{tabular}{|c|c|c|c|}
\hline
\multirow{2}{*}{Number of particles} & \multicolumn{3}{c|}{Method} \\ \cline{2-4} 
 & RK4 & IMR & mHam \\ 
\hline 
$10^4$ & 12 & 12 & 120 \\ 
\hline
$10^5$ & 80 & 105 & 1080 \\ 
\hline
$10^6$ & 780 & 780 & 10320 \\ 
\hline
\end{tabular} 
\caption{Run-time (in seconds) for 1000 integration time steps with the three methods and increasing number of particles. All runs were performed with $q/m=10^4$ and $\Delta t=10^{-4}$ on 360 CPUs.}
\label{tab:FMtime}
\end{table}

The modified Hamiltonian method shows a slightly inferior scaling than the IMR and RK4 methods. As we discussed in \citetalias{bacchini2018a}, however, in production runs we foresee an implementation based on a combination of the methods presented here. While the RK4 method is expected to be generally discarded due to rapid degradation of accuracy, our results indicate that the IMR and (modified) Hamiltonian methods have similar accuracy, with the latter exhibiting desirable energy-conservation properties (e.g.\ zero energy errors in absence of electric fields), at the price of a higher computational cost. Hence, a dynamic switch between IMR and Hamiltonian schemes could be envisioned, that only selects the latter method when energy errors exceeding a chosen tolerance are detected. It is not straightforward to predict how such a combined approach would scale in massively parallel runs. However, under the assumption that the majority of particles can be treated with the IMR scheme, we can infer that the additional cost of the particles treated with the modified Hamiltonian scheme would not influence scaling and overall efficiency dramatically.

\section{Discussion and summary}
\label{sec:discussionsummary}
We presented a generalized framework for the numerical integration of charged particle trajectories in general relativity. The algorithm includes the full effect of the Lorentz force combined with the action of the spacetime curvature. We compared the performance of four numerical integrators, namely the standard explicit fourth-order Runge-Kutta (RK4) method, the second-order implicit midpoint rule (IMR) method, a new second-order implicit Hamiltonian method, and a ``modified'' Hamiltonian integrator that is suitable for simulations of test particles in GRMHD. The Hamiltonian integrator (and its modified version) is a direct extension of that presented in \cite{bacchini2018a} (Part I) which was constructed for the integration of pure geodesic motion. For testing purposes, we applied all schemes to several electromagnetic configurations in the Schwarzwschild, Kerr, and Kerr-Newman spacetimes. As an example of a practical application of the particle integrators, we simulated ensembles of test particles in the electromagnetic fields of a GRMHD simulation with \texttt{BHAC} (\citealt{porth2017}).

For charged particles in the Schwarzschild and Kerr spacetimes (Section \ref{sec:extfields}), we observed large numerical errors associated to the RK4 scheme. With this method, the trajectories of particles travelling in the Wald solution (Section \ref{sec:wald}) and in a dipole  solution (Section \ref{sec:DP}) rapidly experience a spurious damping of the gyration around magnetic field lines. Energy errors accumulating unboundedly throughout the simulations ultimately cause a spurious escape from bound orbits. Chaotic orbits appear to be more severely affected than regular paths, causing energy errors to grow faster. The IMR, Hamiltonian, and modified Hamiltonian schemes, instead, proved reliable in preserving the gyration motion, as well as keeping the particle orbit bounded until the end of the runs. The Hamiltonian scheme, formulated assuming the availability of an analytic four-potential, showed exceptional performance with the exact conservation (to machine precision) of both energy and angular momentum, which are invariants of the motion. The modified Hamiltonian scheme, by construction, preserves energy exactly in the limit of vanishing electric fields and for pure geodesic motion, and generally performs as well as the IMR scheme.

For a more quantitative comparison with known analytic solutions, we tested the integrators against unstable spherical orbits in the Kerr-Newman spacetime (Section \ref{sec:KN}). We derived a set of such orbits by relying on the integrability of orbits in this spacetime, and monitored the deviation of the simulated orbits from the initial radius $r_0$. Such a deviation is triggered by numerical errors, that destabilize the naturally unstable orbits until the particles escape to infinity. Less accurate methods cause an earlier development of the deviation, and hence an earlier release of the particles from the bound motion. Additionally, we checked for the conservation of angular momentum and the Carter constant, which together with the energy represent the invariants of the motion (aside from the norm of four-velocity, which is automatically preserved in the chosen 3+1 split framework). We found that, in all cases, the energy-conserving Hamiltonian scheme performs much better than the RK4, IMR, and modified Hamiltonian schemes, triggering the release of the particle from the orbit much later through the simulation. For the smallest time step used, we observed a factor $\sim$10 of difference between the release times of the Hamiltonian and the RK4 schemes, indicating a much higher stability of the former with respect to the accumulation of numerical errors. The error in the conserved quantities reflects this property, with the Hamiltonian scheme conserving energy, angular momentum, and the Carter constant to machine precision at all times. Similar results were found for pure geodesic motion in \citetalias{bacchini2018a}.

In order to exemplify the application of the integrators in a more practical context, we assessed the performance of all schemes in simulating the motion of particles in grid-defined electromagnetic fields, generally employed in GRMHD codes. We repeated the simulation of test orbits from the previous Sections and monitored the effect of interpolation of $D^i$ and $B^i$ or $A_\mu$ (depending on the scheme) on the accuracy of the results. We found that the use of grid-defined fields only mildly affects the performance of the RK4 and IMR schemes, while it substantially changes the behavior of the Hamiltonian scheme. In the most general case, the latter inevitably loses its energy-conserving character, by exhibiting an error in the energy of the same order of that introduced by the IMR scheme. Angular momentum, instead, is still conserved to machine accuracy due to the formulation of the equations of motion. On the contrary, the modified Hamiltonian scheme retains exact energy conservation, in the case of vanishing electric fields, even when grid-defined quantities are employed. However, it does not preserve angular momentum to machine precision, with errors of the same order of those observed for the IMR scheme. The IMR, Hamiltonian, and modified Hamiltonian schemes perform better than the RK4 scheme, which displays a secular growth of energy errors and spurious release of the particle from bound orbits. The interpolation step is an additional (and inevitable) source of inaccuracy for all schemes, with the error in the conserved quantities decreasing approximately by one order of magnitude when doubling the spatial resolution of the grid.

As an astrophysically-relevant example application, we carried out a set of test particle runs with the GRMHD code \texttt{BHAC} (\citealt{porth2017}), where we implemented all particle integrators discussed here. \texttt{BHAC} employs the dynamic fields $D^i$ and $B^i$ as electromagnetic variables, in place of the full four-potential $A_\mu$. As a consequence, the original Hamiltonian method cannot be employed, and we rely instead on the modified Hamiltonian scheme, which was constructed for this purpose. The GRMHD simulation was initialized with a Fishbone-Moncrief equilibrium (\citealt{fishbonemoncrief1976}) for a magnetized plasma torus around a Kerr black hole. During the GRMHD evolution, the magneto-rotational instability causes the launching of a highly magnetized jet from the polar region, and generates turbulence, shocks, and flux tubes in the accretion flow (Figure \ref{fig:FMfar}; see \citealt{porth2017}). This evolved state was employed as a fixed background for the simulation of up to $10^6$ particles with initial uniform distribution and random velocity. We applied normalization parameters such that the largest charge-to-mass ratio of the particles is $q/m=10^6$ in code units (the physical value for protons is $q/m\sim10^8$  in code units). During the simulation, we observed that the particle Lorentz factor, initially randomly distributed at $t=0$, matches that of the fluid at $t=5$, with faster particles found close to the infall region around the event horizon and in the accelerating jet. Due to the ideal nature of the GRMHD description of the plasma, where no parallel electric fields can exist, we did not observe (correctly) the acceleration of particles to non-thermal energies. The analysis of individual trajectories showed particles following the magnetic field lines, and remaining trapped inside the closed field lines of flux tubes in the turbulent accretion region. Due to the higher strength of the magnetic field, in this region the gyro-motion of the particles becomes relegated to length scales that are much smaller than the system size. In the accelerating jet, where the Lorentz factor is higher, the gyro-radius increases and the particles can be observed traveling along helical paths.

The run-time analysis shows that the IMR method has a computational cost only slightly larger (by a factor $\sim$1.5) than the RK4 method. The Hamiltonian scheme, instead, is considerably more expensive (by a factor $\sim$10) than the IMR scheme. The scaling with the increase in the number of particles shows superlinearity, indicating that the ideal workload per processor can be further increased. The results are encouraging in the perspective of running massively parallel simulations with large ensembles of particles on superclusters (up to $10^8$), e.g.\ for statistics of particle acceleration and for PiC simulations.

The results from the test runs and from the proof-of-principle simulations of test particles in GRMHD indicate that relatively simple implicit methods like the IMR scheme are reliable and adequate for the numerical simulation of charged particles in curved spacetimes combined with electromagnetic fields. Explicit methods such as the RK4 scheme prove unreliable due to unbounded accumulation of errors in the conserved quantities and the incapability to describe gyrating motion (\citealt{qin2013}). The new Hamiltonian method and its modified version presented here provide more physically accurate results than the IMR scheme, conserving integrals of the motion to very high accuracy. For simulations of particles in test electromagnetic fields, where the four-potential is analytically available, the second-order Hamiltonian scheme exhibits high stability and accuracy surpassing those of higher-order methods such as the RK4. When only the dynamic fields $D^i$ and $B^i$ are available, as is the case in typical GRMHD simulations, such conservation properties are only partially retained in the modified Hamiltonian scheme, with preservation of energy ensured only in the limit of vanishing electric fields. Therefore, in such simulations, the application of the IMR scheme may prove equally satisfactory, and the modified Hamiltonian scheme could serve as a robust backup strategy. In future extensions to full GR-PiC algorithms, the modified Hamiltonian scheme could also be employed to construct globally energy-conserving methods such as those developed for special relativistic PiC simulations (\citealt{lapentamarkidis2011}; \citealt{markidislapenta2011}).

The analysis presented here has direct applications to the study of particle acceleration in astrophysical scenarios such as the magnetosphere of compact objects. Supermassive black holes such as Sgr A* at the galactic center, or the active nucleus of the M87 galaxy, are characterized by a strongly nonthermal radiation spectrum. Particle methods are the ideal tool for exploring the nonthermal features associated with particle acceleration mechanisms, which cannot be accurately captured by GRMHD methods. A combination of resistive GRMHD simulations, which are necessary to allow for acceleration mechanisms such as magnetic reconnection (Ripperda et al., in prep.), and particle methods, that can accurately model the particle motion in these environments, is the key strategy for future multi-scale simulations of black holes and neutron stars. In this way, it is possible to self-consistently explore previously unreachable time, length, and energy scales, with a direct impact on our understanding of the microscopic dynamics of accretion flows around compact objects.

Despite the promising results shown by proof-of-principle test particle simulations such as those presented here, care is needed in interpreting results from test particles. In a GRMHD calculation, no information is available on scales smaller than the grid spacing. If the particle gyromotion is such that the Larmor radius is smaller than the typical cell size, errors will arise in the acceleration and trajectory of test particles, due to unresolved physics. A possible approach (aside from increasing the spatial resolution) could be then to retain information only from test particles whose gyration is resolved by the grid spacing, e.g.\ by checking the local gyrofrequency computed from the GRMHD fields.

A second limitation is the lack of feedback of the particles onto the electromagnetic fields, which is retained in a fully kinetic code. The test particles can, however, be employed as a diagnostic tool, both for identifying acceleration sites (that indicate dissipative processes, such as reconnection) and for investigating the acceleration mechanisms. In this context, test particles have been successfully applied in special relativistic resistive MHD runs, showing that aspects of the underlying acceleration mechanisms can be reproduced in good agreement with PiC runs (see \citealt{ripperda2018b}). PiC simulations relevant for accretion flow plasmas in locally flat spacetime regions (e.g.\ \citealt{rowan2017}) and in presence of strong spacetime curvature (\citealt{levinsoncerutti2018}; \citealt{parfrey2018}) could provide a way to quantify the differences between test particle runs and their fully kinetic counterparts. Ultimately, the true nature of microscopic plasma processes in accretion disk environments will have to be studied with kinetic codes, or a combination of MHD and kinetic methods. Coupled PiC-MHD runs have recently been performed in Newtonian setups (\citealt{daldorff2014}; \citealt{markidis2014}; \citealt{toth2016}; \citealt{chen2017}; \citealt{makwana2018}; \citealt{markidis2018}), where a PiC simulation box is embedded in an MHD calculation, in regions where collisionless physics is supposed to be important. The implementation of test particles in curved spacetimes in a framework such as \texttt{BHAC} is a necessary first step in this direction.

\section*{Acknowledgements}
BR and FB were supported by projects GOA/2015-014 (2014-2018 KU Leuven) and the Interuniversity Attraction Poles Programme by the Belgian Science Policy Office (IAP P7/08 CHARM). FB is also supported by the Research Fund KU Leuven and Space Weaves RUN project. BR and OP are supported by the ERC synergy grant ``BlackHoleCam: Imaging the Event Horizon of Black Holes'' (Grant No. 610058). LS acknowledges support from DoE DE-SC0016542, NASA Fermi NNX-16AR75G, NASA ATP NNX-17AG21G, NSF ACI-1657507, and NSF AST-1716567. The computational resources and services used in this work were provided by the VSC (Flemish Supercomputer Center), funded by the Research Foundation Flanders (FWO) and the Flemish Government - department EWI. FB would like to thank Eva Hackmann for precious insight into the unstable particle orbits in the Kerr-Newman spacetime, and Ziri Younsi and Martin Kolo\v{s} for useful discussions throughout the development of this work.



\appendix

\section{Discrete energy-conserving Hamiltonian scheme}
\label{app:ham}
Starting from the Hamiltonian
\begin{equation}
  H(\textbf{x},\pivec) = \alpha\sqrt{1 + \gamma^{ij} \left(\pi_i-\frac{q}{m}A_i\right) \left(\pi_j-\frac{q}{m}A_j\right)} - \beta^k\left(\pi_k-\frac{q}{m}A_k\right) - \frac{q}{m}A_0,
 \label{eq:hamiltonianapp}
\end{equation}
where $A_i$ is a function of $(x^1,x^2,x^3)$, for a system of 3 equations for $x^i$ and 3 equations for $\pi_i$, we define discrete operators $\Delta_i^x$ and $\Delta^i_\pi$ such that
\begin{equation}
 \frac{\Delta H(\xvec,\pivec)}{\Delta t} = \frac{\Delta^x_i H(\textbf{x},\pivec)}{\Delta  x^i}\frac{\Delta x^i}{\Delta t}+\frac{\Delta_\pi^i H(\textbf{x},\pivec)}{\Delta \pi_i}\frac{\Delta \pi_i}{\Delta t} = 0,
 \label{eq:hamenergyconddiscapp}
\end{equation}

Here, $\Delta$ indicates finite differencing over a full time step $\Delta t$, i.e.\ $\Delta x^i=x^{i,n+1}-x^{i,n}$, $\Delta \pi_i=\pi_i^{n+1}-\pi_i^{n}$, and $\Delta H(\xvec,\pivec)=H(\xvec^{n+1},\pivec^{n+1})-H(\xvec^n,\pivec^n)$. In order for the operators $\Delta_i^x$ and $\Delta^i_\pi$ to respect the condition \eqref{eq:hamenergyconddiscapp}, their action on a generic function $f(\xvec,\pivec)$ must be such that a discrete chain rule applies,
\begin{equation}
 \Delta f(\xvec,\pivec) = f(\xvec^{n+1},\pivec^{n+1})-f(\xvec^n,\pivec^n) = \Delta^x_i f(\xvec,\pivec) \Delta x^i + \Delta_\pi^i f(\xvec,\pivec) \Delta \pi_i.
\end{equation}

In order to construct suitable discrete operators, we refer to the work by \cite{fengqin} and to the expressions given in \citetalias{bacchini2018a} and we write, for each component of $x^i$ and $\pi_i$, the discretized Hamiltonian equations
\begin{equation}
 \begin{aligned}
 \frac{\Delta x^{1}}{\Delta t} = \frac{1}{6\Delta\pi_1} & \left[ 
 H(x^{1,n+1},x^{2,n},x^{3,n},\pi_1^{n+1},\pi_2^{n},\pi_3^{n})-H(x^{1,n+1},x^{2,n},x^{3,n},\pi_1^{n},\pi_2^{n},\pi_3^{n}) \right. \\ 
 & +H(x^{1,n+1},x^{2,n+1},x^{3,n+1},\pi_1^{n+1},\pi_2^{n+1},\pi_3^{n+1})-H(x^{1,n+1},x^{2,n+1},x^{3,n+1},\pi_1^{n},\pi_2^{n+1},\pi_3^{n+1}) \\
 & +H(x^{1,n},x^{2,n+1},x^{3,n+1},\pi_1^{n+1},\pi_2^{n+1},\pi_3^{n+1})-H(x^{1,n},x^{2,n+1},x^{3,n+1},\pi_1^{n},\pi_2^{n+1},\pi_3^{n+1}) \\
 & +H(x^{1,n+1},x^{2,n},x^{3,n+1},\pi_1^{n+1},\pi_2^{n},\pi_3^{n+1})-H(x^{1,n+1},x^{2,n},x^{3,n+1},\pi_1^{n},\pi_2^{n},\pi_3^{n+1}) \\
 & +H(x^{1,n},x^{2,n},x^{3,n},\pi_1^{n+1},\pi_2^{n},\pi_3^{n})-H(x^{1,n},x^{2,n},x^{3,n},\pi_1^{n},\pi_2^{n},\pi_3^{n}) \\
 & \left. +H(x^{1,n},x^{2,n},x^{3,n+1},\pi_1^{n+1},\pi_2^{n},\pi_3^{n+1})-H(x^{1,n},x^{2,n},x^{3,n+1},\pi_1^{n},\pi_2^{n},\pi_3^{n+1})\right],
 \end{aligned}
 \label{eq:poshamdisc1}
\end{equation}
\begin{equation}
 \begin{aligned}
 \frac{\Delta x^2}{\Delta t} = \frac{1}{6\Delta\pi_2} & \left[
 H(x^{1,n+1},x^{2,n+1},x^{3,n},\pi_1^{n+1},\pi_2^{n+1},\pi_3^{n})-H(x^{1,n+1},x^{2,n+1},x^{3,n},\pi_1^{n+1},\pi_2^{n},\pi_3^{n})\right. \\
 & +H(x^{1,n},x^{2,n+1},x^{3,n},\pi_1^{n},\pi_2^{n+1},\pi_3^{n})-H(x^{1,n},x^{2,n+1},x^{3,n},\pi_1^{n},\pi_2^{n},\pi_3^{n}) \\
 & +H(x^{1,n},x^{2,n},x^{3,n},\pi_1^{n},\pi_2^{n+1},\pi_3^{n})-H(x^{1,n},x^{2,n},x^{3,n},\pi_1^{n},\pi_2^{n},\pi_3^{n}) \\
 & +H(x^{1,n+1},x^{2,n+1},x^{3,n+1},\pi_1^{n+1},\pi_2^{n+1},\pi_3^{n+1})-H(x^{1,n+1},x^{2,n+1},x^{3,n+1},\pi_1^{n+1},\pi_2^{n},\pi_3^{n+1}) \\
 & +H(x^{1,n+1},x^{2,n},x^{3,n},\pi_1^{n+1},\pi_2^{n+1},\pi_3^{n})-H(x^{1,n+1},x^{2,n},x^{3,n},\pi_1^{n+1},\pi_2^{n},\pi_3^{n}) \\
 & \left. +H(x^{1,n+1},x^{2,n},x^{3,n+1},\pi_1^{n+1},\pi_2^{n+1},\pi_3^{n+1})-H(x^{1,n+1},x^{2,n},x^{3,n+1},\pi_1^{n+1},\pi_2^{n},\pi_3^{n+1})\right],
 \end{aligned}
 \label{eq:poshamdisc2}
\end{equation}
\begin{equation}
 \begin{aligned}
 \frac{\Delta x^3}{\Delta t} = \frac{1}{6\Delta\pi_3} & \left[
 H(x^{1,n+1},x^{2,n+1},x^{3,n+1},\pi_1^{n+1},\pi_2^{n+1},\pi_3^{n+1})-H(x^{1,n+1},x^{2,n+1},x^{3,n+1},\pi_1^{n+1},\pi_2^{n+1},\pi_3^{n}) \right. \\
 & + H(x^{1,n},x^{2,n+1},x^{3,n+1},\pi_1^{n},\pi_2^{n+1},\pi_3^{n+1})-H(x^{1,n},x^{2,n+1},x^{3,n+1},\pi_1^{n},\pi_2^{n+1},\pi_3^{n}) \\
 & +H(x^{1,n},x^{2,n+1},x^{3,n},\pi_1^{n},\pi_2^{n+1},\pi_3^{n+1})-H(x^{1,n},x^{2,n+1},x^{3,n},\pi_1^{n},\pi_2^{n+1},\pi_3^{n}) \\
 & +H(x^{1,n},x^{2,n},x^{3,n+1},\pi_1^{n},\pi_2^{n},\pi_3^{n+1})-H(x^{1,n},x^{2,n},x^{3,n+1},\pi_1^{n},\pi_2^{n},\pi_3^{n}) \\
 & +H(x^{1,n+1},x^{2,n+1},x^{3,n},\pi_1^{n+1},\pi_2^{n+1},\pi_3^{n+1})-H(x^{1,n+1},x^{2,n+1},x^{3,n},\pi_1^{n+1},\pi_2^{n+1},\pi_3^{n}) \\
 & \left. +H(x^{1,n},x^{2,n},x^{3,n},\pi_1^{n},\pi_2^{n},\pi_3^{n+1})-H(x^{1,n},x^{2,n},x^{3,n},\pi_1^{n},\pi_2^{n},\pi_3^{n})\right],
 \end{aligned}
 \label{eq:poshamdisc3}
\end{equation}
\begin{equation}
 \begin{aligned}
 \frac{\Delta\pi_1}{\Delta t} = \frac{1}{6\Delta x^1} & \left[ 
 H(x^{1,n+1},x^{2,n},x^{3,n},\pi_1^{n},\pi_2^{n},\pi_3^{n})-H(x^{1,n},x^{2,n},x^{3,n},\pi_1^{n},\pi_2^{n},\pi_3^{n}) \right. \\
 & +H(x^{1,n+1},x^{2,n+1},x^{3,n+1},\pi_1^{n},\pi_2^{n+1},\pi_3^{n+1})-H(x^{1,n},x^{2,n+1},x^{3,n+1},\pi_1^{n},\pi_2^{n+1},\pi_3^{n+1}) \\
 & +H(x^{1,n+1},x^{2,n+1},x^{3,n+1},\pi_1^{n+1},\pi_2^{n+1},\pi_3^{n+1})-H(x^{1,n},x^{2,n+1},x^{3,n+1},\pi_1^{n+1},\pi_2^{n+1},\pi_3^{n+1}) \\
 & +H(x^{1,n+1},x^{2,n},x^{3,n+1},\pi_1^{n},\pi_2^{n},\pi_3^{n+1})-H(x^{1,n},x^{2,n},x^{3,n+1},\pi_1^{n},\pi_2^{n},\pi_3^{n+1}) \\
 & +H(x^{1,n+1},x^{2,n},x^{3,n},\pi_1^{n+1},\pi_2^{n},\pi_3^{n})-H(x^{1,n},x^{2,n},x^{3,n},\pi_1^{n+1},\pi_2^{n},\pi_3^{n}) \\
 & \left. +H(x^{1,n+1},x^{2,n},x^{3,n+1},\pi_1^{n+1},\pi_2^{n},\pi_3^{n+1})-H(x^{1,n},x^{2,n},x^{3,n+1},\pi_1^{n+1},\pi_2^{n},\pi_3^{n+1})\right],
 \end{aligned}
 \label{eq:velhamdisc1}
\end{equation}
\begin{equation}
 \begin{aligned}
 \frac{\Delta\pi_2}{\Delta t} = \frac{1}{6\Delta x^2} & \left[
 H(x^{1,n+1},x^{2,n+1},x^{3,n},\pi_1^{n+1},\pi_2^{n},\pi_3^{n})-H(x^{1,n+1},x^{2,n},x^{3,n},\pi_1^{n+1},\pi_2^{n},\pi_3^{n}) \right. \\ 
 & +H(x^{1,n},x^{2,n+1},x^{3,n},\pi_1^{n},\pi_2^{n},\pi_3^{n})-H(x^{1,n},x^{2,n},x^{3,n},\pi_1^{n},\pi_2^{n},\pi_3^{n}) \\
 & + H(x^{1,n},x^{2,n+1},x^{3,n},\pi_1^{n},\pi_2^{n+1},\pi_3^{n})-H(x^{1,n},x^{2,n},x^{3,n},\pi_1^{n},\pi_2^{n+1},\pi_3^{n}) \\
 & +H(x^{1,n+1},x^{2,n+1},x^{3,n+1},\pi_1^{n+1},\pi_2^{n},\pi_3^{n+1})-H(x^{1,n+1},x^{2,n},x^{3,n+1},\pi_1^{n+1},\pi_2^{n},\pi_3^{n+1}) \\
 & +H(x^{1,n+1},x^{2,n+1},x^{3,n},\pi_1^{n+1},\pi_2^{n+1},\pi_3^{n})-H(x^{1,n+1},x^{2,n},x^{3,n},\pi_1^{n+1},\pi_2^{n+1},\pi_3^{n}) \\
 & \left. + H(x^{1,n+1},x^{2,n+1},x^{3,n+1},\pi_1^{n+1},\pi_2^{n+1},\pi_3^{n+1})-H(x^{1,n+1},x^{2,n},x^{3,n+1},\pi_1^{n+1},\pi_2^{n+1},\pi_3^{n+1}) \right],
 \end{aligned}
 \label{eq:velhamdisc2}
\end{equation}
\begin{equation}
 \begin{aligned}
 \frac{\Delta\pi_3}{\Delta t} = \frac{1}{6\Delta x^3} & \left[
 H(x^{1,n+1},x^{2,n+1},x^{3,n+1},\pi_1^{n+1},\pi_2^{n+1},\pi_3^{n})-H(x^{1,n+1},x^{2,n+1},x^{3,n},\pi_1^{n+1},\pi_2^{n+1},\pi_3^{n}) \right. \\
 & +H(x^{1,n},x^{2,n+1},x^{3,n+1},\pi_1^{n},\pi_2^{n+1},\pi_3^{n})-H(x^{1,n},x^{2,n+1},x^{3,n},\pi_1^{n},\pi_2^{n+1},\pi_3^{n}) \\
 & +H(x^{1,n},x^{2,n+1},x^{3,n+1},\pi_1^{n},\pi_2^{n+1},\pi_3^{n+1})-H(x^{1,n},x^{2,n+1},x^{3,n},\pi_1^{n},\pi_2^{n+1},\pi_3^{n+1}) \\
 & +H(x^{1,n},x^{2,n},x^{3,n+1},\pi_1^{n},\pi_2^{n},\pi_3^{n})-H(x^{1,n},x^{2,n},x^{3,n},\pi_1^{n},\pi_2^{n},\pi_3^{n}) \\
 & +H(x^{1,n+1},x^{2,n+1},x^{3,n+1},\pi_1^{n+1},\pi_2^{n+1},\pi_3^{n+1})-H(x^{1,n+1},x^{2,n+1},x^{3,n},\pi_1^{n+1},\pi_2^{n+1},\pi_3^{n+1}) \\
 & \left. +H(x^{1,n},x^{2,n},x^{3,n+1},\pi_1^{n},\pi_2^{n},\pi_3^{n+1})-H(x^{1,n},x^{2,n},x^{3,n},\pi_1^{n},\pi_2^{n},\pi_3^{n+1}) \right].
 \end{aligned}
 \label{eq:velhamdisc3}
\end{equation}
One can verify that with such definitions of $\Delta_i^x$ and $\Delta_\pi^i$, the condition \eqref{eq:hamenergyconddiscapp} is respected. The equations above are nonlinear and must be solved with an iterative method. It is clear that difficulties in the solution can arise whenever the increments, $\Delta x^i = x^{i,n+1}-x^{i,n}$ and $\Delta\pi_i = \pi_i^{n+1}-\pi_i^n$, tend to zero. This issue must be handled carefully throughout the computation, as the results can get severely affected by the behavior of the solution around these critical points.

In order to avoid numerical singularities in the system of equations \eqref{eq:poshamdisc1}-\eqref{eq:velhamdisc3}, we can rewrite the difference equations in a more convenient form. In a procedure similar to that adopted in \citetalias{bacchini2018a}, a series of manipulations leads to the alternative expression of the discrete position equation,
\begin{equation}
 \frac{\Delta x^{i}}{\Delta t} = \frac{1}{6}\sum^6 \left[ \alpha\frac{\gamma^{ii}(\pi_i^{n+1}+\pi_i^n-2qA_i/m) + 2\gamma^{ij}(\pi_j-qA_j/m)+2\gamma^{ik}(\pi_k-qA_k/m)}{\sqrt{1+\ce{^\pi_i\Pi^{n+1}}}+\sqrt{1+\ce{^\pi_i\Pi^n}}} -\beta^i \right], \qquad j,k\neq i,
 \label{eq:poshamdiscsimple}
\end{equation}
where
\begin{equation}
 \Pi=\gamma^{jk}(\pi_j-qA_j/m)(\pi_k-qA_k/m),
\end{equation}
and the notation $\ce{^\pi_i\Pi^\tau}$ indicates that in evaluating $\Pi$, the component $\pi_i$ should be taken at time level $\tau=n,n+1$. Each term of the summation in equation \eqref{eq:poshamdiscsimple} evaluates the other components $\pi_{j\neq i}$, as well as all components $x^j$, at the time levels specified by each difference term in equations \eqref{eq:poshamdisc1}-\eqref{eq:velhamdisc3}. This form of the difference equations avoids singularities associated to $\Delta\pi_i=0$, and is therefore more convenient to use in an iterative solution procedure.

A similar simplification procedure, applied to the momentum equations \eqref{eq:velhamdisc1}-\eqref{eq:velhamdisc3}, yields
\begin{equation}
\begin{aligned}
 \frac{\Delta\pi_i}{\Delta t} = \frac{1}{6} \sum^6 & \left\{ -\frac{1}{2}\left(\sqrt{1+\ce{^x_i \Pi^{n+1}}}+\sqrt{1+\ce{^x_i \Pi^{n}}}\right) \frac{\ce{^x_i \alpha^{n+1}}-\ce{^x_i \alpha^{n}}}{\Delta x^i} \right. \\
 & -\frac{\ce{^x_i \alpha^{n+1}}+\ce{^x_i \alpha^{n}}}{2}
 \frac{[\pi_j+(q/m)\ce{^x_i\it{A}^{n+1}_j}][\pi_k+(q/m)\ce{^x_i\it{A}^{n+1}_k}] + [\pi_j+(q/m)\ce{^x_i\it{A}^{n}_j}][\pi_k+(q/m)\ce{^x_i\it{A}^{n}_k}]}{\sqrt{1+\ce{^x_i \Pi^{n+1}}}+\sqrt{1+\ce{^x_i \Pi^{n}}}} \frac{\ce{^x_i \gamma^j^k^{,n+1}}-\ce{^x_i \gamma^j^k^{,n}}}{\Delta x^i} \\
 & + \left(\pi_j-\frac{q}{m}\frac{\ce{^x_i \it{A}^{n+1}_j} + \ce{^x_i \it{A}^{n}_j}}{2}\right) \frac{\ce{^x_i \beta^j^{,n+1}}-\ce{^x_i \beta^j^{,n}}}{\Delta x^i} \\
 & + \frac{q}{2m}\left[ 
 \frac{1}{2}(\ce{^x_i \alpha^{n+1}}+\ce{^x_i \alpha^{n}})
 (\ce{^x_i \gamma^j^k^{,n+1}}+\ce{^x_i \gamma^j^k^{,n}})
 \frac{2\pi_k-q(\ce{^x_i \it{A}^{n+1}_k}+\ce{^x_i \it{A}^{n}_k})/m}
 {\sqrt{1+\ce{^x_i \Pi^{n+1}}}+\sqrt{1+\ce{^x_i \Pi^{n}}}}
 - (\ce{^x_i \beta^j^{,n+1}}+\ce{^x_i \beta^j^{,n}}) \right] 
 \frac{\ce{^x_i \it{A}^{n+1}_j}-\ce{^x_i \it{A}^{n}_j}}{\Delta x^i}\\
 & \left. +\frac{q}{m}\frac{\ce{^x_i \it{A}^{n+1}_0}-\ce{^x_i \it{A}^{n}_0}}{\Delta x^i}\right\},
  \label{eq:velhamdiscsimple}
\end{aligned}
\end{equation}
where the same notation of equation \eqref{eq:poshamdiscsimple} applies here to $\alpha$, $\beta^i$, $\gamma^{ij}$, and $A_i$. Here, the factor $1/\Delta x^i$ does not vanish, thus problems related to singularities may still arise. However, for $\Delta x^i = x^{i,n+1}-x^{i,n}\rightarrow 0$, the incremental ratios in the equations above reduce to analytic derivatives, as
\begin{equation}
\begin{aligned}
 \frac{\Delta\pi_i}{\Delta t} = \frac{1}{6} \sum^6 & \left\{ -\frac{1}{2}\left(\sqrt{1+\ce{^x_i \Pi^{n+1}}}+\sqrt{1+\ce{^x_i \Pi^{n}}}\right) 
 \ce{^x_i (\partial_i\alpha)^n} \right. \\
 & -\frac{\ce{^x_i \alpha^{n+1}}+\ce{^x_i \alpha^{n}}}{2}
 \frac{[\pi_j+(q/m)\ce{^x_i\it{A}^{n+1}_j}][\pi_k+(q/m)\ce{^x_i\it{A}^{n+1}_k}] + [\pi_j+(q/m)\ce{^x_i\it{A}^{n}_j}][\pi_k+(q/m)\ce{^x_i\it{A}^{n}_k}]}{\sqrt{1+\ce{^x_i \Pi^{n+1}}}+\sqrt{1+\ce{^x_i \Pi^{n}}}} \ce{^x_i (\partial_i\gamma^j^k)^n} \\
 & + \left(\pi_j-\frac{q}{m}\frac{\ce{^x_i \it{A}^{n+1}_j} + \ce{^x_i \it{A}^{n}_j}}{2}\right) \ce{^x_i (\partial_i\beta^j)^n} \\
 & + \frac{q}{2m}\left[ 
 \frac{1}{2}(\ce{^x_i \alpha^{n+1}}+\ce{^x_i \alpha^{n}})
 (\ce{^x_i \gamma^j^k^{,n+1}}+\ce{^x_i \gamma^j^k^{,n}})
 \frac{2\pi_k-q(\ce{^x_i \it{A}^{n+1}_k}+\ce{^x_i \it{A}^{n}_k})/m}
 {\sqrt{1+\ce{^x_i \Pi^{n+1}}}+\sqrt{1+\ce{^x_i \Pi^{n}}}}
 - (\ce{^x_i \beta^j^{,n+1}}+\ce{^x_i \beta^j^{,n}}) \right] 
 \ce{^x_i (\partial_i\it{A}_j)^n}\\
 & \left. +\frac{q}{m}\ce{^x_i (\partial_i\it{A}_0)^n} \right\},
  \label{eq:velhamcontsimple}
\end{aligned}
\end{equation}
hence the solution procedure must be handled by substituting (\ref{eq:velhamdiscsimple}) with (\ref{eq:velhamcontsimple}) when the difference $x^{i,n+1}-x^{i,n}<\delta$, where $\delta$ is a prescribed tolerance. A typical choice for such a threshold is $\delta\sim\sqrt{\varepsilon}$, where $\varepsilon$ is the chosen round-off precision (\citealt{press}).

\section{Unstable spherical orbits in the Kerr-Newman spacetime}
\label{app:KN}
We consider the equations of motion for each coordinate of the Kerr-Newman spacetime (\citealt{hackmannxu2013}),
\begin{equation}
 \rho^2 u^0 = \frac{r^2+a^2}{\Delta}\mathcal{R} - a\mathcal{T},
 \label{eq:utKN}
\end{equation}
\begin{equation}
 \left(\rho^2 u^r\right)^2 = \mathcal{R}^2 - (r^2 + K)\Delta,
 \label{eq:urKN}
\end{equation}
\begin{equation}
 \left(\rho^2 u^\theta\right)^2 = K - a^2\cos^2\theta - \frac{\mathcal{T}^2}{\sin^2\theta},
 \label{eq:uthKN}
\end{equation}
\begin{equation}
 \rho^2 u^\varphi = \frac{a\mathcal{R}}{\Delta} - \frac{\mathcal{T}}{\sin^2\theta},
 \label{eq:uphiKN}
\end{equation}
where $\mathcal{R} = (r^2+a^2)E-aL-qQr/m$, $\mathcal{T} = aE\sin^2\theta-L+qP\cos\theta/m$, and $K=C+(aE-L)^2$. For pure geodesic motion in Kerr spacetime, a set of unstable spherical photon orbits was derived by \cite{teo2003}. Similarly, in the case of charged particles around Kerr-Newman black holes, spherical orbits can be found by imposing $u^r=du^r/d\tau=0$. If this condition is fulfilled, the particle is confined on a sphere of radius $r_0$ along an orbit that can be stable or unstable against radial perturbations.

One can verify that the equations above completely characterize the particle motion in terms of the quantities $E$, $L$, $a$, $Q$, $P$, $K$, $q/m$. With reference to Figures 5-8 in \cite{hackmannxu2013}, we can fix the quantities $K$, $a$, $Q^2+P^2$, $qQ/m$ and derive a set of values for $E$, $L$, $r_0$ that correspond to an unstable spherical orbit. To do so, we choose an appropriate value of $L$ and we solve the condition $u^r=du^r/d\tau=0$ from equation \eqref{eq:urKN} for $E$ and $r_0$. In general, this produces multiple solutions, so we choose those corresponding to unstable spherical orbits at $r_0>r_+$. With the complete set of parameters, we can initialize the particle motion at position $(r_0,\theta_0,0)$. The initial $\theta_0$ can be calculated as a turning point of the motion from equation \eqref{eq:uthKN} by imposing $u^\theta=0$. With this particular choice of $r_0$ and $\theta_0$, we have the initial components $u_r=u_\theta=0$. Then, using equations \eqref{eq:utKN} and \eqref{eq:uphiKN}, we obtain an initial value for $u_\varphi$.

\bibliographystyle{apalike}

\label{lastpage}
\end{document}